\documentclass[preprint2]{aastex}
\usepackage{style1}
\usepackage{url}
\usepackage{verbatim}
\RequirePackage{color}
\usepackage{amsmath,amssymb}
\usepackage{epsfig}
\usepackage{graphicx}
\usepackage[english]{babel}
\newcommand{\emaila}{dstaicova@phys.uni-sofia.bg}
\newcommand{\emailb}{fiziev@phys.uni-sofia.bg}
\definecolor{red}{RGB}{250,0,00}
\definecolor{gold}{rgb}{0.85,.66,0}
\definecolor{purple}{RGB}{193, 56, 193}

\usepackage[colorlinks]{hyperref}
\begin{document}

\title{The Spectrum of Electromagnetic Jets from Kerr Black Holes and Naked
Singularities in the Teukolsky Perturbation Theory}

\shorttitle{Electromagnetic Jets from Kerr Black Holes and Naked
Singularities}

%\shortauthors{}

\author{Denitsa Staicova \altaffilmark{1} and Plamen Fiziev \altaffilmark{1,2}}
\altaffiltext{1}{Department of Theoretical Physics, Sofia University
"St. Kliment Ohridski",  5 James Bourchier Blvd., 1164 Sofia, Bulgaria}
\altaffiltext{2}{BLTF, JINR, Dubna, 141980 Moscow Region, Russia}
\emaila

\emailb

\maketitle

\hskip 1.truecm

\begin{abstract}

We give a new theoretical basis for examination of the presence of the Kerr black hole (KBH)
or the Kerr naked singularity (KNS) in the central engine of different astrophysical
objects around which astrophysical jets are typically formed:
X-ray binary systems, gamma ray bursts (GRBs), active galactic nuclei (AGN), etc.

Our method is based on the study of the exact solutions of the Teukolsky master equation
for electromagnetic perturbations of the Kerr metric.
By imposing original boundary conditions on the solutions
so that they describe a collimated electromagnetic outflow,
we obtain the spectra of possible {\em primary jets} of radiation,
introduced here for the first time.
The theoretical spectra of primary electromagnetic jets are calculated numerically.

Our main result is a detailed description of the qualitative change of the behavior of primary electromagnetic jet
frequencies under the transition from the KBH to the KNS,
considered here as a bifurcation of the Kerr metric.
We show that quite surprisingly the novel spectra describe linearly stable primary electromagnetic jets
from both the KBH and the KNS. Numerical investigation of the dependence of these primary jet spectra
on the rotation of the Kerr metric is presented and discussed.

\end{abstract}

\keywords{Jets, GRB, Black holes, Naked singularities, Teukolsky
master equation}

\section{Introduction}
The formation of astrophysical jets is one of the frontiers in our
understanding of physics of compact massive objects. Collimated outflows are
observed in many types of objects: brown dwarfs, young stars, X-ray binary systems (\citeauthor{binary}),
neutron stars (NS), GRBs, AGN (for a brief review see \citeauthor{jets1}), even galactic clusters,
and observational data continue to surprise us with the universality of jet formation in the Universe.
For example, according to \citeauthor{Fermi}, observations surprisingly hint for independence
of AGN jet-formation from the type of their host galaxy.
Despite the differences in the temporal and dimensional scales and the suggested emission
mechanisms of the jets, evidences imply that there is some kind of common process
serving as an engine for their formation. The true nature of this engine, however, remains elusive.

An example of the difficulties in front of models of an engine like this, can be seen
in GRB physics. GRBs are cosmic explosions whose jets are
highly collimated ($\theta_{jet}\sim 2^\circ-5^\circ$), highly variable in time
($\sim$ seconds), extreme in their energy output ($\sim 10^{51} - 10^{56}$ erg)
and appearing on different redshifts (currently observed up to $z^{obs}_{max}=8.2$) ( \citeauthor{Meszaros},
   \citeauthor{Zhang}, \citeauthor{Burrows}, and for a review on the online GRB repository see \citeauthor{Evans} \citeyear{Evans}).
Although there are already several hundred GRBs observed by current missions and a lot of details are well studied,
the theory of GRBs remains essentially incomplete,
since we lack both an undisputable model of their central engine
and a mechanism of creation of the corresponding jets.

Currently, GRBs are divided into two classes --
short and long -- based on their duration (with a limit $T_{90}\simeq 2s$, for details see \citeauthor{short_long}). This
temporal division seemed to correlate with their origin and spectral
characteristics, leading to the theory that short GRBs are likely
products of binary mergers of compact objects (black hole (BH) and neutron star (NS) or NS and NS), while long GRBs are an outcome of the collapse of massive stars\footnote{The major evidence for the
collapsar model, the connection of long GRBs with supernovae (SN) was confirmed by the
observations of GRB 980425/SN 1998bw and GRB060218/SN2006aj (\citeauthor{Mirabal}) -- events that
started as a GRB and then evolved spectrally into an SN. But the question remains why not
every long GRB is accompanied by SN.}. This clear distinction, however, is
questioned by observations of bursts with common for the two classes
properties (for the latest observation of such GRB: \citeauthor{Antonelli}, for a review \citeauthor{short_long_b}) suggesting the existence of an intermediate, third class of GRBs and by
statistical analysis suggesting that the two classes are not qualitatively
different (\citeauthor{GRB_two}, \citeauthor{short_long_c} and  \citeauthor{Lv}) -- both evidences of common producing mechanism for the two classes. Furthermore, the possibility to explain short GRBs by merger of binary systems of BH-NS or NS-NS was recently questioned by the lack of detection of gravitational waves in 22 cases of short GRBs (\citeauthor{LIGO}, \\ \citeauthor{Andromeda}). At present we are not able to
explain definitely these observational results. Either we have no binary mergers of compact objects in sGRB's,
or the very large distances to the observed events prevent observation of corresponding gravitational waves by the existing detectors.
The absence of gravitational waves from merger of binaries, if confirmed by future careful investigations,
may  seriously challenge the current paradigm in the sGRB field.

The dominating theoretical model in the GRB physics describing the evolution of {\em already formed}
jet is the ''fireball model`` \hspace{0px} introduced by \citeauthor{Piran}. In this model series of shells of
relativistic particles are accelerated by unknown massive object called
the ''central engine``. The observed lightcurve is produced by internal
collisions of the shells and by their propagation in the circumburst medium. This
model, however, doesn't explain the process behind the formation of the shells
or the collimation of the jet.

The discrepancies between this model and the observational data were outlined recently in \citeauthor{Lyutikov},
most important of them -- the yet unexplained afterglow decay plateaus (and the
sharp decay of the light curve afterwards) and the discovered by the mission
SWIFT flares (additional sharp maxima superimposed over the decaying afterglow
lightcurve) observed in different epochs of the burst and with different
fluences (for a detailed study of flares see \citeauthor{Maxham}). Although both
plateaus and flares imply energy injections by the central engine, the flares
that may appear up to $10^6s$ after the trigger with fluences sometimes
comparable with that of the prompt emission, question the very nature of these
energy injections. Different theories and simulations are being explored to
explain these problems (for recent studies see \citeauthor{jets}, \citeauthor{cannonball} ,
\citeauthor{Fan}, \citeauthor{jets_t} \citeyear{jets_t}, \citeauthor{jets_a} \citeyear{jets_a}),   but the obvious conclusion is that we still lack
clear understanding of the central engine of GRB.

Theoretically, the most exploited models of central engines include a Kerr black hole
(KBH) in super-radiant mode(\citeauthor{superradiance}\!,\! \citeauthor{superradiance4} \citeyear{superradiance4}) --
the wave analogue of the Penrose process or the Blandford-Znajek process (\citeauthor{Blandford-Znajek} \citeyear{Blandford-Znajek},  \citeauthor{Blandford})
based on electromagnetic extraction of energy from the KBH -- the electromagnetic analogue of Penrose process.
Both processes have their strong and weak sides, but the main problem is that they cannot provide enough energy
release in a very short time interval, typical of the GRB. The Penrose process seems to be
not efficient enough (it offers significant acceleration only for already relativistic matter),
as shown long time ago by Wald (\citeyear{superradiance7}).
This problem is also discussed in \citeauthor{Bardeen}, who argue that the energy gap between a bound stable orbit
around the KBH and an orbit plunging into the ergo region is so big,
''energy extraction cannot be achieved unless hydrodynamical forces or superstrong radiation reactions
can accelerate fragments to speed more than $0.5c$ during the infall``.
As for the Blandford-Znajek process which offers a good explanation of the collimation we observe, it requires extremely intensive magnetic fields to accelerates the jets to the energies observed in GRBs (theoretical estimations show that $B \sim 10^{15}\div 10^{16}$ G is needed for a jet with $E \geq 10^{51}$ erg ). Even when the magnetic fields are sufficiently strong ( as in \citeauthor{BZ_field} ), numerical simulations imply that the BZ process is not efficient enough (\citeauthor{BZ process}) for the formation of the powerful jets seen in GRBs (\citeauthor{Blandford-Znajek_b},  \citeauthor{Lee}, \citeauthor{Blandford-Znajek_d}, Barkov \& Komissarov: \citeyear{Barkov}, \citeyear{Barkov2} , \citeauthor{Lei} \citeyear{Lei},\\ \citeauthor{Krolik} \citeyear{Krolik},\! \citeauthor{Komissarov} \citeyear{Komissarov}).

Furthermore, the most speculated in the theory GRB engines include a rotating black hole,
but the observational data on GRBs do not provide definitive clarification on this assumption for now.
The main problem is that we cannot observe the central engine directly
not only because of the huge distance to the object and its small size.
The usual techniques to study its strong gravitational field are not applicable for the GRB case,
because of the properties of their physical progenitors and their highly non-trivial behavior:
1) We are not able to observe the motion of objects  in the vicinity of the very central engine of GRB.
2) Instead of a quiet phase after the hypothetical formation of the KBH, we observe late time engine activity (flares) which is hardly compatible with the KBH model.
3) The visible jets are formed at a distance of 20-100 event horizon radii (\citeauthor{jets_d},\citeauthor{jets_b}, \citeauthor{jets_c}.)
where one cannot distinguish the exterior field of a Kerr black hole from the exterior field
of another rotating massive compact matter object solely by measuring the two parameters of the metric
-- $M$ and $a$ (for a more detailed discussion see \citeauthor{Fiziev0902.1277}, \citeauthor{Fiziev2010a}).

In this situation, the only way to reveal the true nature of the physical object behind the central engine
is to study the spectra of perturbations of its gravitational field.
This background field is either described exactly by the Kerr metric (\citeauthor{Kerr}), if the central engine is a KBH or a KNS, or
at least approximately modeled by the Kerr metric with the same $M$ and $a$,
if the central engine is some other object. In any case we can use the Kerr metric to study
the jet problem in the above conditions.

Different types of central engines yield different boundary conditions for the perturbations we intend to consider
and these boundary conditions generate different spectra specific for the given object.
Thus finding appropriate spectra to fit our observations enables us to uncover the true nature of the central engine.

Similar method for discovering of the BH horizon was proposed for the first time in \citeauthor{Detweiler},
and studied more recently in \citeauthor{Dreyer}, \citeauthor{QNM9}, Chirenti \& Rezzolla \citeyear{Chirenti_a},   \citeyear{Chirenti_b}.

In the present article we give a theoretical basis for application of the same ideas
for studying the nature of the central engine of GRBs, AGN, etc.,
based on the spectra of their jets (\citeauthor{Fiziev0902.1277}).

The main new hypotheses studied in the present article is that the central engine generates
powerful {\em  primary jets of radiation}:
electromagnetic, gravitational and maybe neutrino primary jets.
We investigate a simple mechanism of formation and collimation
of such primary jets by the rotating gravitational field of the central engine,
using a new class of solutions to the Teukolsky master equation.
Further on, one can imagine that the primary jets, if powered enough,
inject energy into the surrounding
matter and start the process of shock waves and other well studied effects, investigated
in the MHD models of jets, accelerating  the existing particles, creating pairs
of particles-antiparticles and so on. This way we are trying to make a new step
toward filling the existing gap between modern detailed MHD numerical
simulations of jet dynamics after the jets are created
and possible mechanisms of the very jet creation. The present paper has to be considered
as a first step in this new direction of investigation.

Here we examine the electromagnetic-primary-jet spectra (see Section 2)
of the KBH and the KNS supposing that the gravitational field of the central engine
is described {\em exactly} by the Kerr metric.

The study of the spectra of some types of perturbations of rotating black holes has already
serious theoretical and numerical basis, particularly concerning the quasi-normal modes (QNM) of black holes.
The QNM case examines linearized perturbations of the Kerr metric described by the Teukolsky
angular and radial equations (the TAE and the TRE accordingly, see in Section 2)
with two independent additional requirements as follows:

A. Boundary condition on the solutions to the TRE: only ingoing waves in the horizon, only outgoing waves to radial
infinity -- the so called {\em black hole boundary condition} (BHBC).

B. Regularity condition is usually imposed on the solutions to the TAE
(see in \citeauthor{Teukolsky2} and also in \citeauthor{Chandrasekhar}).

The complex frequencies obtained in the case of QNM are the solutions of the eigenvalue problem corresponding
to these two conditions. They represent the ''ringing`` which governs the
behavior of the black hole in late epochs and they depend only on the two metric
parameters $M$ and $a$.

A major technical difficulty when searching for QNM is that one solves
connected  problem with two complex spectral parameters -- the separation constants $E$ and $\omega$.
This was first done by \citeauthor{Teukolsky_Press} and later developed through the method of
continued fractions by \citeauthor{QNM1}. For more recent results, see also \citeauthor{QNM8}.

In the recent articles (\citeauthor{FS1},\\ \citeauthor{FS2}),
we illustrate the strength of purely gravitational effects in the formation of primary jets of radiation.
The idea of such physical interpretation is derived from the visualization of the singular solutions 
to the TAE which shows collimated outflows.
Our toy model is based on linearized electromagnetic (spin-one) perturbations of
the Kerr black holes with modified additional conditions:
While we keep the BHBC imposed on the TRE, we drop the regularity condition on the TAE
and we work with specific {\em singular} solutions of the TAE instead.
To achieve this, we impose a polynomial condition on the solutions of the TAE.
This new angular condition reflects the change of the physical problem at hand: now we describe
a primary jet of radiation as a solution to the TAE with an angular singularity
on one of the poles of the KBH or KNS.

It is easy to impose the polynomial condition using the properties of the confluent Heun functions,
which enter in the exact analytical solutions of the Teukolsky equations
(for more on the use of confluent Heun functions in Teukolsky equations see:
\citeauthor{Fiziev0902.1277}, \citeauthor{Fiziev0906.5108} and \citeauthor{Fiziev0908.4234}).

Here we focus on the case of electromagnetic perturbations ($\mid s \mid=1 $ )
which seems to be more relevant to the GRB theory, since their observation
is due to electromagnetic waves of different frequencies -- from optical to gamma-rays at high energies.
Similar situation exists for jets from other astrophysical objects.
For them radio frequencies may be also relevant, say for jets from AGNs.
In this paper, we elaborate our theoretical basis and we report our latest
numerical results and their theoretical implications.

Analogous results can be obtained for gravitational and neutrino primary jets.
Since such jets are still not observed in the Nature, we will not include them
in the present article and we will discuss only {\em electromagnetic primary jets}.
For brevity, from now on we will omit ``electromagnetic'' from electromagnetic primary jets,
where it doesn't lead to confusion. 

We study in detail the {\em complex} spectra of primary-jet frequencies and
their dependence over the rotational parameter of the Kerr metric, $a$.
The parameter $a$ in our calculations varies from $a=0$ to $a=M$ and then from $a=M$ to $a>>M$.
The case $a\in(0,M)$ corresponds to the {\em normally spinning} Kerr spacetime;
and the case $a\in(M,\infty)$, to the {\em overspinning} Kerr spacetime.

\begin{figure*}[htb]
\begin{center}
$\begin{array}{lcl}
a) \epsfxsize=50mm \epsffile{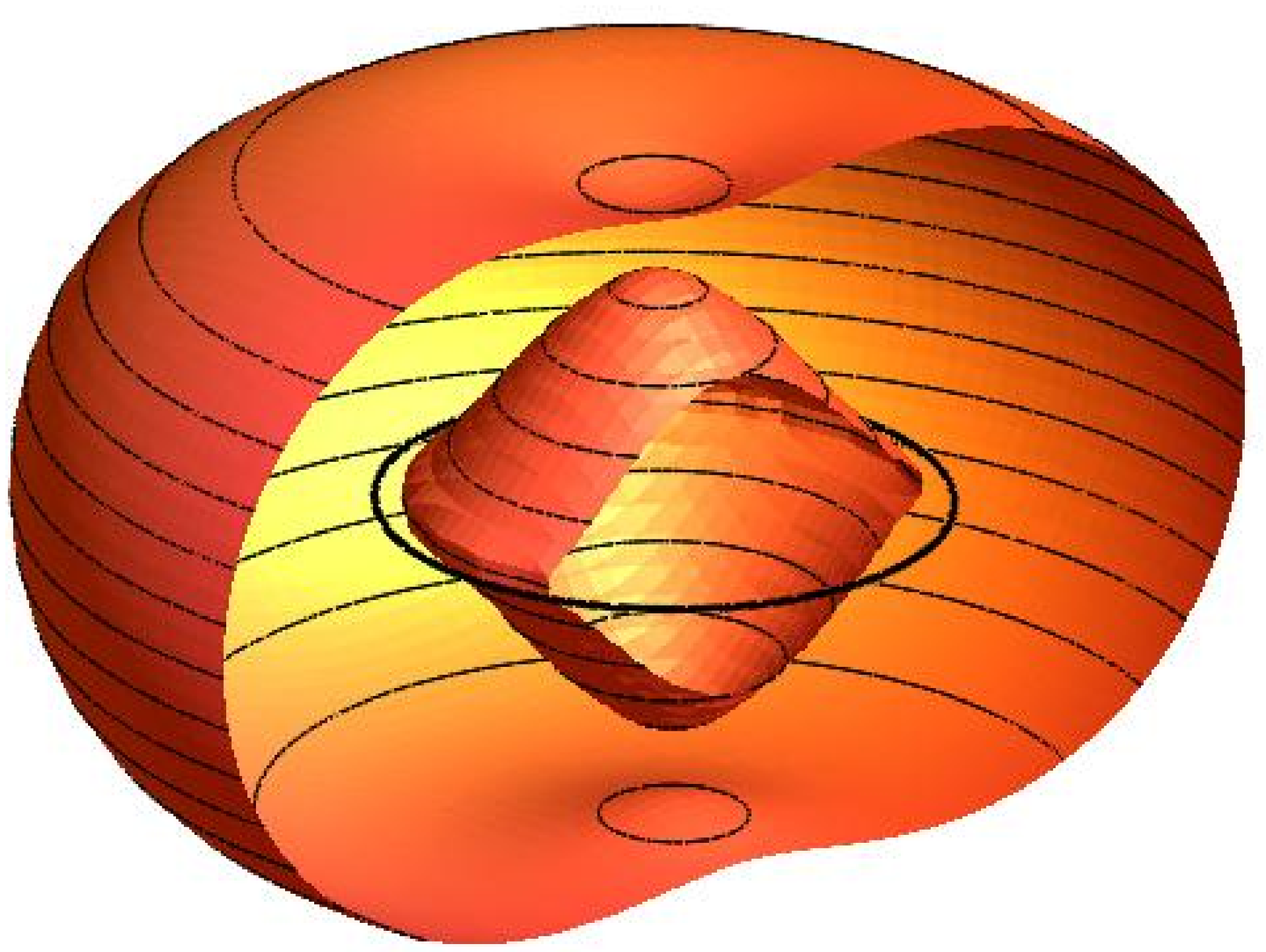} &
b) \epsfxsize=50mm \epsffile{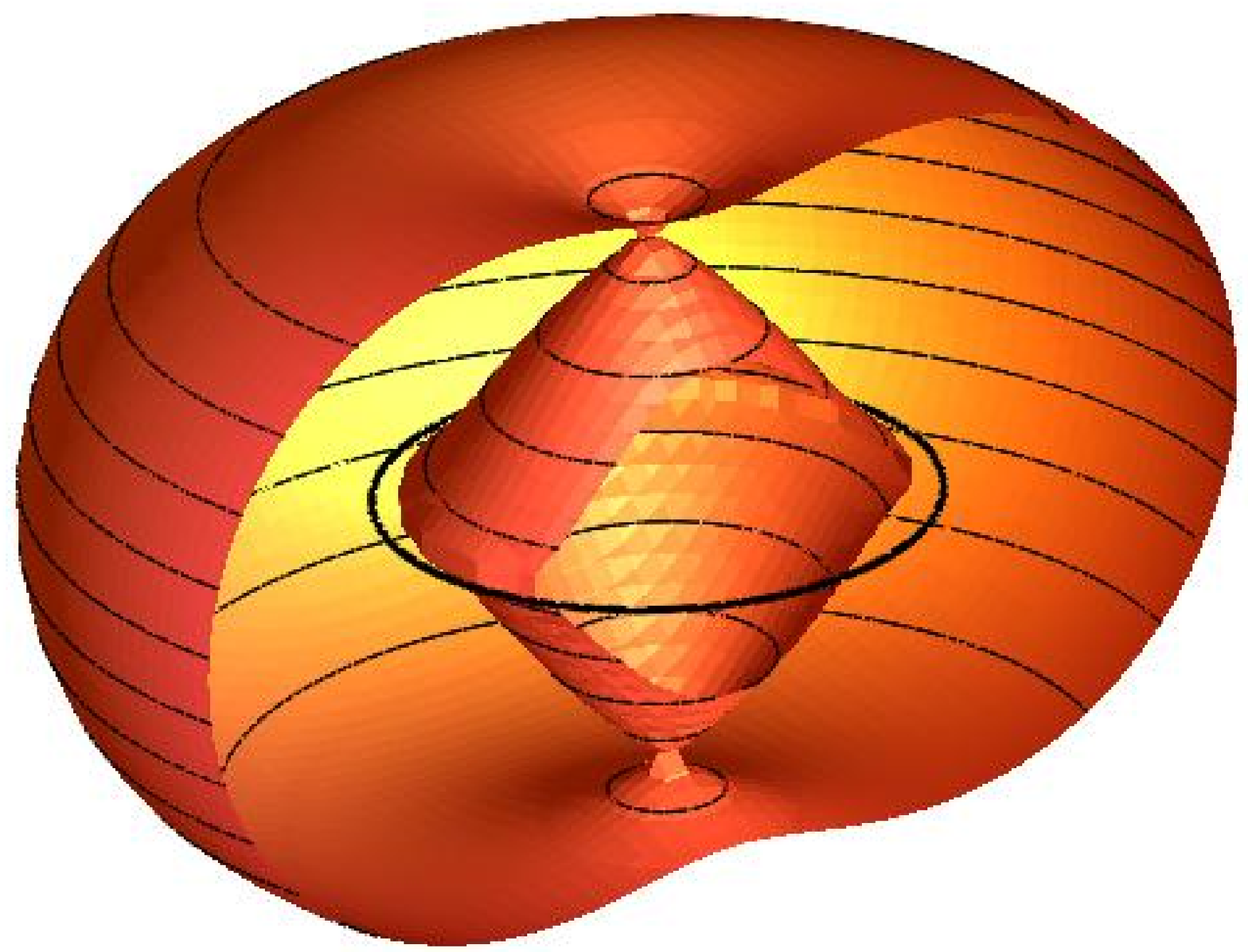} &
c) \epsfxsize=50mm  \epsffile{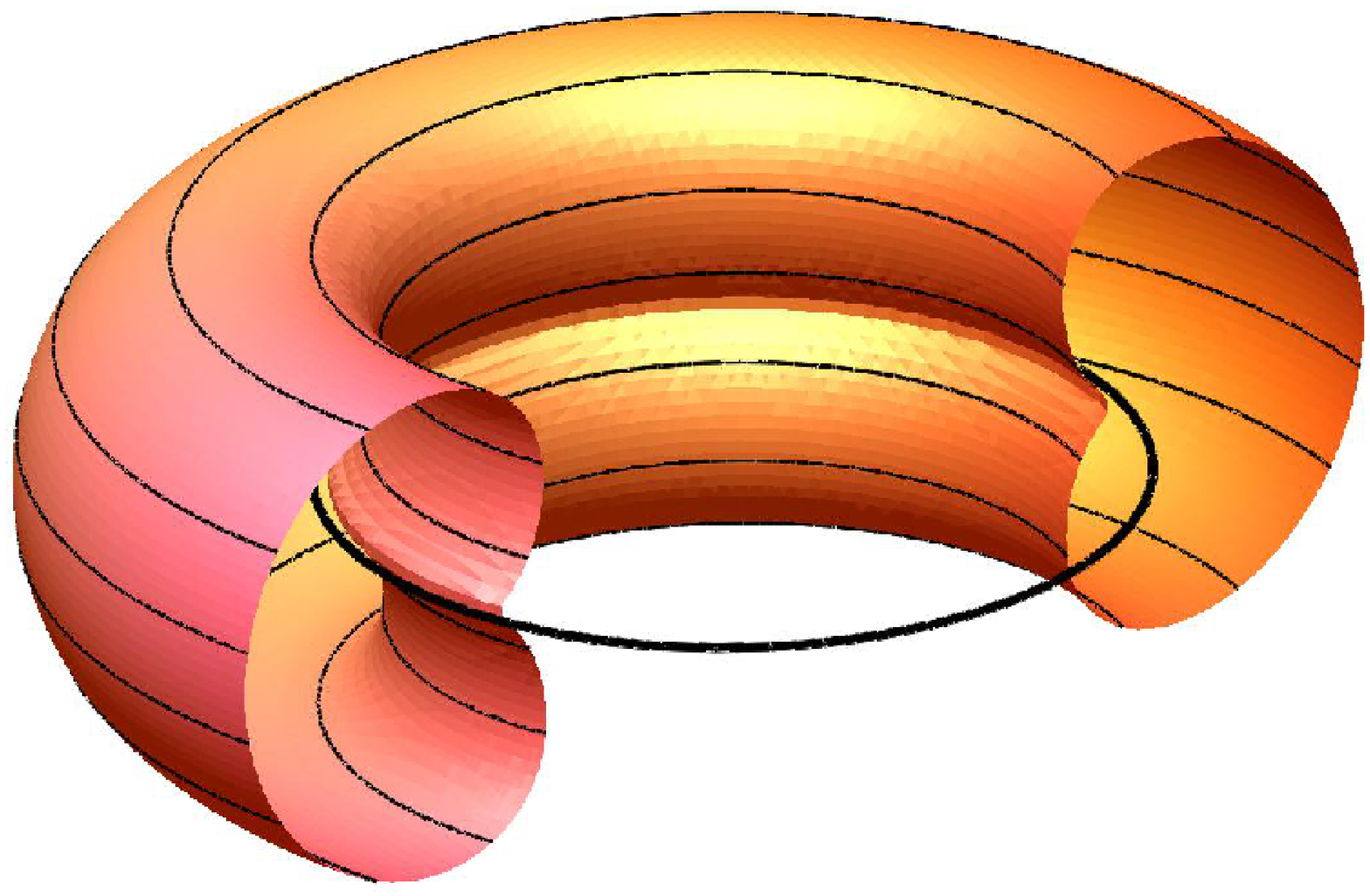} \\
\end{array}$
\end{center}
\caption{A plot of the ergo surfaces for a) KBH: $a<M$, b) extremal case: $a=M$ and 
c) KNS: $a>M$. One can clearly see the ring singularity of the Kerr metric and 
the change of topology due to the transition from ergospheres to ergo torus} 
\label{ergo}
\end{figure*}

Much like in flat spacetime, we can consider different physical problems in the Kerr spacetime.
Imposing BHBC for $0<a<M$ we obtain the well studied case of the KBH.
Under proper boundary conditions which have to be specified for $a\in(M,\infty)$,
we obtain different physical problems related with the Kerr naked singularity (KNS).
The case $a=M$ under BHBC corresponds to the extremal KBH.
This case is beyond the scope of the present article
since it requires a slightly different mathematical treatment.

Our main result is the description of the qualitative change of the behavior of the primary jet
frequencies under the transition from the KBH to the KNS.
Moreover, we discover for the first time that the stability condition in our 
primary jet spectra (positive imaginary part ensuring
damping of the perturbations with time) remains fulfilled even in the KNS regime ($a>M$).
This happens even though, the imaginary part of those frequencies tends to zero as $a\rightarrow M$ -- after this critical point,
with the increase of $a$, the imaginary part of the frequencies stay positive and grows until it reaches a proper constant value for $a>>M$.

In our numerical study of primary jet spectra, there exist two \textit{special} frequencies ($n= 0,1$)
 which have real parts that coincide with the critical frequency of
superradiance in the QNM case. In our case, the frequencies,
however, are complex, and the imaginary part of one of them is of
the same order of amplitude as the real part, yielding an
exponential damping of the maybe superradiance-like emission of primary
jets, created by the KBH or the KNS. The other frequency has a very
small (but positive) imaginary part (for $a < M$). We track the
non-trivial change of this special superradiance-like frequency with
the change of $a$. The relation $\omega_{n=0,1,m}(a)$ is best fitted
by formula (3.4a) in \citeauthor{Fiziev0908.4234} for $n = N = 0,1$,
$m = 0, \pm 1, \pm 2$ in the whole range of $a$ (except for a few
points for small $a$). This provides an analytical formula for these
modes of the spectra.

As for the other frequencies, they possibly form an infinite set as in the case of QNM.
We show that the primary-jet frequencies differ from the QNM ones due
to the different condition imposed on the solutions to the TAE.
The dependence of the primary-jet frequencies
on the parameter $a$ is also presented here for the first time.
The results are discussed in comparison with the case of QNM.

The article is organized as follows: Section 2 develops the mathematical theory behind our spectra,
Section 3 presents the numerical results we obtained, in Section 4 we summarize our results
and we discuss their possible physical implications.

\section{ A toy mathematical model of formation of electromagnetic primary jets by the central engine}

The main problem of most GRB models is that they do not address the formation of the collimated outflow itself, but instead they model the propagation of the jet afterwards. 

The propagation of relativistic jets made of particles can be best understood by full 3-d general
relativistic MHD simulations. In numbers of MHD simulations, however, the energy and/or the angle of collimation of the matter accelerated by the central engine are
just initial parameters (some recent examples \citeauthor{1008.4899}, \citeauthor{1009.1224}). While this approach can bring information about the environment of the burst or the properties of the jet, it doesn't allow us to study the origin of the jet i.e. the central engine. 
More and more studies try to obtain those parameters by evolving real initial
conditions like the mass of the accretion disk/torus around the central object(s) or the masses
of the binaries in the merger (see for example \citeauthor{1007.4203},  \citeauthor{1001.3074}, \citeauthor{1009.0161} ), but
there are still many problems. The powerful late flares and the huge energies of some GRBs, so hard to be fit by the models,
force us to rethink our understanding of the physics of compact massive objects.
What is clear, however, is that precisely such object and its
strong gravitational field should play a definite role in the central engine of powerful events like GRBs.
Only extreme objects of that kind are able to accelerate and collimate matter to the observed high energies
in so short interval of time, using the physics we know. 

Trying to uncover possible mechanism able to work as a central engine, 
we wanted to get back to the basics -- finding the purely gravitational effects 
that one can observe due to a compact massive object. Such effects should be intrinsic properties of those objects, independent of their environment and they are general enough to eventually describe {\em the origin} of  variety of astrophysical jets. 

We started examining the electromagnetic perturbations of the metric of such object 
under boundary conditions different than the already studied in the QNM case.
Although the full physics of the relativistic jets may be outside the scope 
of the linear perturbation theory, understanding the outcome from manipulating the boundary 
conditions in that regime may give us important clues about the real picture. 
Besides, it is clear that in the domain of formation of astrophysical jets 
(i.e. 20-100 gravitational radii far from the central engine), 
the gravitational field is weak enough and its small perturbations can be described by linear perturbation theory.   
Hence, this physics could be at least partially unveiled by studying the spectra of radiation 
of that engine in the frame of linear perturbation theory.

Thus, in our toy model for formation of primary jets of radiation we use
a compact massive rotating object -- a KBH or KNS under appropriate boundary conditions.
Sufficiently far from the horizon its gravitational field is described exactly
or at least approximately by the Kerr metric. The only quantities on which this metric depends are the mass $M$ and
the rotational parameter $a$ related to the angular momentum $J$, by $a=J/M$ (everywhere in the article $c=G=1$).

For $a<M$, the Kerr metric has two {\em real} horizons (\citeauthor{Kerr}) defined by $r_{_\pm\!}\!=\!M \pm \sqrt{M^2\!-\!a^2}$.
In our studies  we will focus on the outer region $r>r_{+}$ which is known to be linearly stable with respect to QNM.
This picture holds for $a<M$ (the normally spinning Kerr spacetime).
For the overspinning Kerr spacetime $a>M$ and $r{_\pm}$ are complex.
When black hole boundary conditions are imposed
for $a<M$ the Kerr metric describes a rotating black hole.
The extremal limit is reached when $a=M$ and in this case $r_{+}=r_{-}=M$.
Under proper boundary conditions for $a>M$ the Kerr metric describes a rotating KNS.
In our work, we will discuss only electromagnetic primary jets from the KBH or the KNS ($s=-1$).

The change of the metric due to the change of the parameter $a$ can
be seen also in the change of the topology of ergo-surfaces of the
Kerr metric (Fig. \ref{ergo}). The ergo-surfaces are defined by
$g_{tt}=1-\frac{2Mr}{r^2+a^2\cos^2(\theta)}=0$ which gives the
stationary limit surfaces $r^\pm_E$, where
$r^{\pm}_E=M\pm\sqrt{M^2-a^2\cos(\theta)^2}$. As seen on Fig.
\ref{ergo}, for $a$ from  $a<M$ to $a>M$ we observe a clear
transition form two ergo-spheres to an ergo-torus.

From a mathematical point of view, it is more natural to work with the dimensionless parameter $b=M/a \in [0,\infty)$.
Obviously the point $b=1$ is a {\em bifurcation} point for the family of the Kerr ergo-surfaces.
When the bifurcation parameter $b$ increases from $0$ to $\infty$,
the ergo-torus ($b\in(0,1)$) bifurcates at the point $b=1$ to two ergo-spheres ($b=(1,\infty)$).
The point $b=0$ corresponds to the ring singularity. At the point $b=\infty$ the outer ergo-sphere transforms
to the Schwarzschild event horizon and the inner ergo-sphere degenerates to the well-known singularity $r=0$ of the Schwarzschild metric.

The evolution of linear perturbations with different spin ($\mid s \mid=0, 1/2,
1, 3/2, 2$) on the background of the Kerr metric was pioneered by Teukolsky in the
70s and led to the famous Teukolsky Master Equation (TME). This differential equation
unifies all physically interesting perturbations ${}_{s}\Psi(t,r,\theta,\phi)$
written in terms of Newman-Penrose scalars and is able to describe completely multitude of
physical phenomena (\citeauthor{Teukolsky}).

The TME in the Boyer-Lindquist coordinates is separable under the substitution:
$\Psi=e^{i(\omega t+m\phi)}S(\theta)R(r)$ where $m=0,\pm 1,\pm 2\dots$ for integer spins and
$\omega=\omega_R+i\omega_I$ is complex frequency
(note that in this article we use the Chandrasekhar notation in which the sign of $\omega$
is opposite to the one Teukolsky used).
This frequency and the parameter $E$ are the two complex constants of the separation.
The stability condition requires  $\omega_I>0$ ensuring that the initial perturbation
will damp with time. Making the separation, we obtain
the Teukolsky angular (TAE) and the Teukolsky radial (TRE) equations which
govern the angular and the radial evolutions of the perturbation, respectively.
The key property of those equations is the fact that they both can be solved in
terms of the confluent Heun function as first written in detail in
\citeauthor{Fiziev0902.1277}.

Explicitly, the TAE and the TRE are:
\begin{multline}
 \Big(\left(1\!-\!u^2\right)S_{lm,u}\Big)_{,u}+\\
\left((a\omega u)^2+2a\omega
su\!+\!{}_sE_{lm}\!-\!s^2-\frac{(m\!+\!su)^2}{1\!-\!u^2}\right)S_{lm}=0,
\label{TAE}
\end{multline}
and
\begin{multline}
{\frac {d^{2}R_{\omega,E,m}}{d{r}^{2}}} + (1+s)  \left( {\frac{1}{r-{\it r_{+}}}}+
{\frac{1}{r-{\it r_{-}}}} \right){\frac {dR_{\omega,E,m}}{dr}} + \\                                                                                                 +\Biggl( {\frac { K ^{2}}{ \left( r-r_{+} \right)  \left( r-r_{-} \right) }}- is \left( {\frac{1}{r-{\it r_{+}}}}+ {\frac{1}{r-{\it r_{-}}}} \right)  K- \\
-\lambda - 4\,i s \omega r \Biggr)
{\frac {R_{\omega,E,m}} {( r-r_{+})( r-r_{-})}}=0
\label{TRE}
\end{multline}
 \noindent where $\Delta=r^2-2Mr+a^2=(r-r{_-})(r-r{_+})$, $K=-\omega(r^2+a^2)-ma$,
$\lambda=E-s(s+1)+a^2\omega^2+2am\omega$ and $u=\cos(\theta)$.

In general, these equations are ordinary linear second-order differential equations with 3
singularities, two of which regular, while $r=\infty$ is
irregular singularity obtained via confluence of two regular singular points. For the TRE, the regular singularities $r_\pm$ are different when $a\neq M$.
At the bifurcation point, $a=M$, $r_+=r_-$ and the TRE has another irregular singularity ($r=M$) instead of the two regular ones.
Hence this is a biconfluent case. Note that the ring singularity $r=0,\,  \theta=\pi/2$ is not a singularity of the TME and does not play any role
in its solutions (as easily seen from  Eq. \eqref{TAE} and Eq. \eqref{TRE}).

When $a\neq M$, these equations can be reduced to the confluent Heun equation and they can be solved in terms of confluent
Heun's function defined as the unique particular solution of following differential
equation:
\begin{multline}
 {\frac {d^{2}}{d{z}^{2}}}H \left( z \right) +  \left( \alpha+{\frac {
\beta\!+\!1}{z}}+{\frac {\gamma+1}{z\!-\!1}} \right) {\frac {d}{dz}}H \left( z
 \right) \! \\ +\! \left(\frac{\mu}{z}\!+\!\frac{\nu}{z\!-\!1}\right) H \left( z
\right)\!=\!0.
\label{Heun}
\end{multline}
The solution called the Heun function is regular in the vicinity of the regular singular point $z=0$ and is normalized
to be equal to $1$ at this point, see the monograph \citeauthor{slav} and some additional references in
\citeauthor{Fiziev0902.1277}, \citeauthor{Fiziev0904.0245}.

From the properties of the confluent Heun function follows that the general solutions of Eq. \eqref{Heun}
can be written, using the \textsc{maple} notation, as
\begin{multline}
H \left( z \right) ={\it C_1}\,\text{HeunC} \left( \alpha,\beta,\gamma
,\delta,\eta,z \right) +  \\ {\it C_2}\,{z}^{-\beta}\text{HeunC} \left(
\alpha,-\beta,\gamma,\delta,\eta,z \right)
\label{HeunSol},
\end{multline} where $\delta=\mu+\nu-\alpha\frac{\beta+\gamma+2}{2}$ and
$\eta=\frac{\alpha(\beta+1)}{2}-\mu-\frac{\beta+\gamma+\beta\gamma}{2}$.

According to  articles Fiziev (\citeyear{Fiziev0902.1277},\! \citeyear{Fiziev0906.5108},\! \citeyear{Fiziev0908.4234}),
Fiziev \& Staicova (\citeyear{FS1},  \citeyear{FS2}) and 
\citeauthor{RBPF} \citeyear{RBPF}) the solutions of the TAE and the TRE for the electromagnetic case $s=\!-1$
can be written in the form:
\begin{align*}
R_1 \left( r \right) ={\it C_1}{{ e}^{i\omega\,r}}\text{HeunC} \left(
\alpha,\beta,\gamma,\delta ,\eta,-{\frac {r-{\it r_{_+}}}{{\it
r_{_+}}-{\it r_{_-}}}} \right)\times\\
 \left( r\!-\!{\it
r_{_+}} \right) ^{{-i\frac { \omega\,({a}^{2} + {{\it r_{_+}}}^{
2})+am }{{\it r_{_+}}-{\it r_{_-}}}}} \left( r\!-\!{\it r_{_-}} \right)
^{{ i\frac {\omega\,({a}^{2}+{{\it r_{_-}}}^{2})+am}{ {\it
r_{_+}}-{\it r_{_-}}}}}
\end{align*}
and
\begin{align*}
R_2 \left( r \right) ={ \it C_2}{{ e}^{i\omega\,r}}\text{HeunC}
\left( \alpha, -\beta,\gamma,\delta ,\eta,{ -\frac {r-{\it
r_{_+}}}{{\it r_{_+}}-{\it r_{_-}}}} \right)\times\\
 \left( r\!-\!{\it
r_{_+}} \right) ^{{i\frac {\omega\,({a }^{2}+{{\it
r_{_+}}}^{2})+am}{{\it r_{_+}}-{\it r_{_-} }}+1}} \left( r\!-\!{\it
r_{_-}} \right) ^{{ i\frac{\omega\,({a}^{2}+{{\it
r_{_-}}}^{2})+am}{{\it r_{_+}}-{\it r_{_-}}}}}
\end{align*}
in the case, $r>r_{+}$, where:\\ $\alpha =-2\,i \left( {\it r_{_+}}-{\it r_{_-}}
 \right) \omega$,  $\beta =-{\frac {2\,i(\omega\,({a}^{2}+{{\it
r_{_+}}}^{2})+am)}{{\it r_{_+}}-{\it r_{_-}}}}-1$,\\
$\gamma ={\frac {2\,i(\omega\,({a}^{2}+{{\it
r_{_-}}}^{2})+am)}{{\it r_{_+}}-{\it r_{_-}}}}-1$,\\
$\delta =-2i\!\left({\it r_{_+}}-{\it r_{_-}}
\right)\!\omega\!\left(1-i
 \left( {\it r_{_-}}+{\it r_{_+}} \right) \omega \right)$,\\
$\eta =\!\frac{1}{2}\frac{1}{{ \left({\it r_{_+}}\!-\!{\it r_{_-}}
\right) ^{2}}}\times\\
\Big( 4{\omega}^{2}{{\it r_{_+}}}^{\!\!4}\!+ 4\left(i \omega
\!-\!2{\omega}^{2}{\it r_{_-}}\right) {{\it r_{_+}}}^{\!\!3}\!+\! \left(
1\!-\!4a\omega\,m\!-\!2{\omega}^{2}{
a}^{2}\!-\!2E \right) \times \\ \left( {{\it r_{_+}}}^{\!\!2}\!+\!{{\it r_{_-}}}^{\!\!2}
\right) \!+\! \\
 4\left(i\omega\,{\it r_{_-}} \!-\!2i\omega\,{\it r_{_+}}\!+\!E\!-\!{\omega}^
{2}{a}^{2}\!-\!\frac{1}{2} \right) {\it r_{_-}}\,{\it r_{_+}}\!-4{a}^{2} \left(
m\!+\!\omega\,a
 \right) ^{2} \Big).$

For the angular function $S(\theta)$, we obtain the following solutions:
\begin{align*}
 S_{{\pm},s,m}(\theta)={\text{e}^{\alpha\,  z_{\pm}(\theta) }}
\left(z_{\pm}(\theta)\right)^{\beta/2}
\left( z_{\mp}(\theta)\right) ^{\gamma/2}\times\\
\text{HeunC}
\left(\alpha,\beta,\gamma,\delta,\eta,z_{\pm}(\theta)\right)
\label{S1}
\end{align*}

\noindent where
\begin{align*}
&\alpha=\pm 4\,a\omega,\beta=s\!\mp \!m\!, \gamma=\!
s\!\pm \!m,\delta=\mp 4\,a\omega s,\\
&\eta=\frac{{m}^{2}-{s}^{2}}{2}\!\pm\!2\,a\omega
s\!-\!{(a\omega)}^{2}\!-\!E\!+\!s^2,\\
&z_{+}(\theta)=\cos^2\left(\frac{\theta}{2}\right),z_{-}
(\theta)=\sin^2\left(\frac{\theta}{2}\right)=1-z_{+}(\theta).
\end{align*}

Note, the confluent Heun function depends on 5 parameters (here $\alpha,\beta,\gamma$, $\delta$ and $
\eta$ ) but the values of those parameters differ for the TAE and the TRE.

To fix the physical problem we want to study, we impose appropriate
boundary conditions on the solutions to the TRE and the TAE.
These boundary conditions yield the spectrum of the separation constants $E$ and $\omega$.

For the solutions of the the TAE, we use the new requirement that the confluent Heun functions should be polynomial.
The polynomiality condition reads:
\begin{align*}
&\frac{\delta}{\alpha}+\frac{\beta+\gamma}{2} + N + 1=0.
&\Delta_{N+1}(\mu)=0 .
\end{align*}
It yields a collimated singular solutions \citeauthor{Fiziev0902.1277}.
Here, the integer $N\geq0$ is the degree of the polynomial and $\Delta_{N+1}(\mu)$ is
three-diagonal determinant specified in \citeauthor{Fiziev0902.1277}, \citeauthor{Fiziev0904.0245}.

The polynomial requirement for the angular solutions fixes the
following relation between $E$ and $\omega$:
\begin{equation}
{}_{s=-1}E_{m}^{\pm}(\omega)\!=\!-{(a\omega)}^{2}-2\,a\omega\,m\!\pm\!2\,\sqrt
{{(a\omega)}^{2}\!+\!a\omega\,m}.
\label{E}
\end{equation}

This simple {\em exact} relation leads to a significant technical advantage.
Instead of working with a complicated connected system of spectral equations (as in the QNM case),
by imposing the polynomial requirement we have to solve only {\em one} spectral equation for the variable $\omega$.
Thus we essentially simplify the calculations and we obtain interesting from physical point of view new results.

Just examining this simple form of the relation $E(\omega)$ and using the
polynomial solutions, we plotted $S(\theta)$ which controls the angular
behavior of the solution and we were able to obtain different types of collimated
outflows, generated by electromagnetic perturbations of the Kerr metric for different values of $\omega$
which not necessarily belong to the jet spectrum (\citeauthor{FS1}).
The existence of such collimation in our results justifies the names ''jet`` solution and ''jet`` spectra that we use.
Unlike the QNM case, the considered here "jet" case naturally generates 
a collimated electromagnetic radiation outflow from KBH and KNS.
In the present article we confirm the above preliminary results obtaining numerically the
primary-jet spectrum of $\omega$ defined by the KBH and the KNS jet-boundary-conditions.

It is important to emphasize that in the case of the angular equation we work with
the {\em singular} solutions of the differential equation. The question of the
possible physics behind the use of such singularity is answered in
\citeauthor{Fiziev0908.4234}. By looking for solution in a
specific factorized form $\Psi=e^{i(\omega t+m\phi)}S(\theta)R(r)$, the function
we obtain in general is not a physical quantity. Instead
it defines a factorized kernel
${}_{s}K_{E,\omega,m}(t,r,\theta,\phi)\sim e^{i(\omega
t+m\phi)}{}_{s}S_{E,\omega,m}(\theta){}_{s}R_{E,\omega,m}(r)$ of the general
integral representation for the physical solutions of the TME:
\begin{multline}
 {}_{s}\Psi(t,r,\theta,\phi)=\\ \sum
_{m=\!-\!\infty}^{\infty}\!\frac{1}{2\pi}\int\!{d\omega}\! \int\!{dE}\,
{}_{s}A_{\omega,E,m}\,{}_{s}K_{E,\omega,m}(t,r,\theta,\phi).
\label{Kernel}
\end{multline}

This form of the mathematical representation of the physical solution is written as the most general
superposition of all particular solutions of the TME and it assumes summation over all admissible values of $E$ and $\omega$.
If one wants to work with the solutions corresponding to certain boundary conditions,
then one has to account for the specific admissible spectra of $E$ and $\omega$ in those cases.
For example, if we have obtained discrete spectra for $\omega$, this leads to the use of singular kernel proportional to $\sum_{n}^{}\delta(\omega-\omega_{n})$, where $\delta$ is the Dirac function.
This reduces the integral over $\omega$ to summation over $\omega_{n}$.
Similarly, for solutions with definite total angular momentum we have $E=l(l+1)$ and the integral
over $E$ is replaced by summation over $l$--integer (or half integer),
because of singular factor $\sum_{l}^{}\delta(E-l(l+1))$ in Eq. \eqref{Kernel}.

From Eq. \eqref{Kernel} it becomes clear that the singular solutions of the angular equation
do not cause physical difficulties, only if we are able to find appropriate amplitudes ${}_{s}A_{\omega,E,m}$
that will make the physical solution ${}_{s}\Psi(t,r,\theta,\phi)$ regular.
Examples of finding such amplitudes are being developed.
It is clear that this formula justifies the use of singular solutions of the TAE (for $s=-1$
the singularity is on one of the poles of the sphere, for $s=1$ -- on the other one).

Knowing $E(\omega)$, we can find numerically the frequencies
$\omega$ by imposing boundary condition on the solutions to the
radial equation. The Kerr spacetime can be considered as a
background for different physical problems. As in flat spacetimes,
one has to fix the physics imposing the corresponding boundary
conditions. A specific peculiarity of Kerr spacetime is that in
general the TRE (Eq. \eqref{TRE}) has 3 different singular points:
$r_+$, $r_-$ and $\infty$. To fix the correct physical problem, we
have to specify the boundary conditions on two of them (the
asymptotic behavior of the solution at these singular points). This
represents the so-called {\em central two-point connection problem}
(as described in \citeauthor{slav}). In principle, we can impose
boundary conditions on different pairs of singular points: on $r_-$
and $r_+$; on $r_-$ and $+\infty$; on $r_-$ and $-\infty$; on $r_+$
and $+\infty$; on $r_+$ and $-\infty$; and on $-\infty$ and
$+\infty$. Each of these choices will fix a specific type of
physical problem posed in the complex domain of variables.
Obviously, these different problems may have different physical
properties, especially, with respect to their linear stability.

Note that choosing a specific central two-point connection problem
we fix the physical problem independently of the values of bifurcation parameter $b$.
This way, we are able to study the bifurcation phenomenon for the given physical problem in the whole range of the parameter $b$.
Since we want to study primary jets from the KBH and the KNS,
we start by imposing BHBC on the solutions to the TRE at the points $r_+$ and $\infty$
for the first case ($a<M$), where the regular singularities are {\em real}.
These boundary conditions are physically well motivated (\citeauthor{Teukolsky}, \citeauthor{Chandrasekhar})
and the central two-point connection problem for the KBH has a clear physical meaning.
The same central two-point connection problem exists in the overspinning case,
in spite of the fact that its two regular singularities $r_{\pm}$ are {\em complex}.
We use the same boundary conditions for the case of primary jets from the KNS,
since the physics of the problem is defined by them.
Thus, we transform the BHBC to naked singularity boundary conditions
using an analytical continuation of the central two-point connection problem.

BHBC can be summarized to:

1. On the horizon ($r \rightarrow r_{+}$), we require only ingoing (in the horizon) waves.
This specifies which of the two solutions of the TRE ($R_{1}$ or $R_{2}$) works in each interval
for the frequency $\omega$. In our case, for any integer $m$,
$R_1$ is valid for $\Re(\omega) \in (- m\Omega_{+}, 0)$, $R_2$ is valid in the complimentary set  $\Re(\omega)\in(-\infty,+\infty) \setminus (- m \Omega_{+},0)$ where
$\Omega_{+}=a/2Mr_{+}$. For $m=0$, the only valid ingoing solution in the whole range $\Re(\omega)\in(-\infty,+\infty)$ is $R_{2}$.

2. At infinity ($r \rightarrow \infty$) we allow only outgoing waves. Explicitly, at infinity we have a linear
combination of an ingoing ($R_{\leftarrow}$)
and an outgoing ($R_{\rightarrow}$) wave:
\begin{equation*}
R=C_{\leftarrow}\,R_{\leftarrow}+C_{\rightarrow}\,R_{\rightarrow}
\end{equation*}
where $C_{\leftarrow}$, $C_{\rightarrow}$ are unknown constants. In order to
have only outgoing waves, we need to have $C_{\leftarrow}=0$. Since this
constant is unknown, we find it indirectly from:
\begin{equation}
C_{\leftarrow}=\frac{R}{R_{\leftarrow}}-C_{\rightarrow}\frac{R_{\rightarrow}}{R_
{\leftarrow}}.
\label{C}
\end{equation}
If in this equation we set
$\lim\limits_{r\to\infty}\frac{R_{\rightarrow}}{R_{\leftarrow}}= 0$, this will eliminate the second term in Eq. \eqref{C}. To do this, we complexify $r$ and $\omega$ and choose the direction in the complex plane $\mathbb{C}_r$ in which this limit tends to zero most quickly. This direction turns out to be $arg(r)=3\pi/2 - arg(\omega)$, i.e., $r=\mid r \mid e^{3/2i\pi-i\,arg(\omega)}$ connecting the arguments of the complexified $r$ and $\omega$ (\citeauthor{QNM9}). Having fixed that, it is enough to solve the spectral condition in the form
\begin{equation}
C_{\leftarrow}=\lim\limits_{\mid r\mid\to\infty}\frac{R}{R_{\leftarrow}}=0
\label{Rbc}
\end{equation}
in order to completely specify the primary jet spectra ${}_{s}\omega^{jet}_{n,m}$.
For primary jets from the KNS we use the same spectral condition in the complex domain (see above).

 Since in this paper we calculate only the case $s=-1$, for primary jets from the KBH and the KNS,
 we will omit the prefix $s=-1$ in front of $\omega_{n,m}$ and also the index ''jet``.

\section{Numerical results}
In Fiziev \& Staicova (\citeyear{FS1}, \citeyear{FS2}), we presented some preliminary
results for the solutions of the TAE. They showed that the
polynomial angular solutions describe collimated structures from
various types for arbitrary $\omega$, even if the BHBC (Eq.
\eqref{Rbc}) is not fulfilled. In this article we will focus on the
solutions of the radial equations, thus fixing the specific spectrum
$\omega_{n,m}$ for primary jets from the KBH and the KNS from Eq.
\eqref{Rbc}.

\begin{figure}[!h]
\begin{center}
 \includegraphics[scale=0.3]{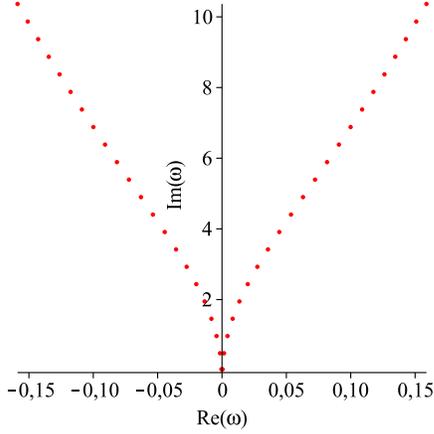}
 % a0_all.eps: 0x0 pixel, 300dpi, 0.00x0.00 cm, bb=20 118 575 673
\caption{Complex plot of the frequencies $\omega_{n}$ for $a = 0$.  The index
 $n = 0, 1, . . .$ increases with the distance to the origin. For every $n > 0$,
there is a pair of 2 roots $\omega^{1,2}_n$ symmetrical with respect to the imaginary axis. In the case $a=0$, the frequencies are independent of $m$}
\label{fig:m0_all}
\end{center}
\end{figure}

\subsection{Numerical methods}

To find the zeros of Eq. \eqref{Rbc}, we use the software package \textsc{maple}
which currently is the only one able to work with Heun's functions.

The roots presented here are found with modified by the team M\"uller algorithm, with
precision to 13 digits. Although the results for different versions of \textsc{maple} may
vary in some cases due to improvements in the numerical algorithm,
key values were checked to have at least 6 stable digits, in most cases -- more than 8 stable digits.
\begin{figure}[!hb]
\vspace{0.0in}
\begin{center}
\includegraphics[trim=5mm 0mm 3mm 3mm,clip,
scale=0.3,keepaspectratio]{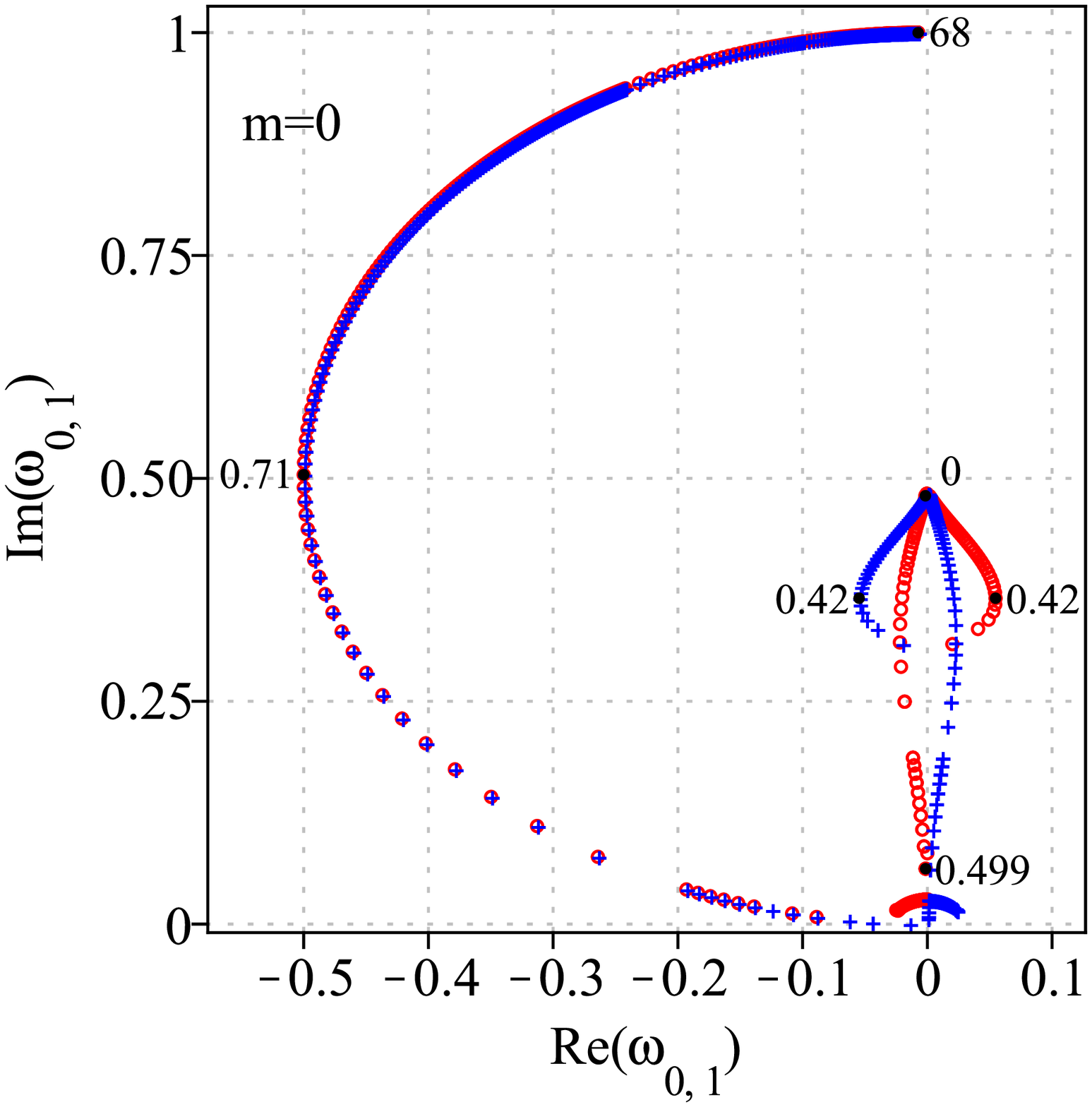}
%\vspace{0.1cm}
\includegraphics[trim=5mm 0mm 3mm
3mm,clip,scale=0.3,keepaspectratio]{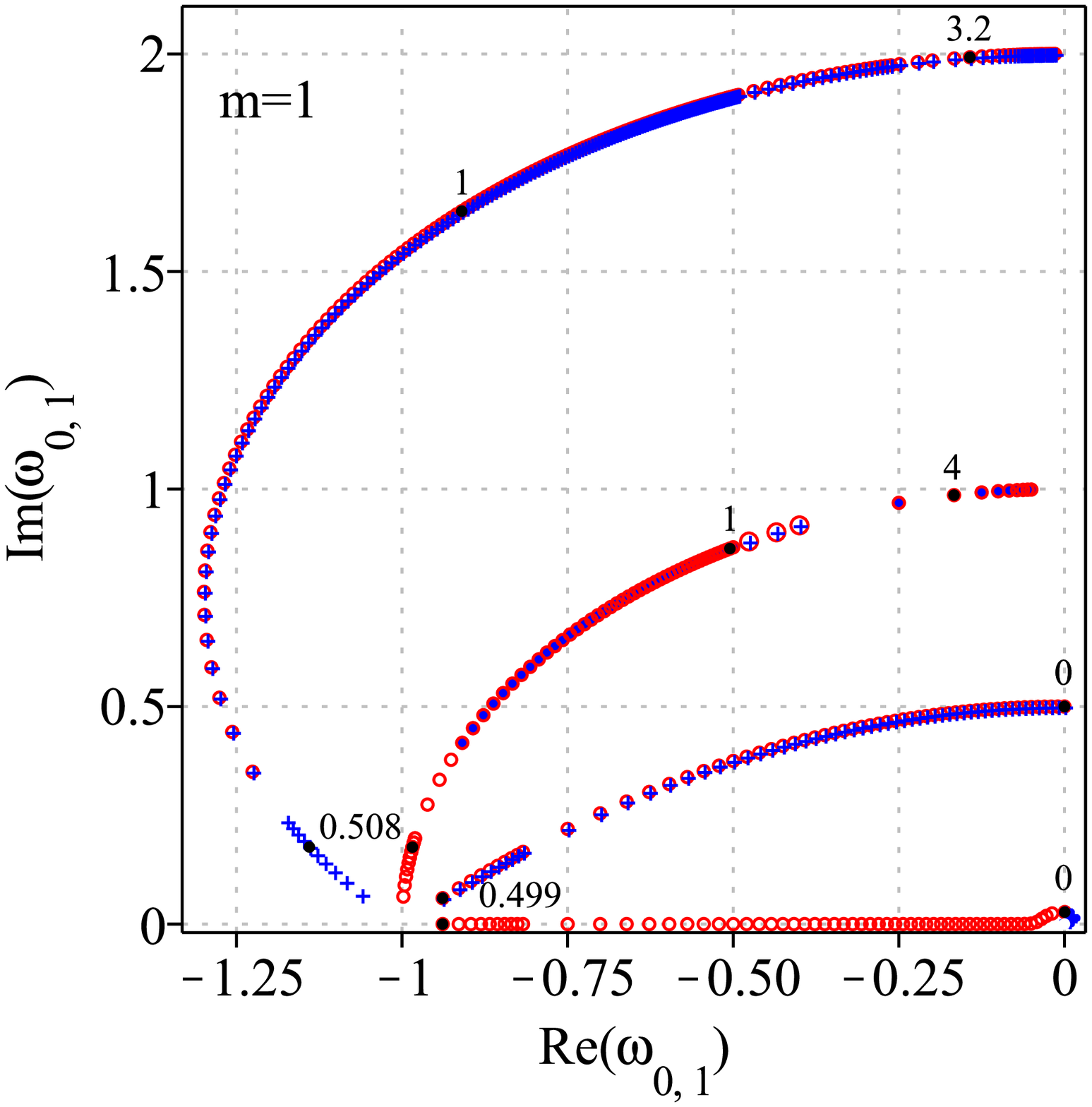}
%\vspace{0.1cm}
\includegraphics[trim=5mm 0mm 3mm
3mm,clip,scale=0.3,keepaspectratio]{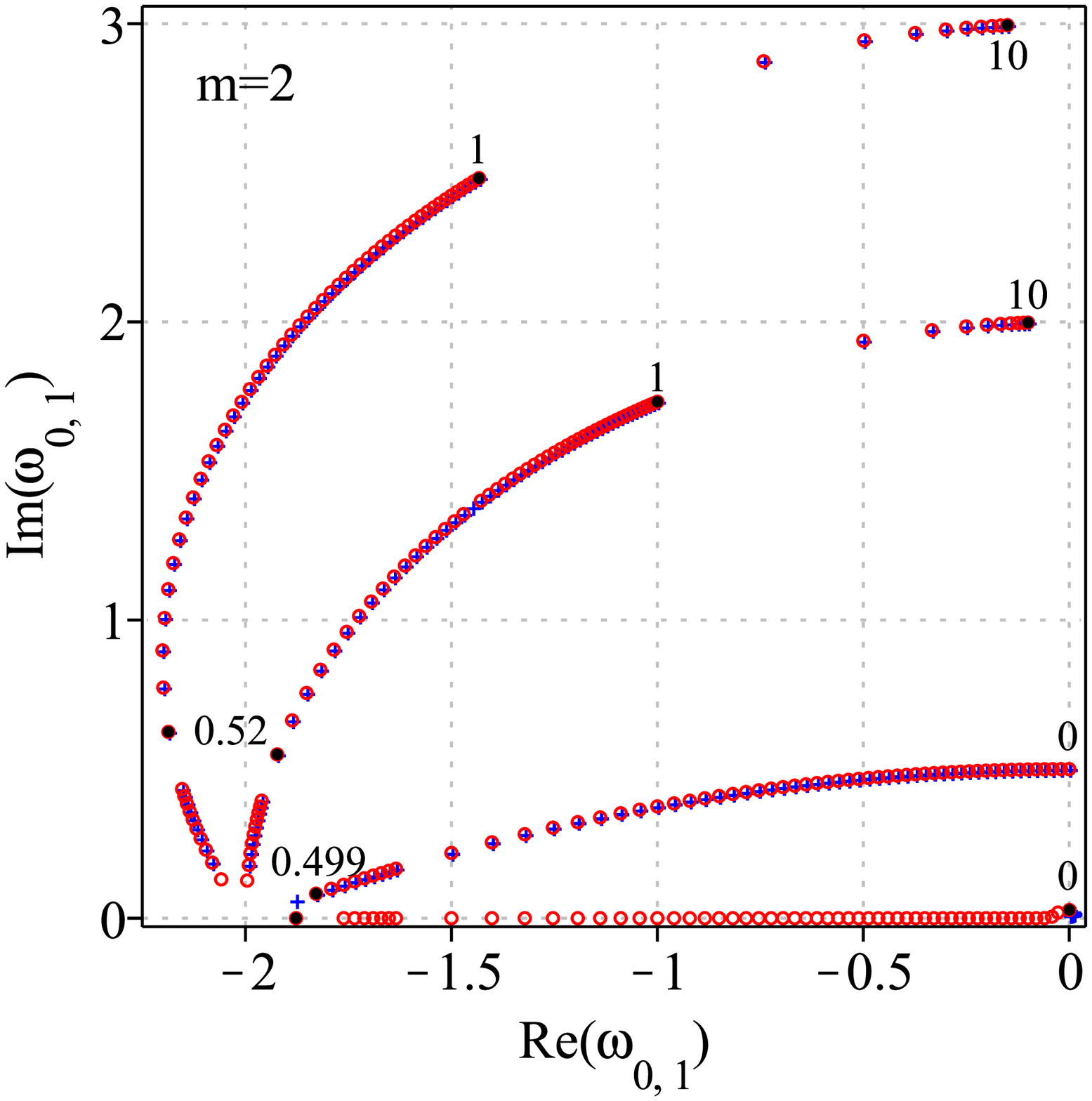}
\vspace{-0.2cm} \caption{Complex plot of the two lowest modes $\omega^{\pm}_{n=0,1;m}(a)$ for $m\!=\!0,1,2$. With red circles we plot $\omega^{+}_{n=0,1,m}$, with blue crosses -- $\omega^{-}_{n=0,1,m}$;
on the curve we mark some values of $a$. It is clear that for
$m\!>\!0$, $\omega^{+}_{1,m}$ and $\omega^{-}_{1,m}$ coincide almost
everywhere. For $m\neq0$, $a<M$, the symmetry with respect to the imaginary axis breaks} \label{cp}
\end{center}
\vspace{-0.35cm}
\end{figure}
The parameters are fixed as follows : $s=-1$, $M=1/2$, $r=110$. This value for the radial variable $r$ is chosen so that it represents the actual numerical infinity -- the closest point at which the frequencies found by our numerical method stop changing significantly for further increase of $r$ (for fixed $a,m,n$). In study of the dependence on $a$, the step for the rotational parameter is $\delta a=0.01$ for $a<M$ and becomes adaptive for $a>M$.

 In our work, we have studied in detail cases with $m=0, \pm 1, 2 $, covering the cases $\mid m\mid<\mid s\mid$, $\mid m\mid=\mid s\mid$, $\mid
m\mid>\mid s\mid$. The complete set of our numerical results is available by request.

\subsection{Summary of the results}

Graphical representation of our results can be seen in figures 2-11.

In the case $a=0$ when no rotation of metric is present we find a set of pairs of frequencies $\omega^{1,2}_{n}$ with equal imaginary
parts and real parts symmetrical  with respect to the imaginary axis ($\Re(\omega^1_n)=-\Re(\omega^2_n)$ and $\Im(\omega^1_n)=\Im(\omega^2_n)$). This set
(Fig. \ref{fig:m0_all}) is maybe infinite
if the yet unknown next frequencies follow the same line as in Fig. \ref{fig:m0_all}.
The set is independent of azimuthal number $m$  as it should be for $a=0$.
Frequencies are numbered with $n$ according to their distance from the origin $\mid \omega_n\mid$,
starting with $n=0$ for the lowest.

Although Fig. \ref{fig:m0_all} shows the case without rotation, it
drastically differs from the results obtained by solving the Regge-Wheeler
Equation for the Schwarzschild metric in the QNM case (\!
\citeauthor{QNM}\!\!\citeyear{QNM},\! \citeauthor{Chandrasekhar},\! \citeauthor{QNM1},\! \citeauthor{QNM2} \!\!\citeyear{QNM2},\! \\ \citeauthor{QNM3}, \citeauthor{QNM4},  \citeauthor{QNM5} \citeyear{QNM5},
\citeauthor{QNM6}, \citeauthor{QNM7}, \citeauthor{QNM9}, \citeauthor{QNM10}, \citeauthor{QNM11},   \citeauthor{Berti09} \citeyear{Berti09}), because
of the change of the conditions for the solutions of the TAE.

Using the so found initial frequencies, we track their evolution with the
change
of the rotational parameter $a$.  We work with both frequencies in the pairs for
each $n$ ($\omega^{1,2}_{n,m}$) and we consider both signs in front of the
square root in ${}_{s=-1}E_{m}^{\pm}(\omega)$ (Eq. (\ref{E})), which we denote  $\omega^{\pm}_{n,m}$, respectively.  On the figures, the number of points we
present is limited by the abilities of the  \textsc{maple} numerical procedures
evaluating the confluent Heun function.

\begin{figure*}[thb]
\begin{center}
$\begin{array}{lcr}
a) \includegraphics[scale=0.25,keepaspectratio]{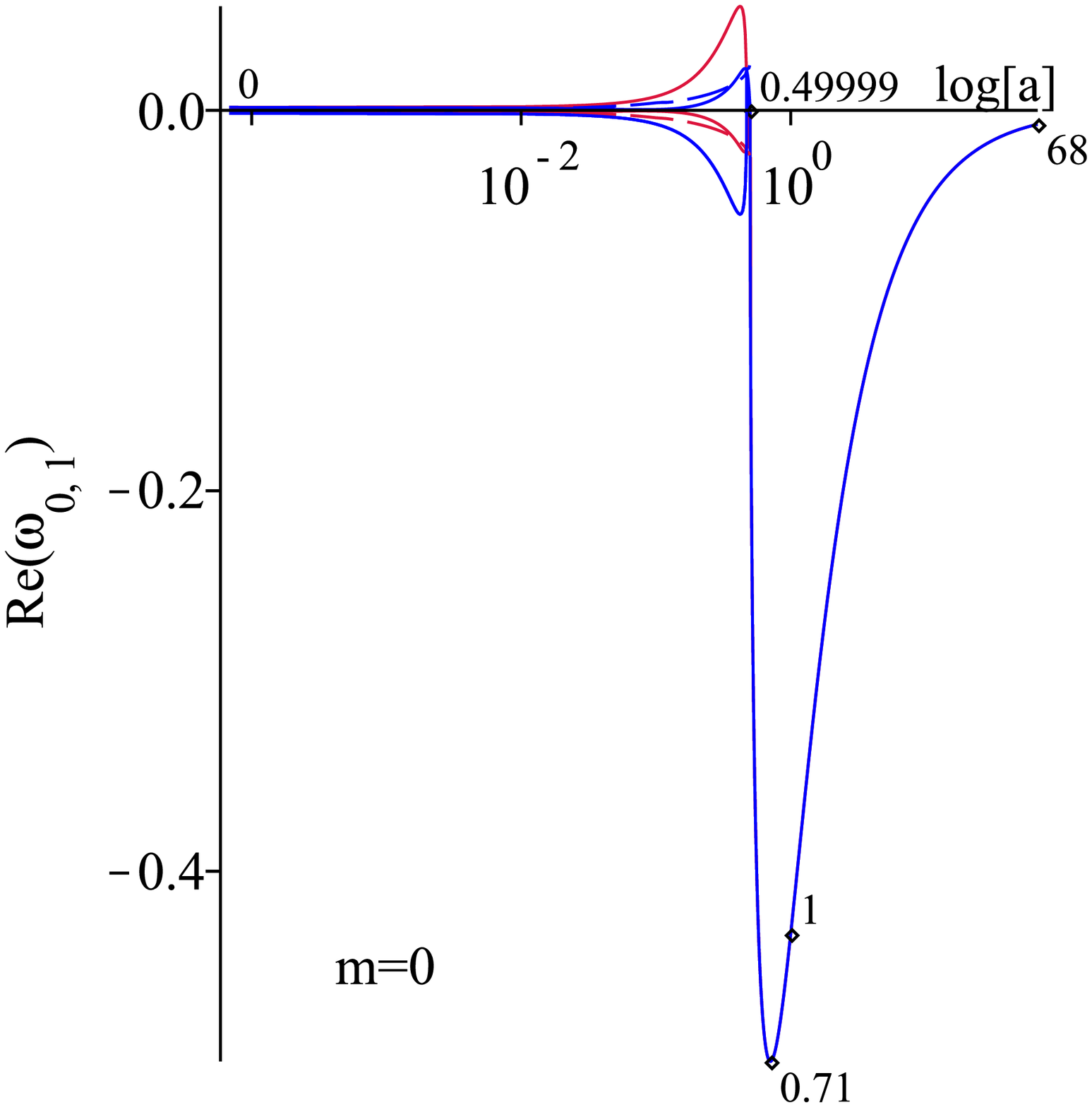}
& b) \includegraphics[scale=0.25,keepaspectratio]{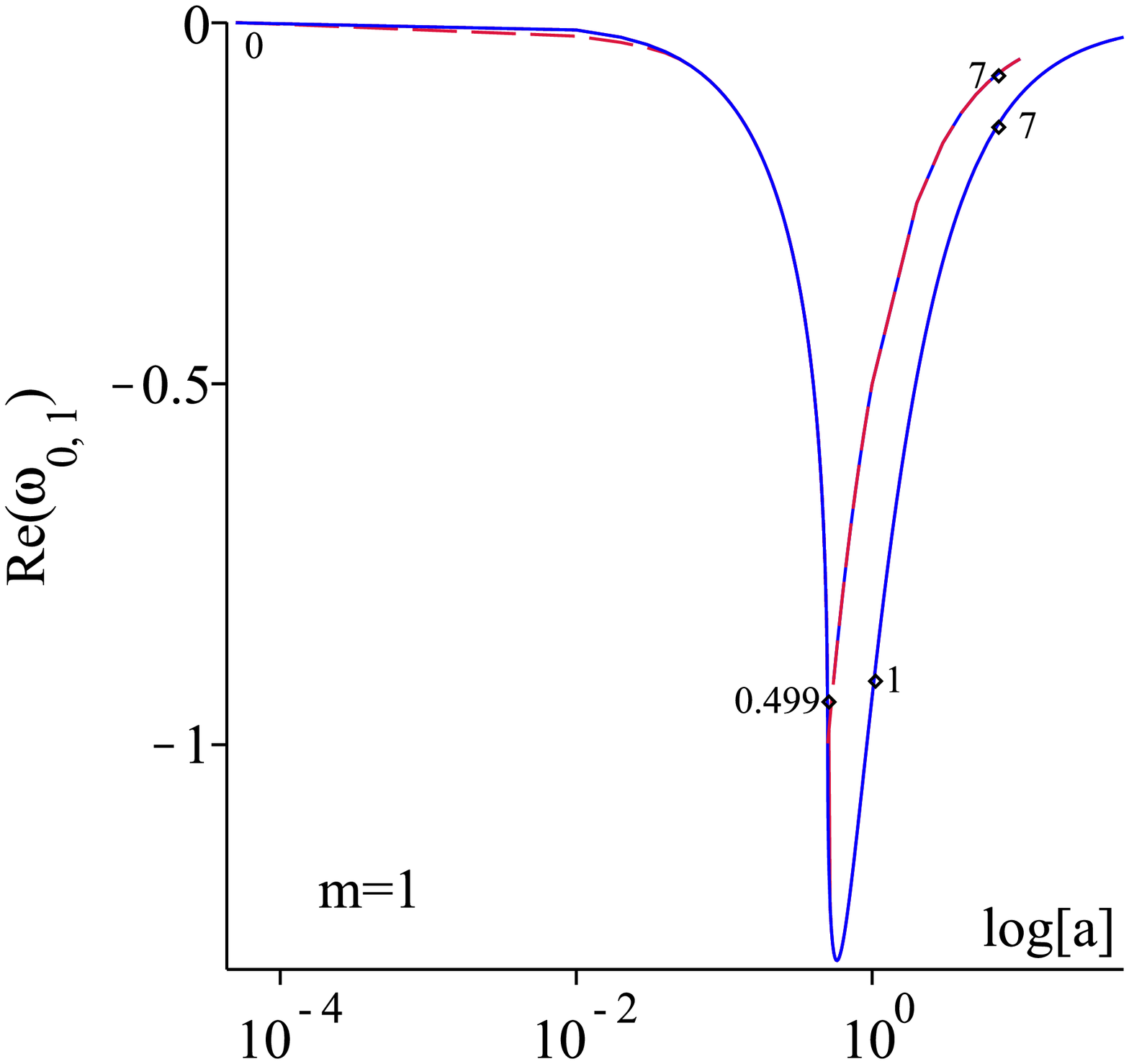}
& c) \includegraphics[scale=0.25,keepaspectratio]{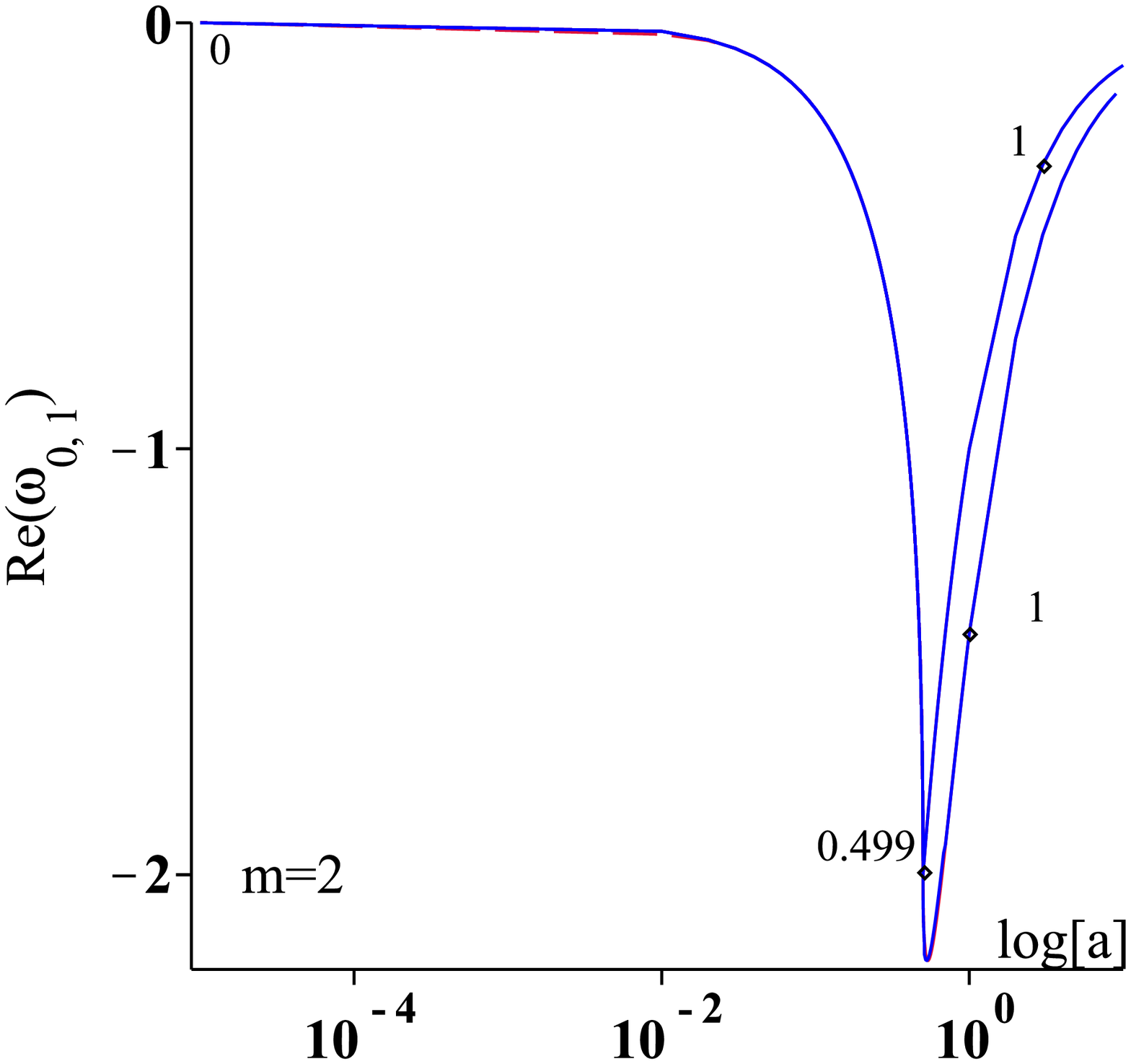}\\
d) \includegraphics[scale=0.25,keepaspectratio]{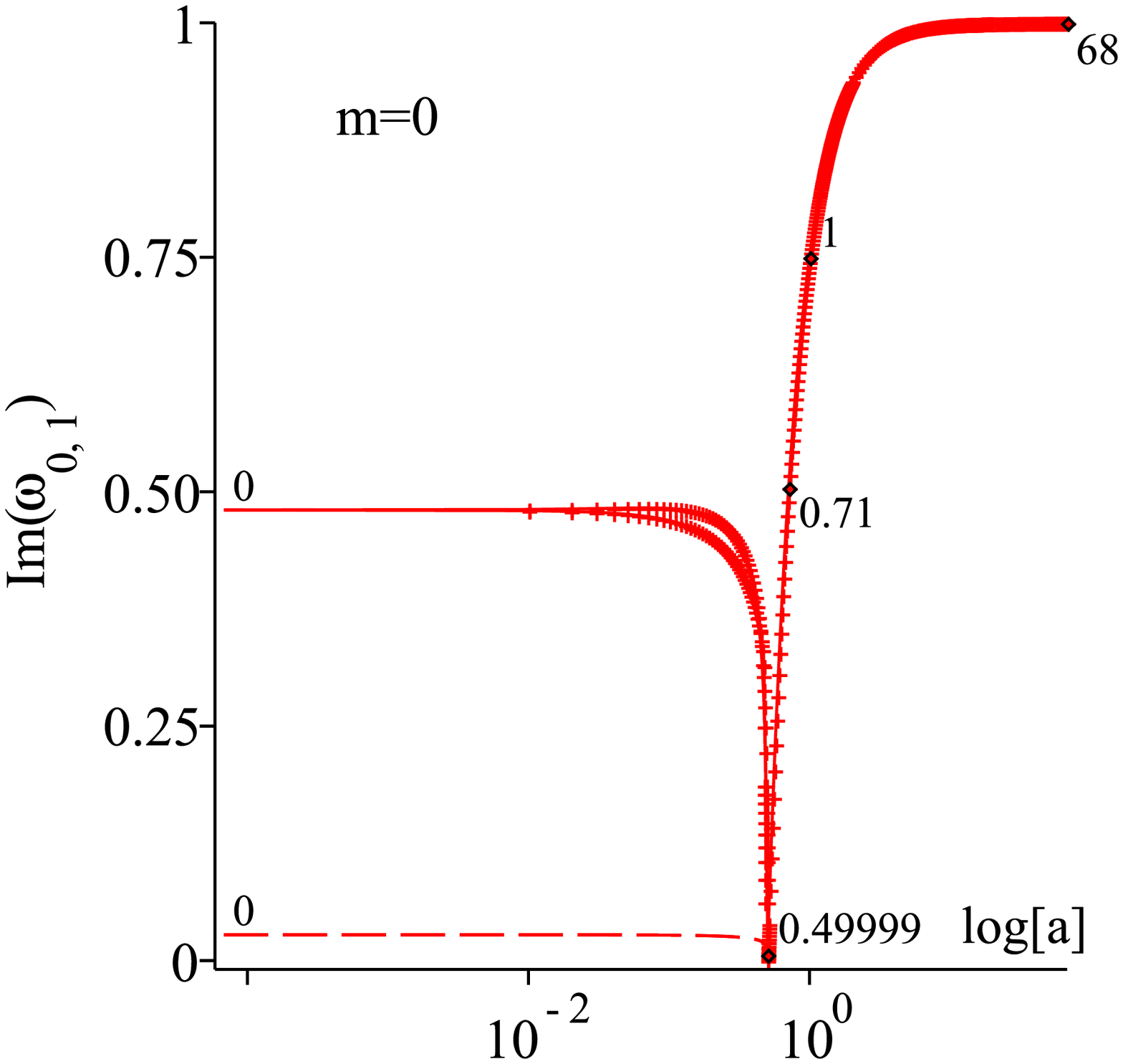}  &
e)  \includegraphics[scale=0.25,keepaspectratio]{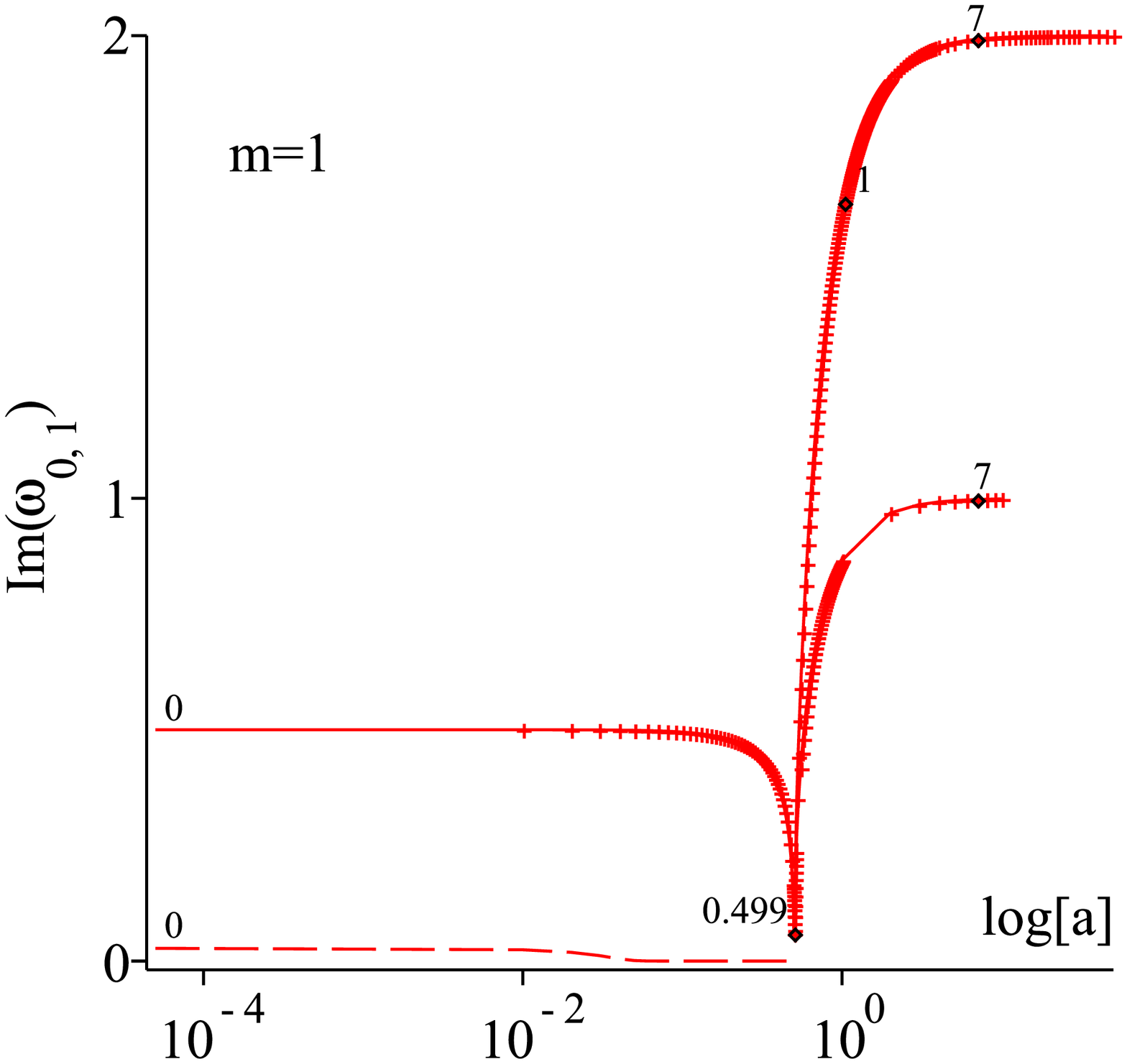} &
f)  \includegraphics[scale=0.25,keepaspectratio]{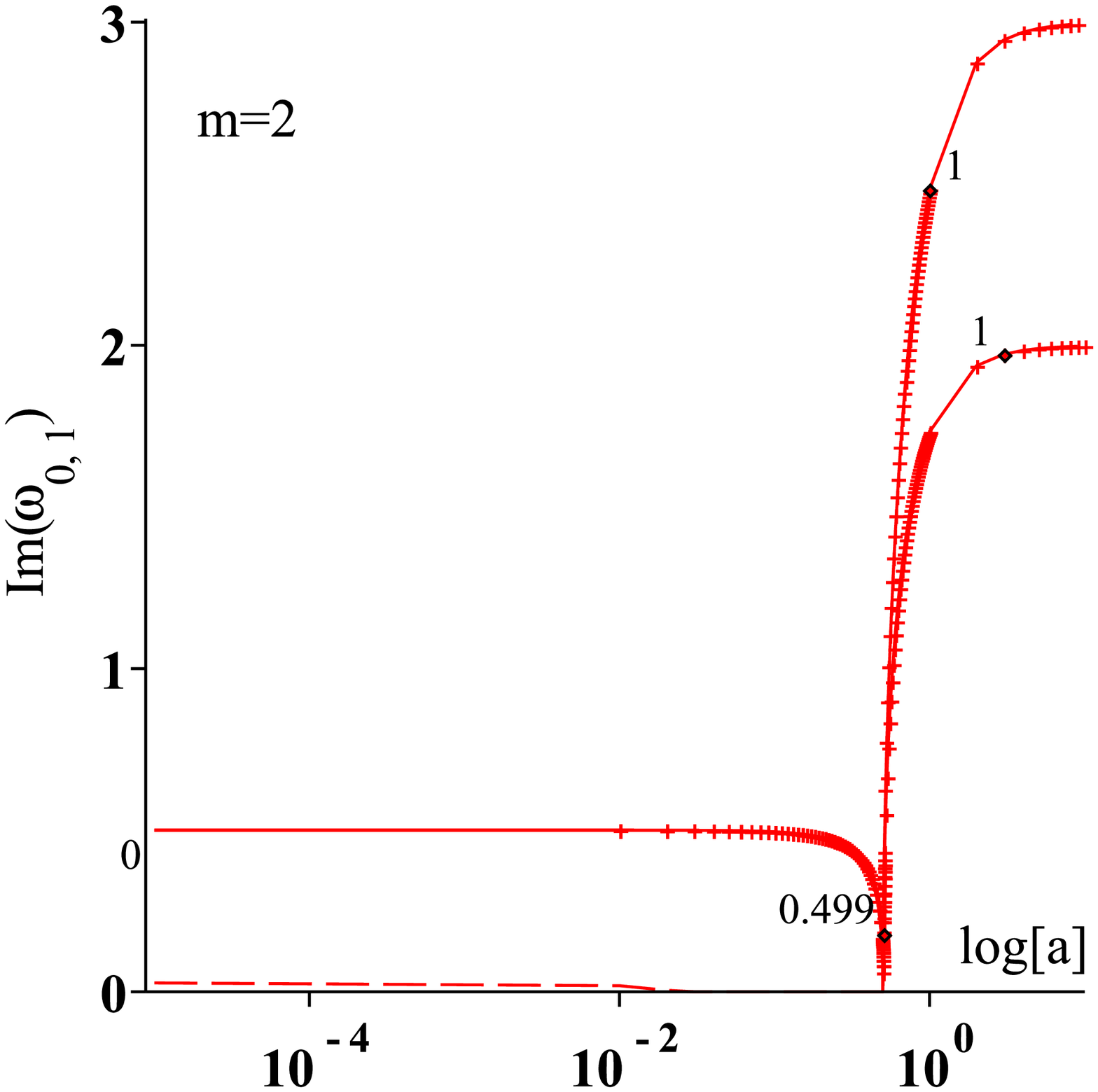} \\
\end{array}$
\end{center}
\caption{Dependence of the real and the imaginary part of two lowest modes ($n=0,1$) on $a$ for $m=0,1,2$.
At the bifurcation point $b=1$ ($a=M=0.5$), $\Im(\omega^{\pm}_{n=0,1;m}) \rightarrow 0$  implying a
critical event at this point. Also $\Re(\omega^{\pm}_{n=0,1;m}) \equiv m \Omega_{+}$ for $a<M$, $m\neq0$. On the plots $\omega^{+}_{0,m}$ is marked with the dashed line,  $\omega^{-}_{0,m}$ has been ommited. The solid lines and the crosses correspond to $\omega^{\pm}_{1,m}$ -- they mostly coincide for $m\neq0$}
\label{m:Re_Im}
\end{figure*}

The relation $\omega_{n,m}(a)$ is nontrivial as can easily be seen on Fig.
\ref{cp} and Fig. \ref{m:Re_Im}.

\begin{figure*}[htbp]
\begin{center}
$\begin{array}{lccccr}
a) \hspace{-15px}
\includegraphics[scale=0.27,keepaspectratio]{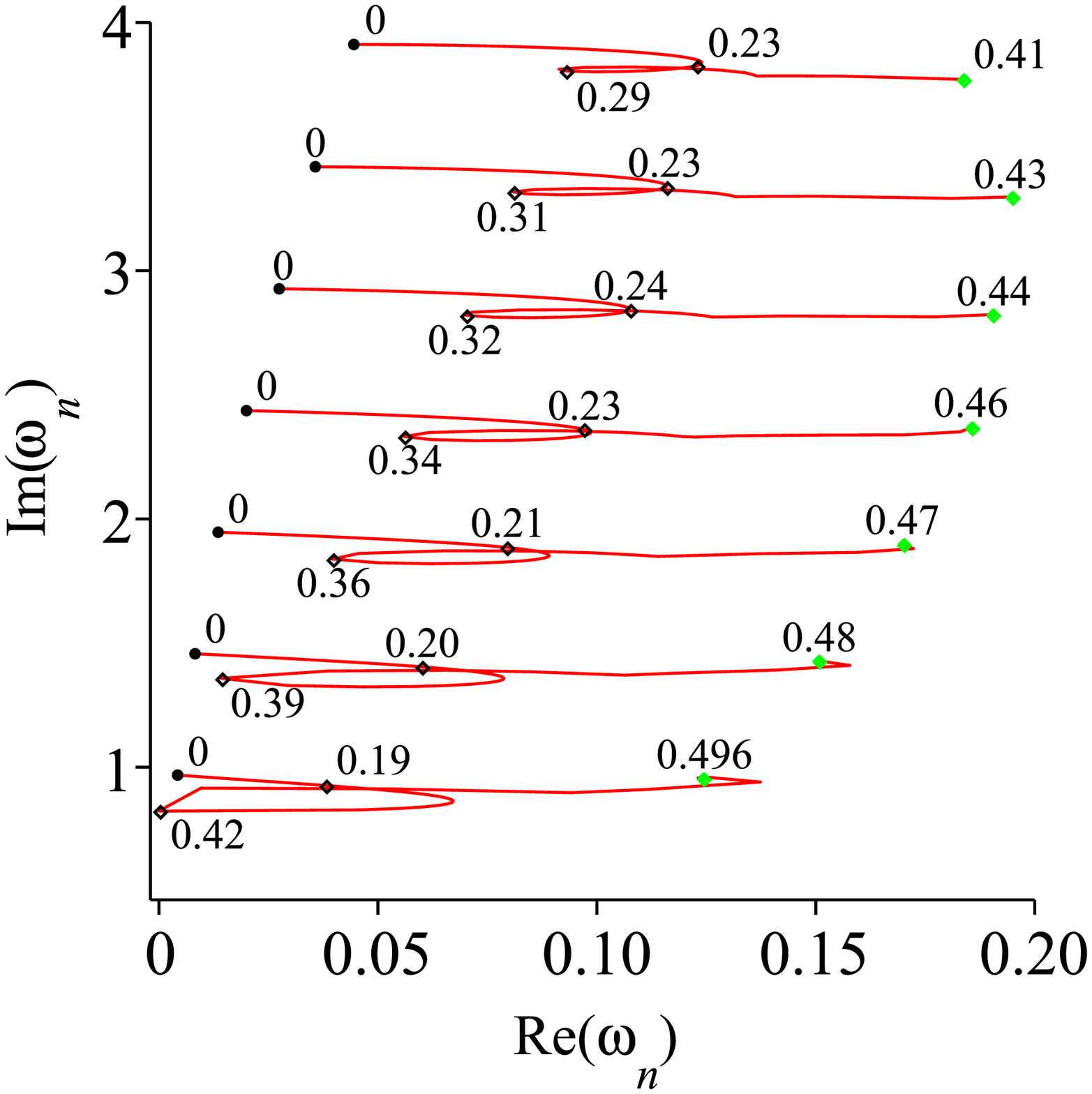} &
b)
\hspace{-15px}\includegraphics[scale=0.27,keepaspectratio]{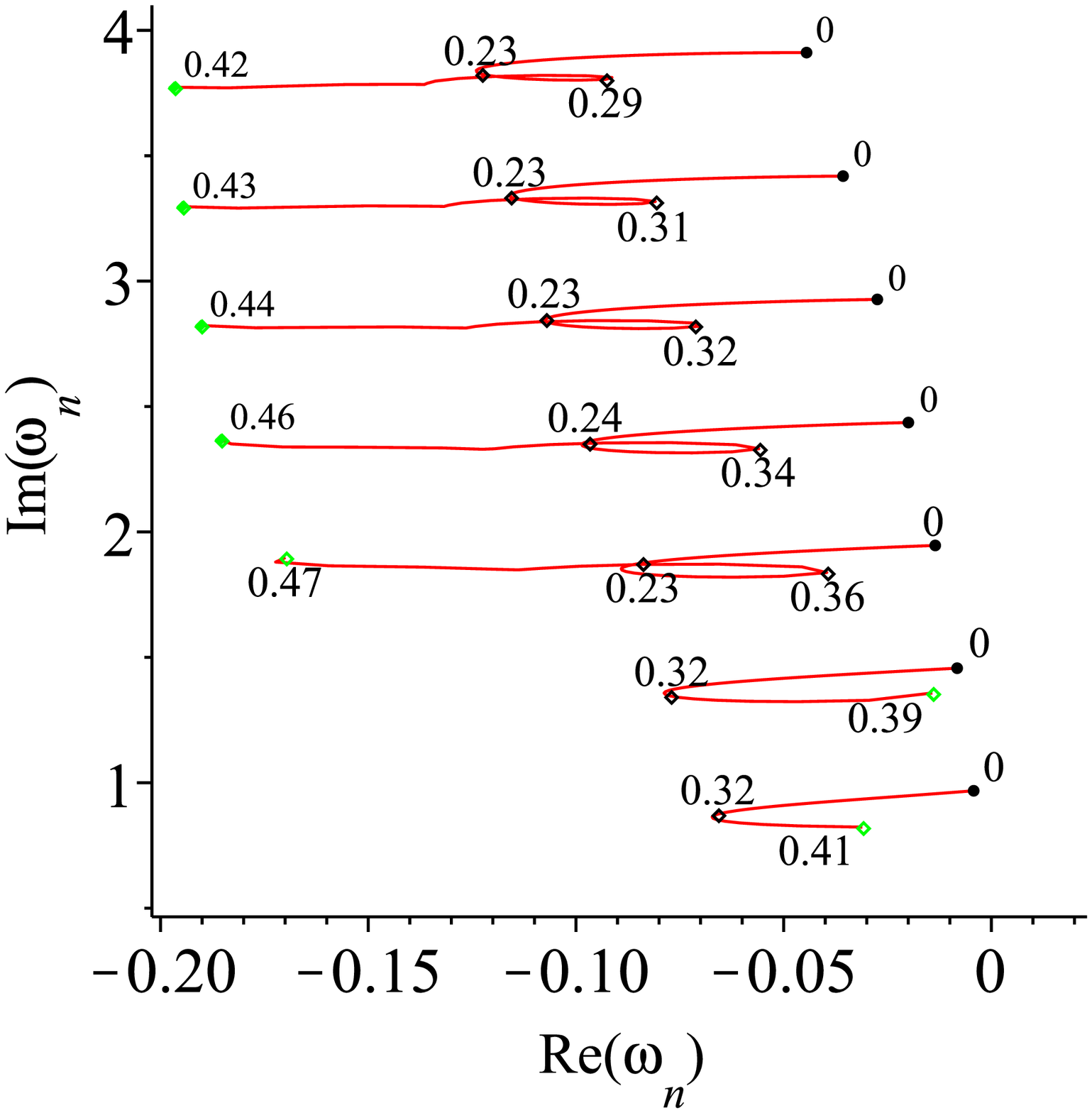}
&
c) \hspace{-15px}
\includegraphics[scale=0.27,keepaspectratio]{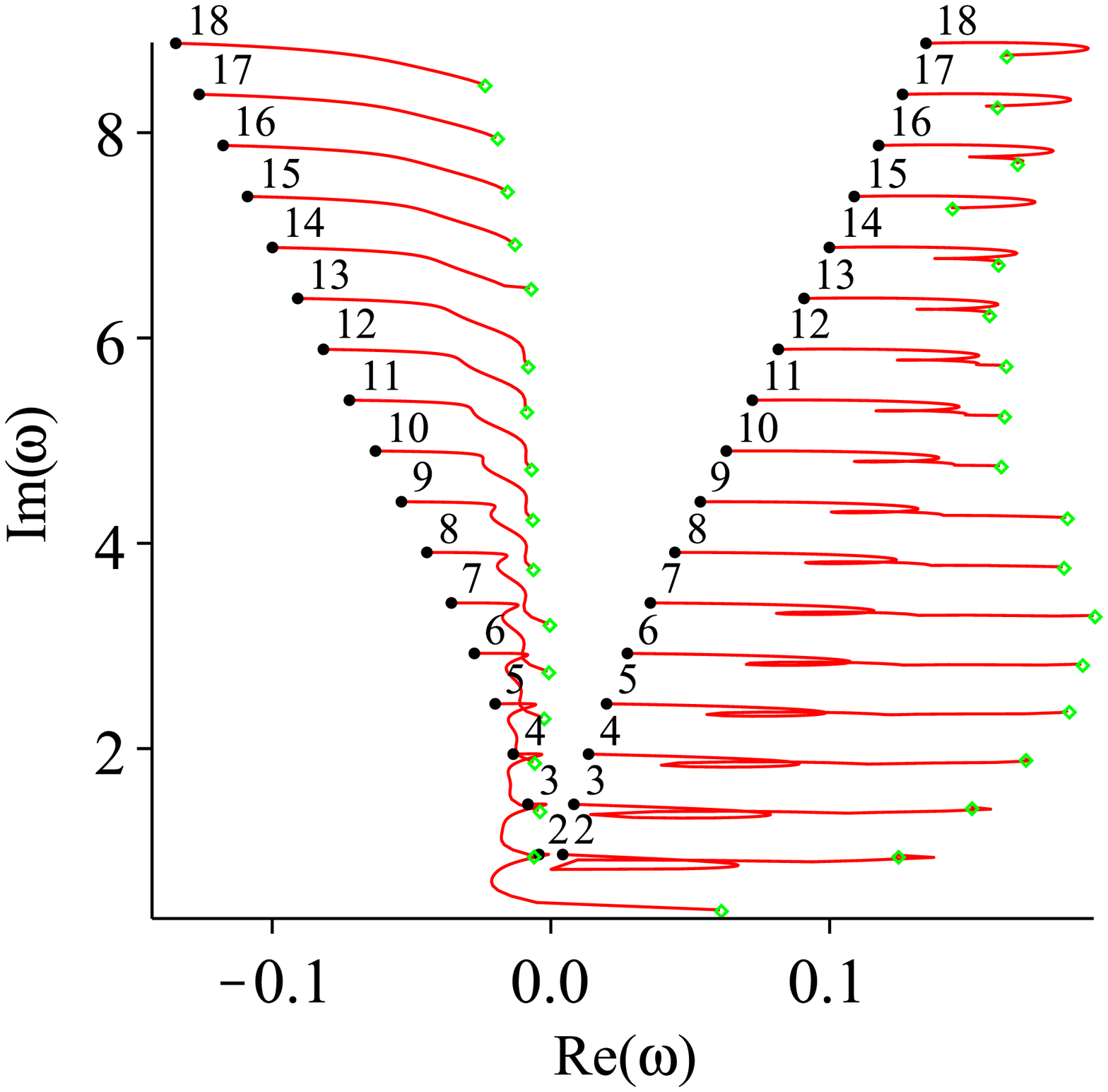}  \\

\end{array}$
\end{center}
\caption{Complex plot of some of the frequencies with different $n$
obtained in the case $m=0$, a) $\omega^{+}_{n,0}$, $n=2-7$, b) $\omega^{-}_{n,0}$, $n=2-7$,
c) $\omega^{+}_{n,0}$ for $n=2-18$ (the black dot stays for
$a=0$, green diamond -- the last $a$ where a root was found). The modes  $n=0,1$
are not plotted. For $m=0$, there is interesting symmetry between figs. a) and b) with respect to the vertical axis (see the text for details) }
\label{m0_loops}
\end{figure*}

The results we obtained can be summarized as follows:

\begin{itemize}
\item The two lowest modes $n=0,1$ we call \textit{special} and they are the only ones we could trace to $a>M$.
\item  For the special modes, for $a<M$ we observe
$\Re(\omega^{+}_{0,m})\approx \Re(\omega^{\pm}_{1,m})$ (for $m \neq 0$ and
$a>0.1$) while $\Im(\omega^{+}_{0,m})<< \Im(\omega^{\pm}_{1,m})$ (for $m\neq0$, $a>0.1$, $\Im(\omega^{\pm}_{0,m})<10^{-7}$;  in the case $m=0$, $\Im(\omega^{\pm}_{0,0})$ decreases
more slowly). Since the real parts of the two modes are almost equal for $a<M$ and
the imaginary part of the $n=0$ mode is negligible, in our figures in that range,
we emphasize only the complex frequencies $\omega^{\pm}_{1,m}$.

It's important to note that $\omega^{-}_{0,m}$ (for
$m>0$, $a<M$) has different behavior from $\omega^{+}_{0,m}$ -- it has very small real
and imaginary parts (compared to all the other modes), and since our numerical precision for it is low, we would not discuss that mode.  For $m<0$ the situation reverses and  $\Re(\omega^{-}_{0,m})\approx \Re(\omega^{\pm}_{1,m})$, while $\omega^{+}_{0,m}$ becomes the peculiar one.

\item For $m>0$ the real part of the special frequencies $\Re(\omega_{n=0,1,m})$ decreases steadily with the increase of $a$, until one of the modes has a minimum at the point $a = M$, the other -- at $a > M$, but close to it -- see Fig. \ref{m:Re_Im} c), e). In the case  $m=0$, $\Re(\omega_{1,0})$ has a maximum for $a<M$ and then has a minimum for $a>M$ (Fig. \ref{m:Re_Im} a) ).

\item The imaginary part of the special frequencies $\Im(\omega_{1,m})$ (Fig. \ref{m:Re_Im} b), d), f))
for all the cases stays positive,
it decreases for $a<M$ and it tends to zero for $a=M$. For $a=M$, the
solutions of the TRE can no longer be expressed in terms of confluent Heun
functions. In this case, one has to use biconfluent Heun function
(\citeauthor{Fiziev0908.4234}). We didn't calculate at this point. When
$a>M$, $\Im(\omega_{n,m})$ increases again until it reaches almost constant
value at high $a$.

\item In the range $a>M$ , $m \neq 0$ we obtain two modes with different values of
the real and imaginary parts, but similar behavior. The real parts
of the modes both have minima, but at different points. The
imaginary part of both modes tends to zero as $a\rightarrow M$, but
for $a>>M$ they reach different constants with
$\Im(\omega_{1,m})-\Im(\omega_{0,m})\approx 1$. These two modes
appear for $m=\pm1,\pm2,\pm3$ (Fig. \ref{cp} and Fig.
\ref{m:Re_Im}). We couldn't find other modes for $a>M$ using our
numerical methods.

\item Modes with $n>1$ demonstrate highly non trivial behavior and strong
dependence of the parameter $a$, even though numerically they cannot be
traced to $a>M$ and in some cases we are able to fix a very limited number of
points.

For $m=0$, those modes persistently demonstrate signs of loops, (see Fig.\ref{m0_loops} and Fig.\ref{m0n1ed}), which seem to disappear for $m\neq0$ (see Fig. \ref{m1_loops}, Fig. \ref{m1sn}).

The real parts of the modes with $n>1$ seem to form a surface whose physical
meaning is yet unknown, while their imaginary parts split in two
with increasing of the rotational parameter (Fig.\ref{m0ed}, Fig.\ref{m1sn}).

Note that the BHBC requires frequencies with
$\Re(\omega_{n,m})>-m\Omega_{+}$ or $\Re(\omega_{n,m})<0$ (for
$m<0$!) to be obtained using the solution $R_{2}(r)$, while
frequencies with $0<\Re(\omega_{n,m})<-m\Omega_{+}$ to be obtained
using  the solution $R_{1}(r)$. Since working numerically with the
solution $R_{1}(r)$ so far has been challenging, we present here
only frequencies obtained with the solution $R_{2}(r)$. Thus, on
Fig. \ref{m1sn}, only frequencies whose real part is above the $n=1$
mode plotted in magenta (or below $\Re(\omega)=0$) satisfy BHBC.
This, however, is the case only for $m\neq0$. For $m=0$, according to the BHBC the solution $R_{2}(r)$ can be used in the whole range $\Re(\omega_{n,0})\in (-\infty,+\infty)$ and all the frequencies we
obtained represent ingoing in the horizon modes.

\item The frequencies $\omega^{+}_{n,m}$ and $\omega^{-}_{n,m}$ are
symmetric for $m=0$ (Fig. \ref{fig_05}) for $n>0$, and they coincide
in the cases $m\neq0$, $n=1$ for $a\geq a_{m}$, where $a_{m}$ tends to zero
for sufficiently big $m$ (for $m=1$, $a_{m}=0.1$, for $m=2$, $a_{m}=0.04$,
for $m=5$, $a_{m}=0.02$, for $m=10$, $a_{m}=0$). For $n>1$,
$\omega^{+}_{n,m}$ and $\omega^{-}_{n,m}$ do not coincide.

\begin{figure}[htb]
\begin{center}
$\begin{array}{lr}
\multicolumn{1}{l}{\mbox{\bf (a)}} & \multicolumn{1}{l}{\mbox{\bf (b)}}\\ \\
\hspace{-0.25in}
\includegraphics[scale=0.2,keepaspectratio]{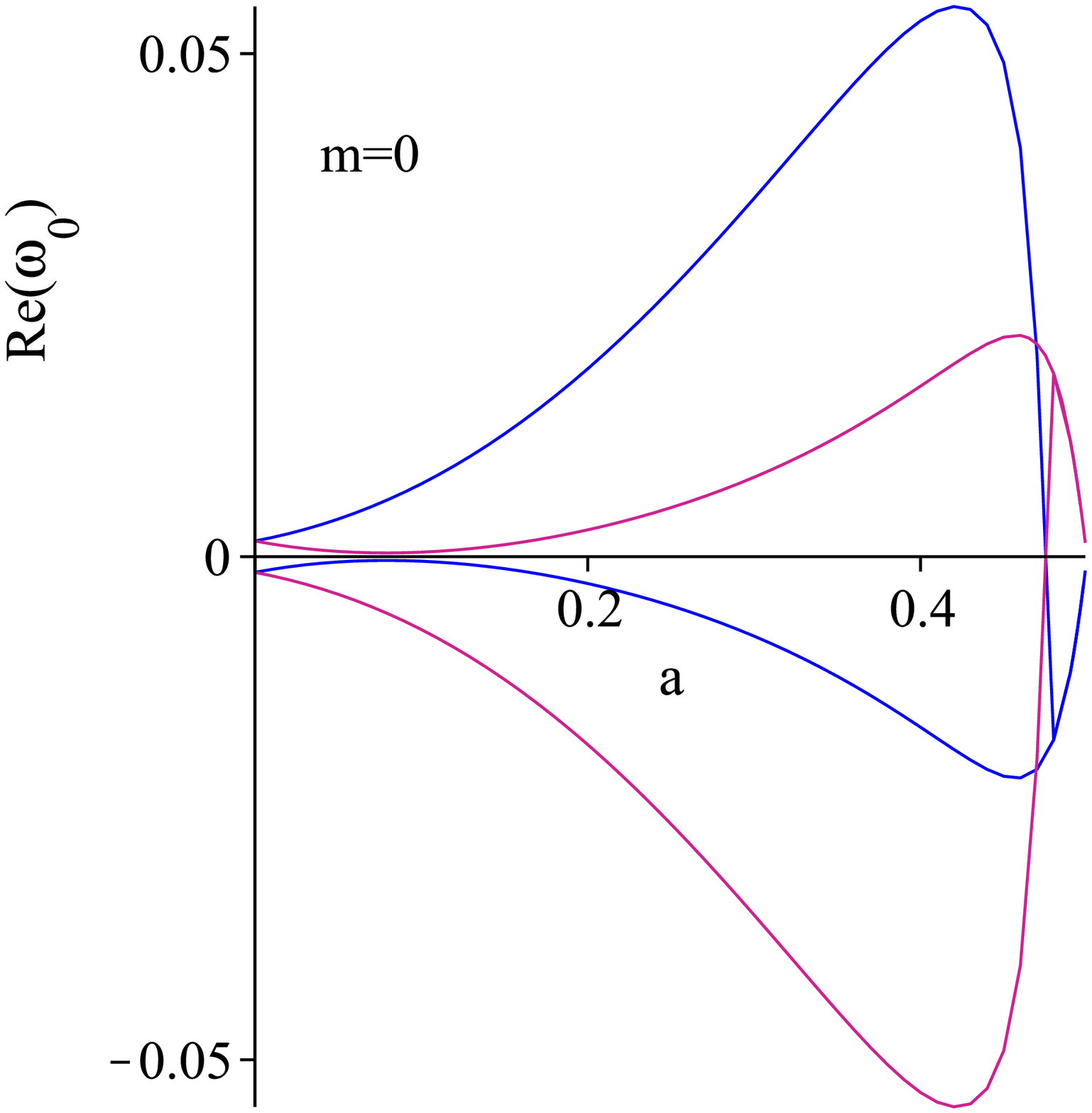}
&\hspace{-0.15in}
\includegraphics[scale=0.2,keepaspectratio]{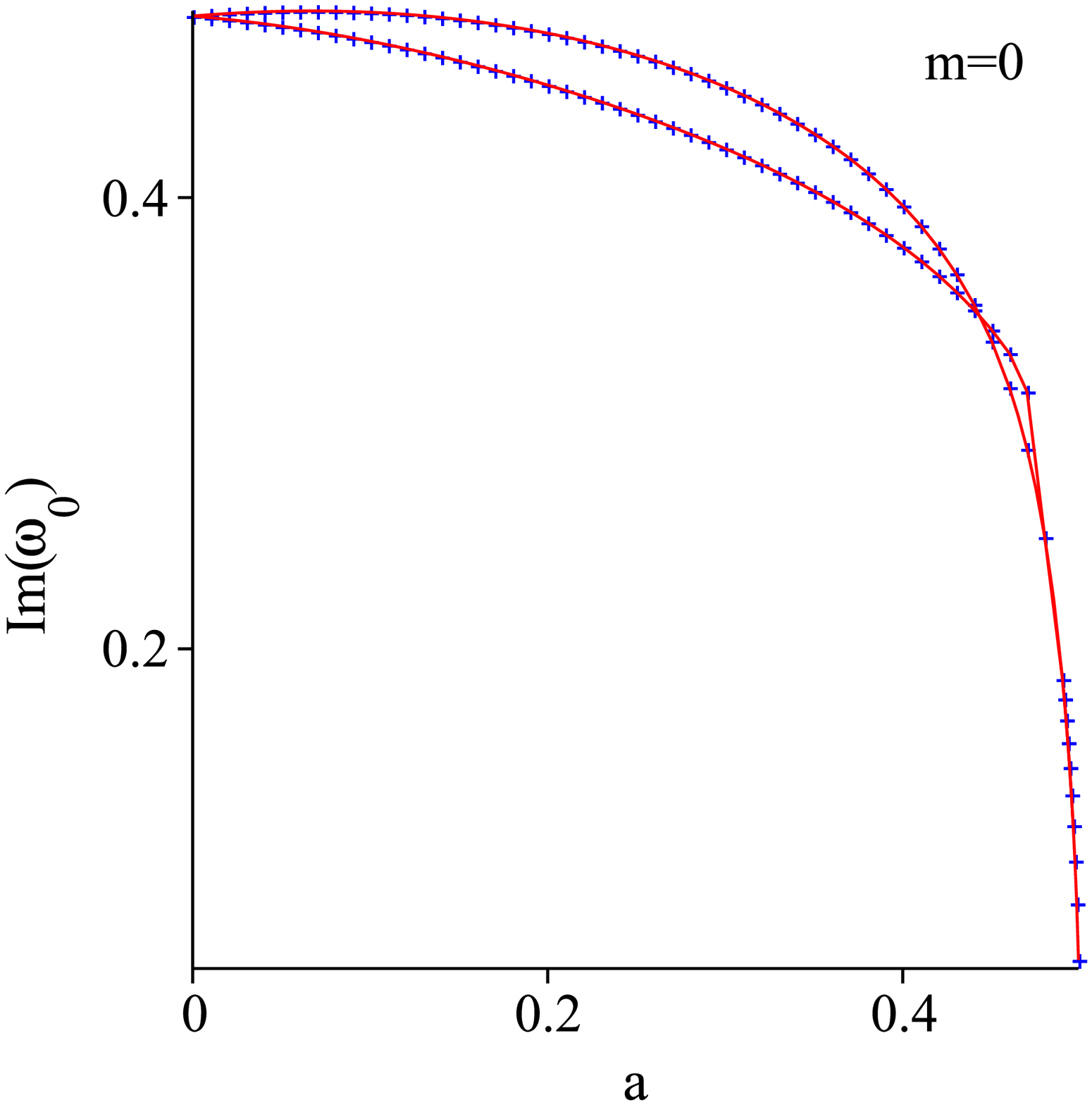}
\hspace{-0.15in} \\
\end{array}$
\end{center}
\caption{A particular case of the symmetry observed in the frequencies $\omega^{\pm}_{1,0}$ for $a<M$. When $m \neq 0$ $\omega^{+}_{n,m}$ and $\omega^{-}_{n,m}$ coincide for $a>a_{m}$}
\label{fig_05}
\end{figure}

\item There is symmetry in our spectra, which is confirmed for $m=\pm 0,
1, 2$ for
all of the modes with $n>0$ (except at some points):
\begin{align*}
\Re(\omega^{1,2}_{n,m})=-\Re(\omega^{2,1}_{n,-m}),\,
\Im(\omega^{1,2}_{n,m})=\Im(\omega^{2,1}_{n,-m}),
\end{align*}
\noindent where $\omega^{1,2}$ correspond to the two frequencies from the pairs observed for $a=0$ (frequencies with positive and negative real part for each $m,n$ on Fig. \ref{fig:m0_all}).

This symmetry allows us to study, for example, only the case $m>0$ if or when
it works numerically better.
\end{itemize}

From the figures, one can clearly see that, for every $m$, the relation
$\omega_{1,m}(a)$ demonstrates systematic behavior characterized by some kind of critical event at the bifurcation point $b=M/a=1$, where the damping of the
perturbation tends to zero ($\Im(\omega_{n,m})\to 0$). Physically, this transition
corresponds to the change of the topology of the ergo-surface visualized on
Fig. \ref{ergo}.

%%%%%%%%%%%%%%%%%%%%%%%%

\begin{figure*}[htb]
\begin{center}
$\begin{array}{lcccr}
\hspace{-0.3in}\epsfxsize=1.65in
\epsffile{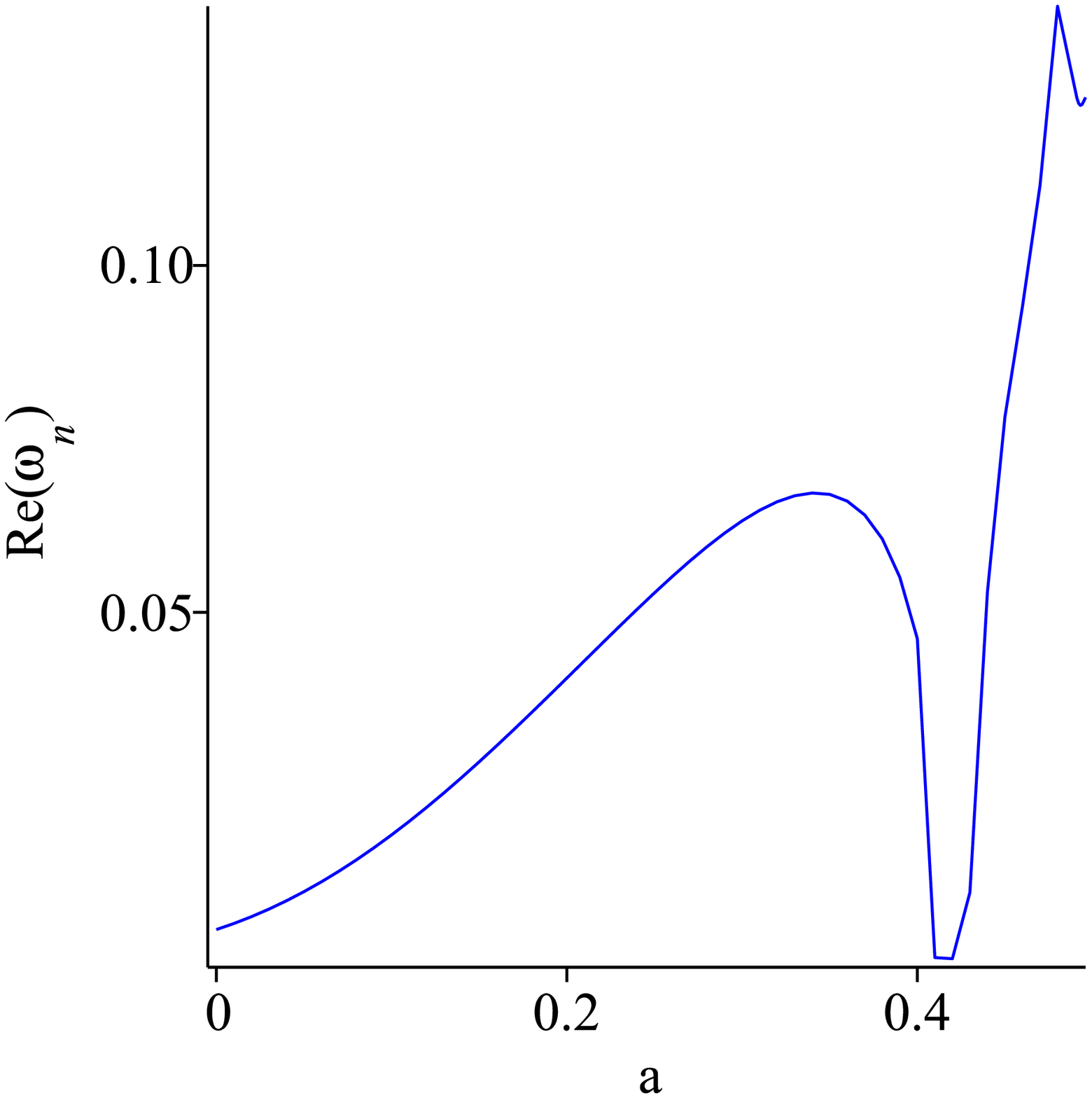} &\hspace{-0.1in}
\epsfxsize=1.65in
    \epsffile{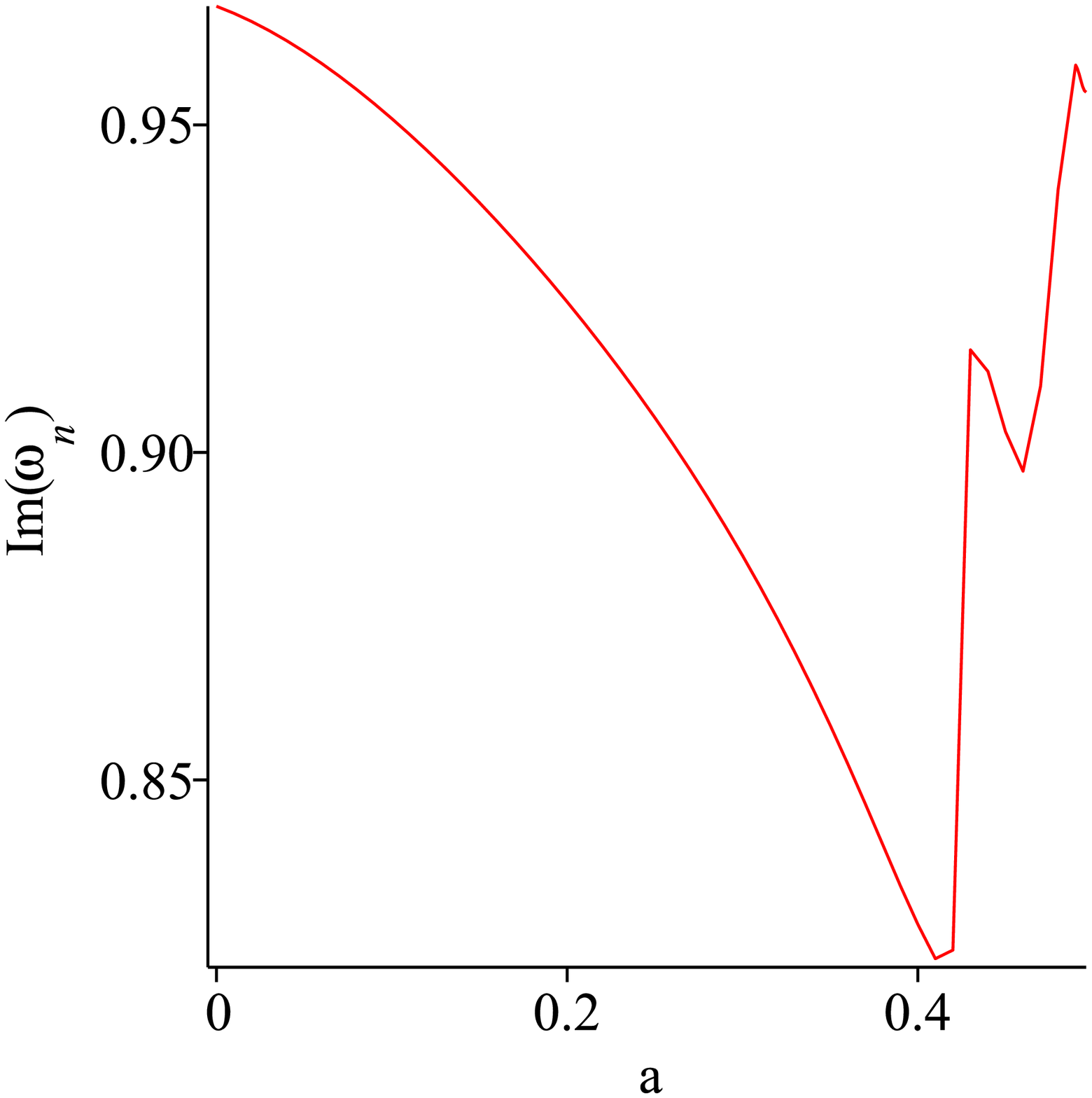} &\hspace{-0.1in}
\epsfxsize=1.65in
\epsffile{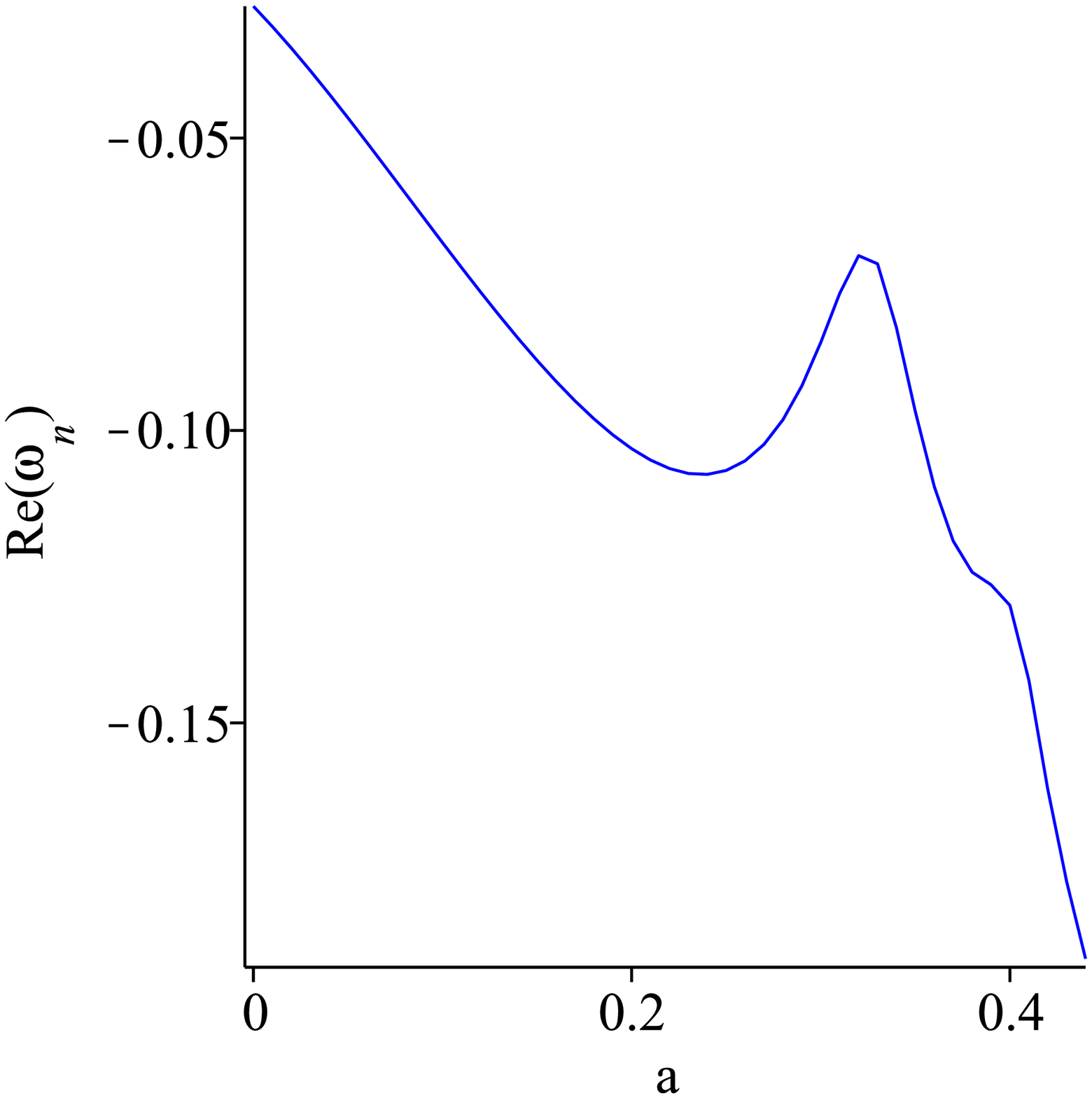} &\hspace{-0.1in}
\epsfxsize=1.65in
    \epsffile{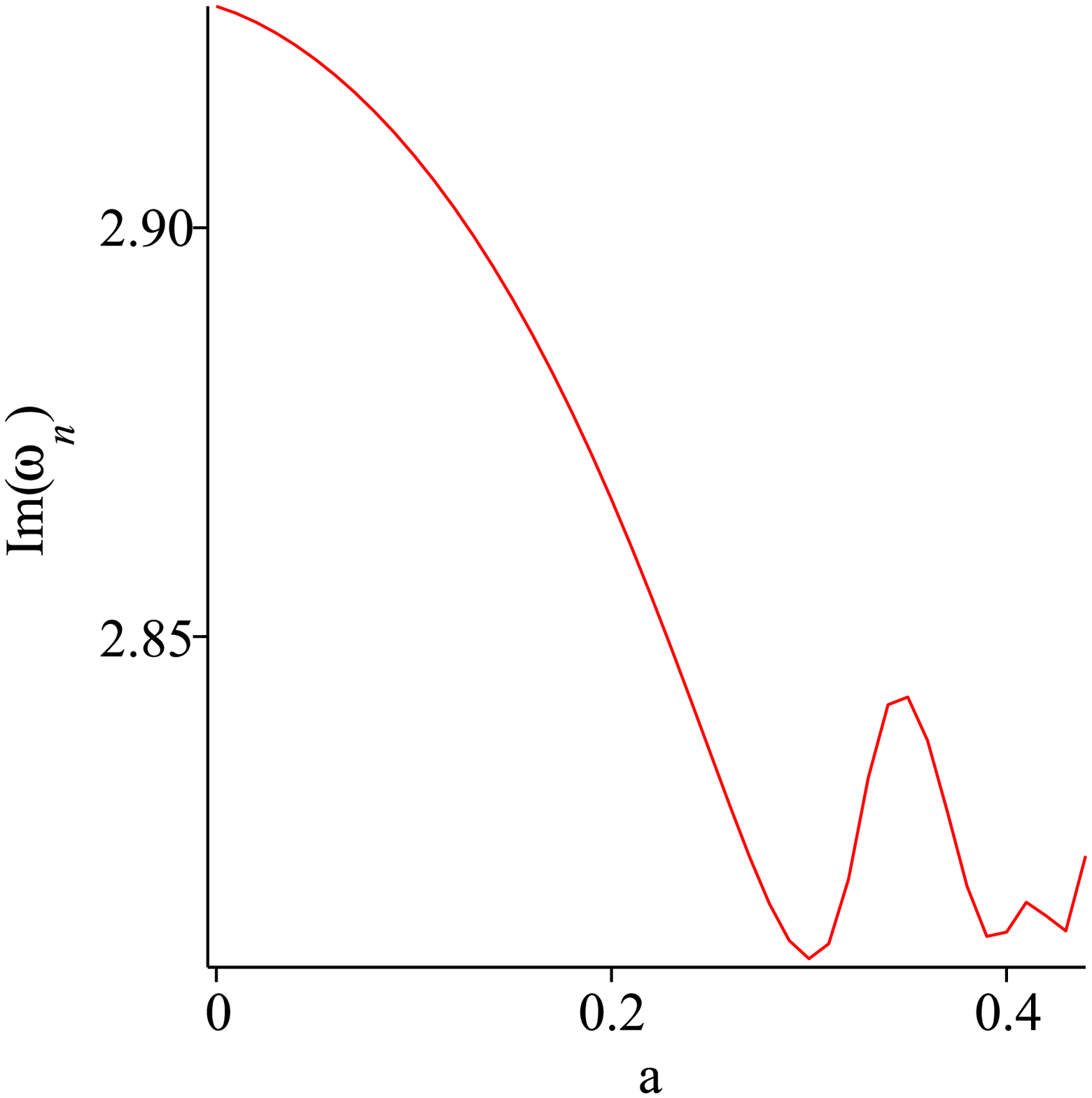} \hspace{-0.3in} \\
a)&b)&c)&d)\\
\end{array}$
\end{center}
\caption{Two cases showing the detailed behavior of modes with $n\!>\!1$ for $m\!=\!0$. a), b)
correspond to $n\!=\!2$; c), d) to $n\!=\!6$ }
\label{m0n1ed}
\end{figure*}

\begin{figure*}[htb]
\begin{center}
$\begin{array}{clclc}
\hspace{-0.3in} \epsfxsize=1.65in
\epsffile{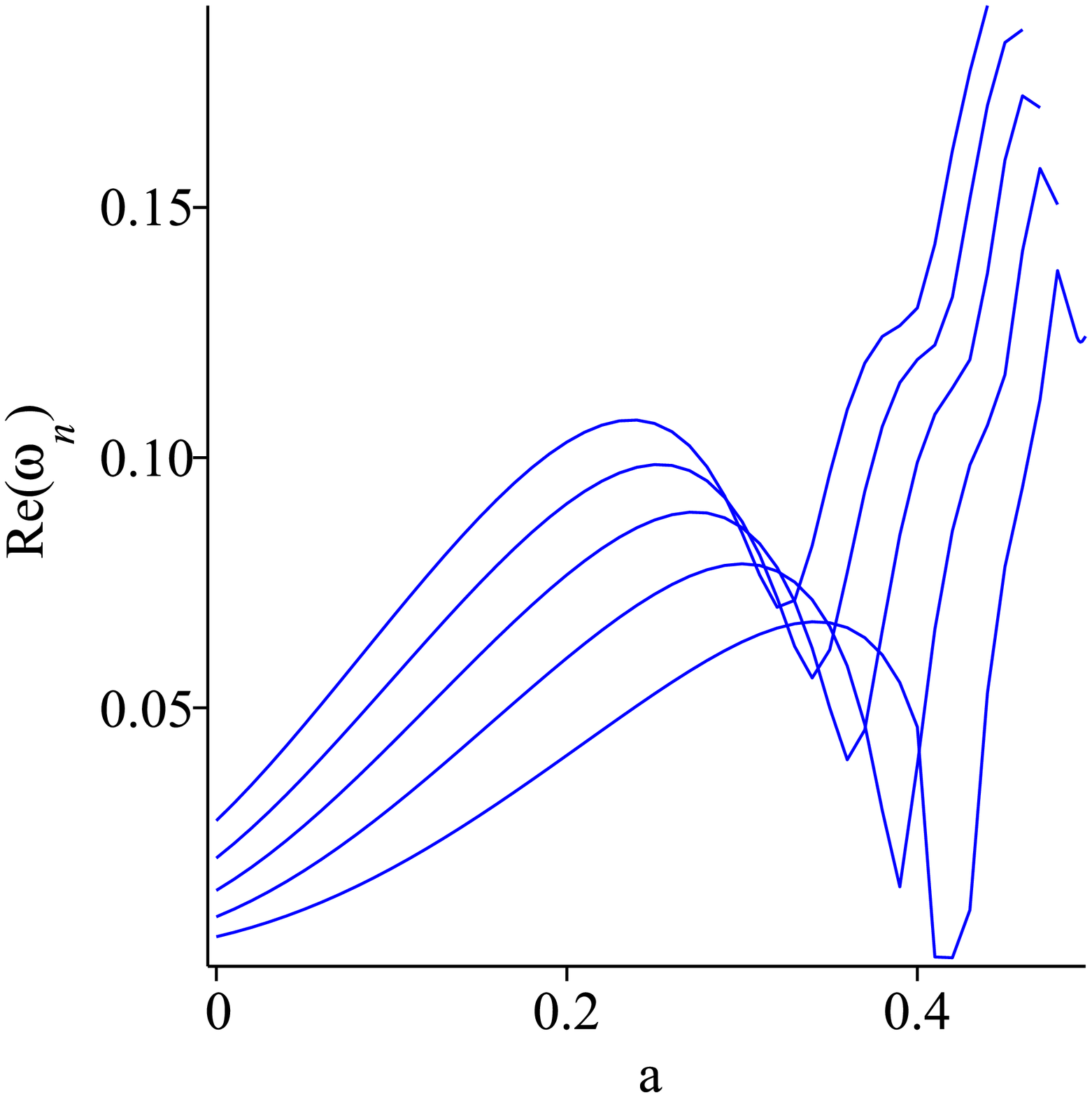} & \hspace{-0.1in}
\epsfxsize=1.65in
\epsffile{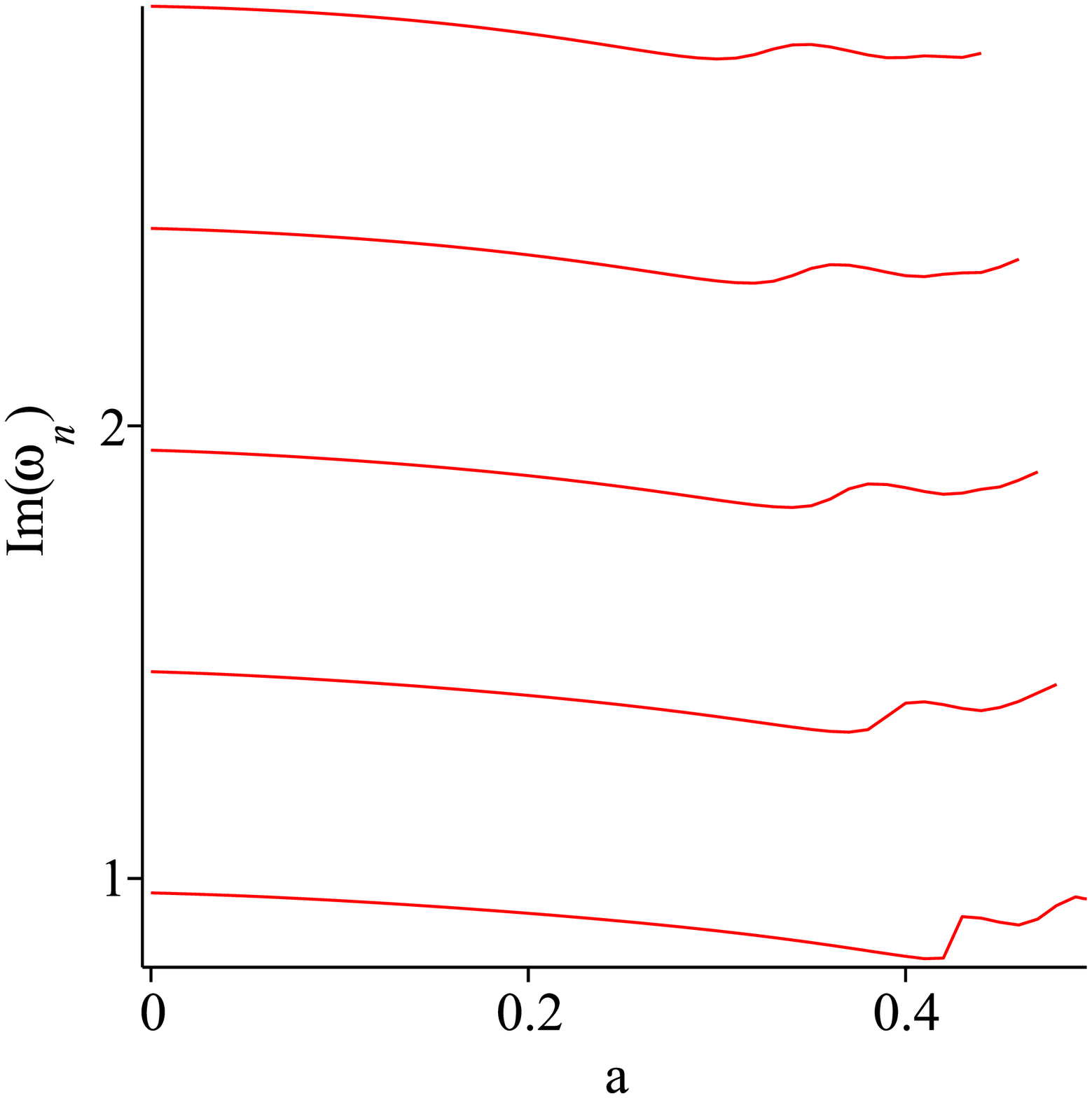} & \hspace{-0.1in}
 \epsfxsize=1.65in
\epsffile{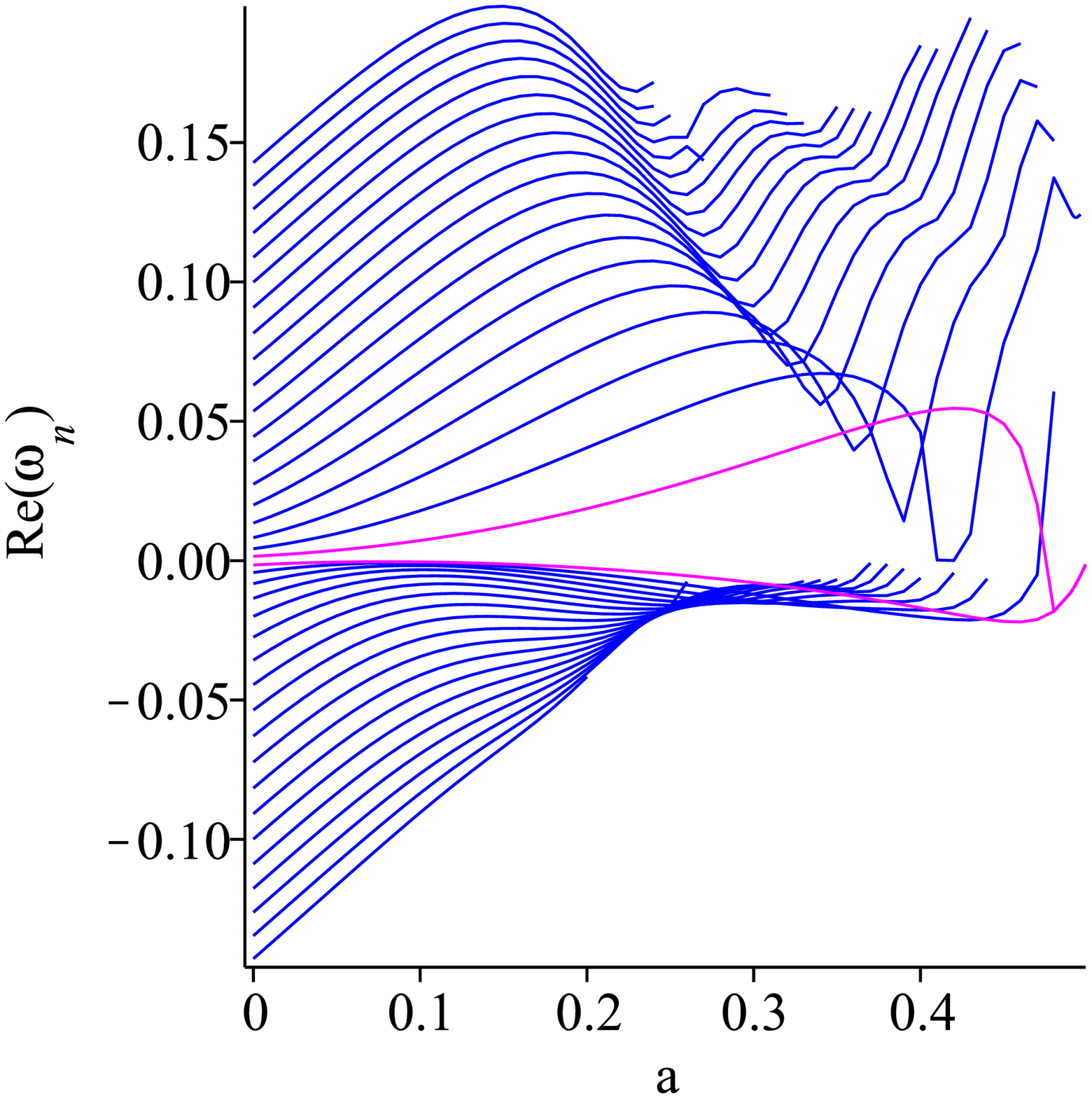} & \hspace{-0.1in}
 \epsfxsize=1.65in
\epsffile{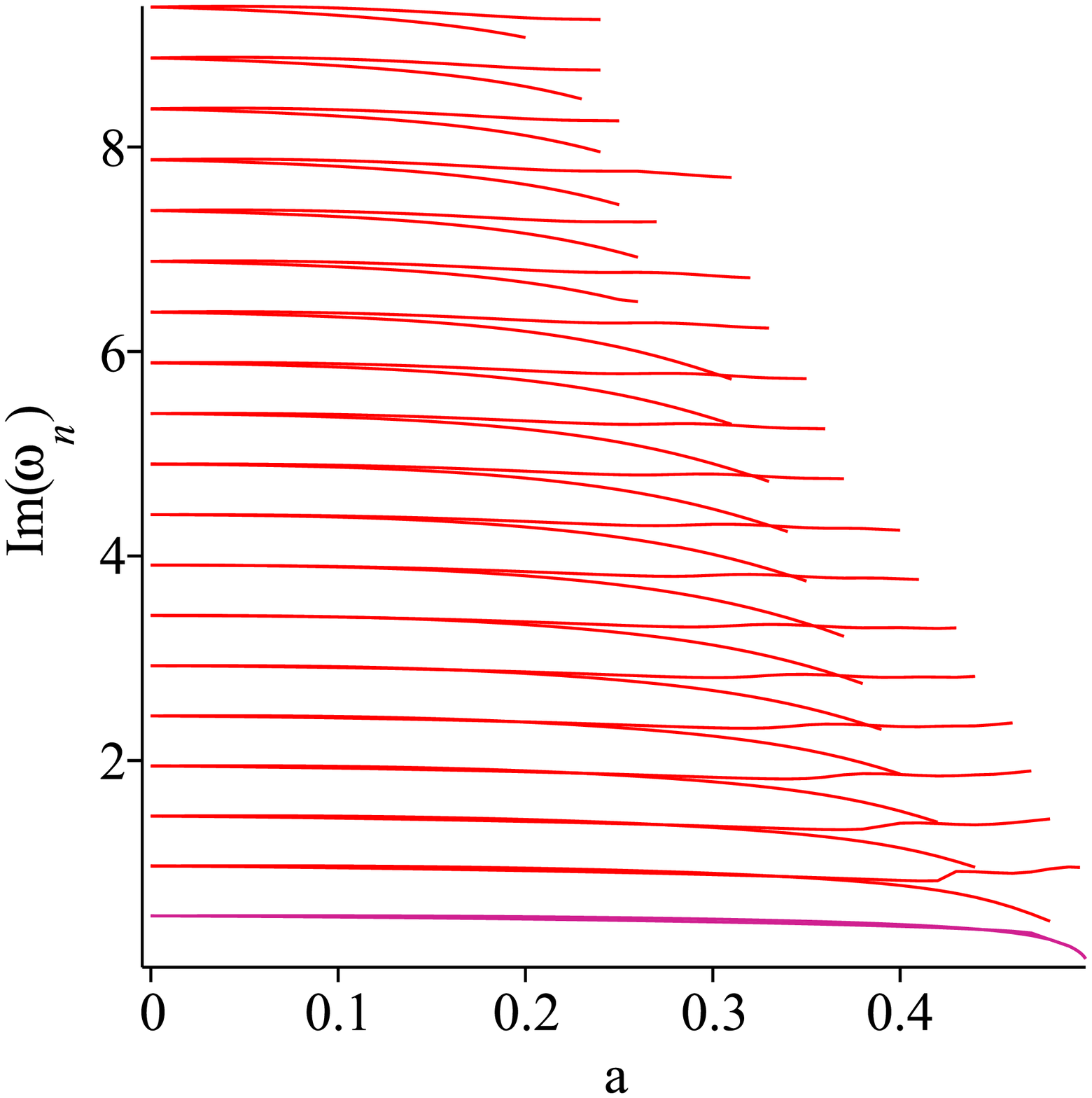} \hspace{-0.3in}\\
a) & b) & c) & d)\\
\end{array}$
\end{center}
\caption{The case $m=0$, modes with $n>1$ a), b): $\Re(\omega^{+}_{n,0})$ and $\Im(\omega^{+}_{n,0})$, $n=2-5$ c), d): $\Re(\omega^{+}_{n,0})$ and $\Im(\omega^{+}_{n,0})$, $n=1-18$. On c) and d) we plotted
the conjugate roots $\omega^{1,2}_{n,0}$ on one plot. The magenta lines correspond to the special mode $n=1$}
\label{m0ed}
\end{figure*}

\begin{figure*}[htb]
\begin{center}
$\begin{array}{lcr}
a)  \epsfxsize=2in
    \epsffile{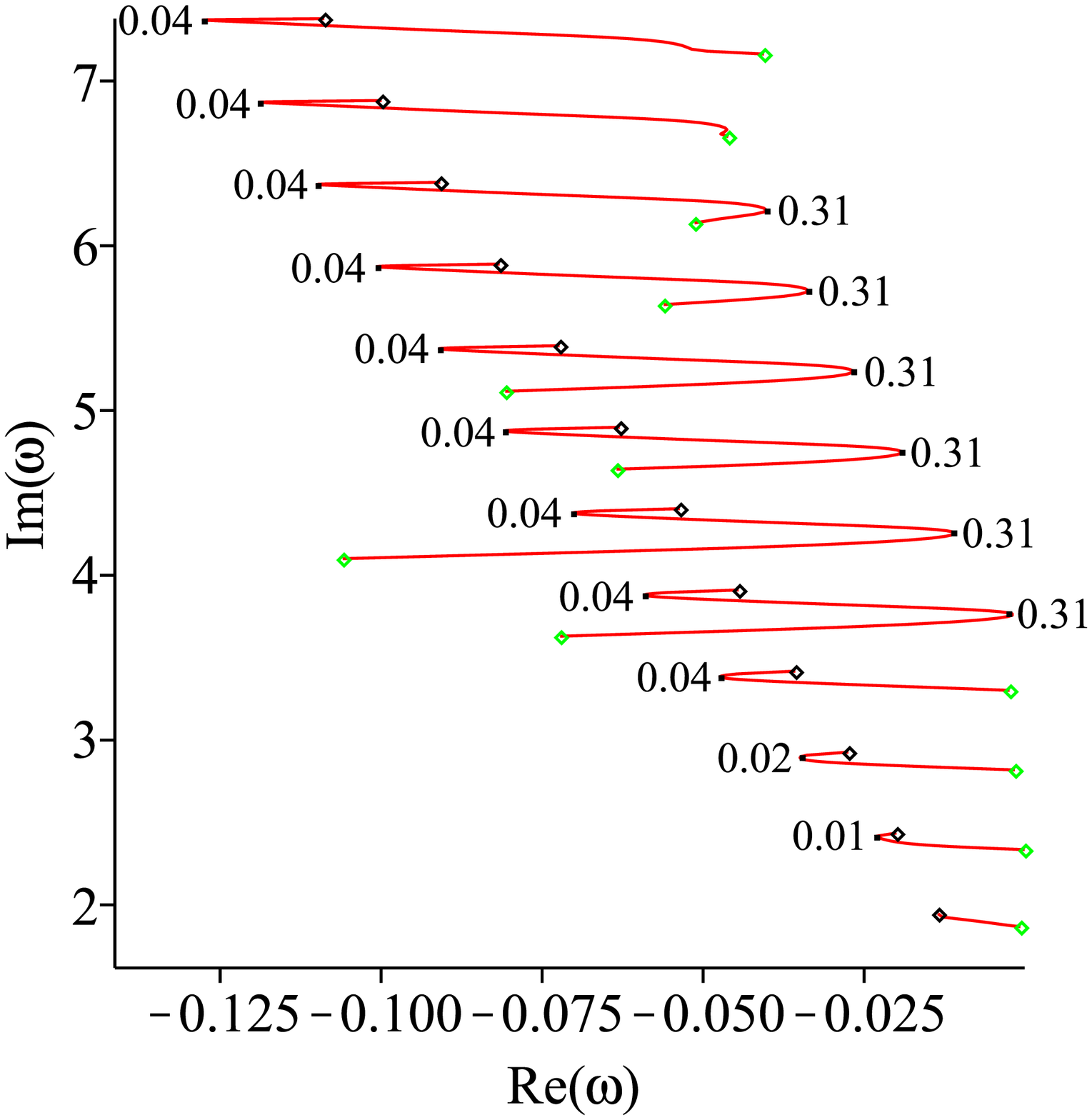} &
b) \epsfxsize=2in
\epsffile{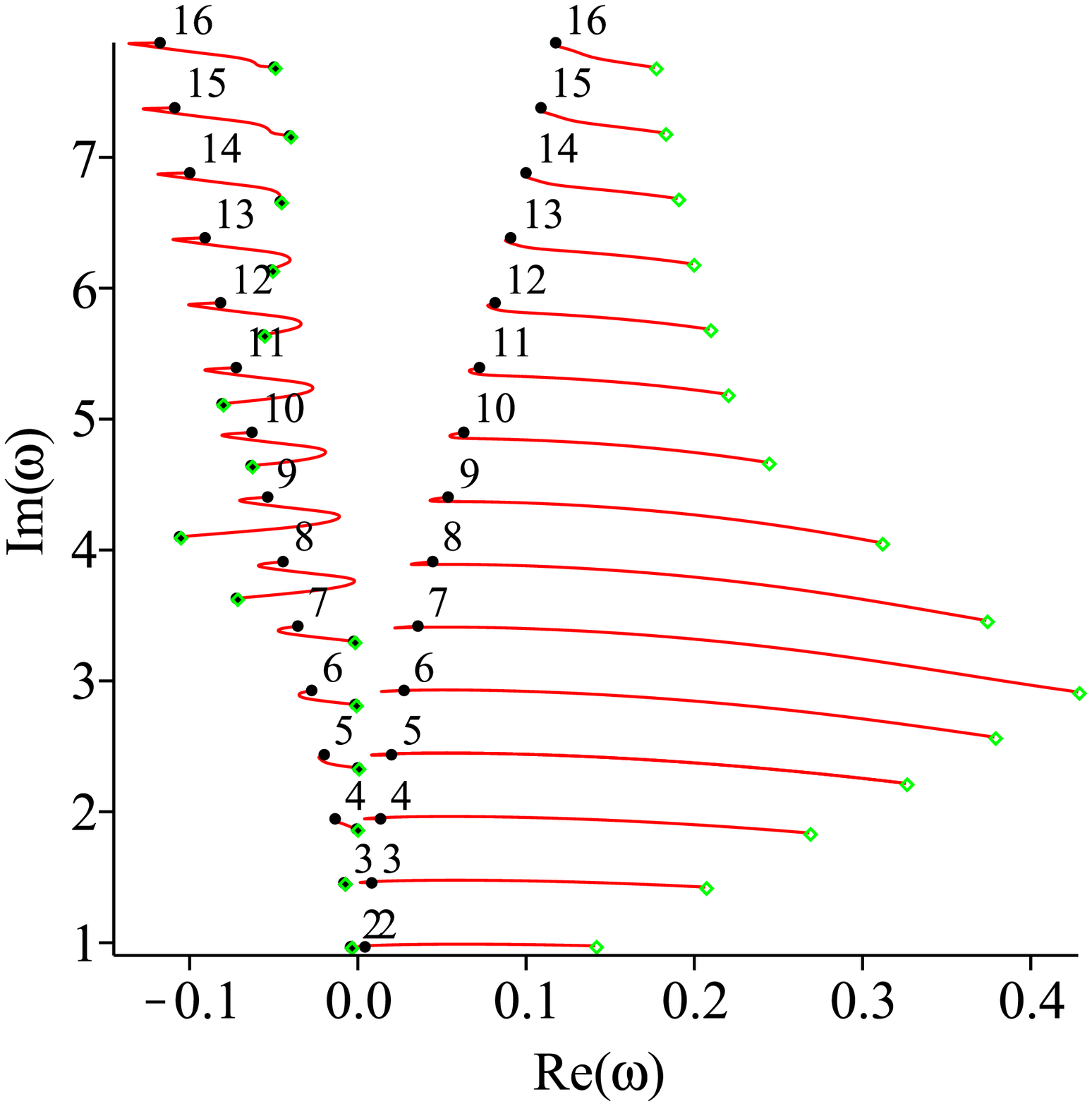} &
c)  \epsfxsize=2in
    \epsffile{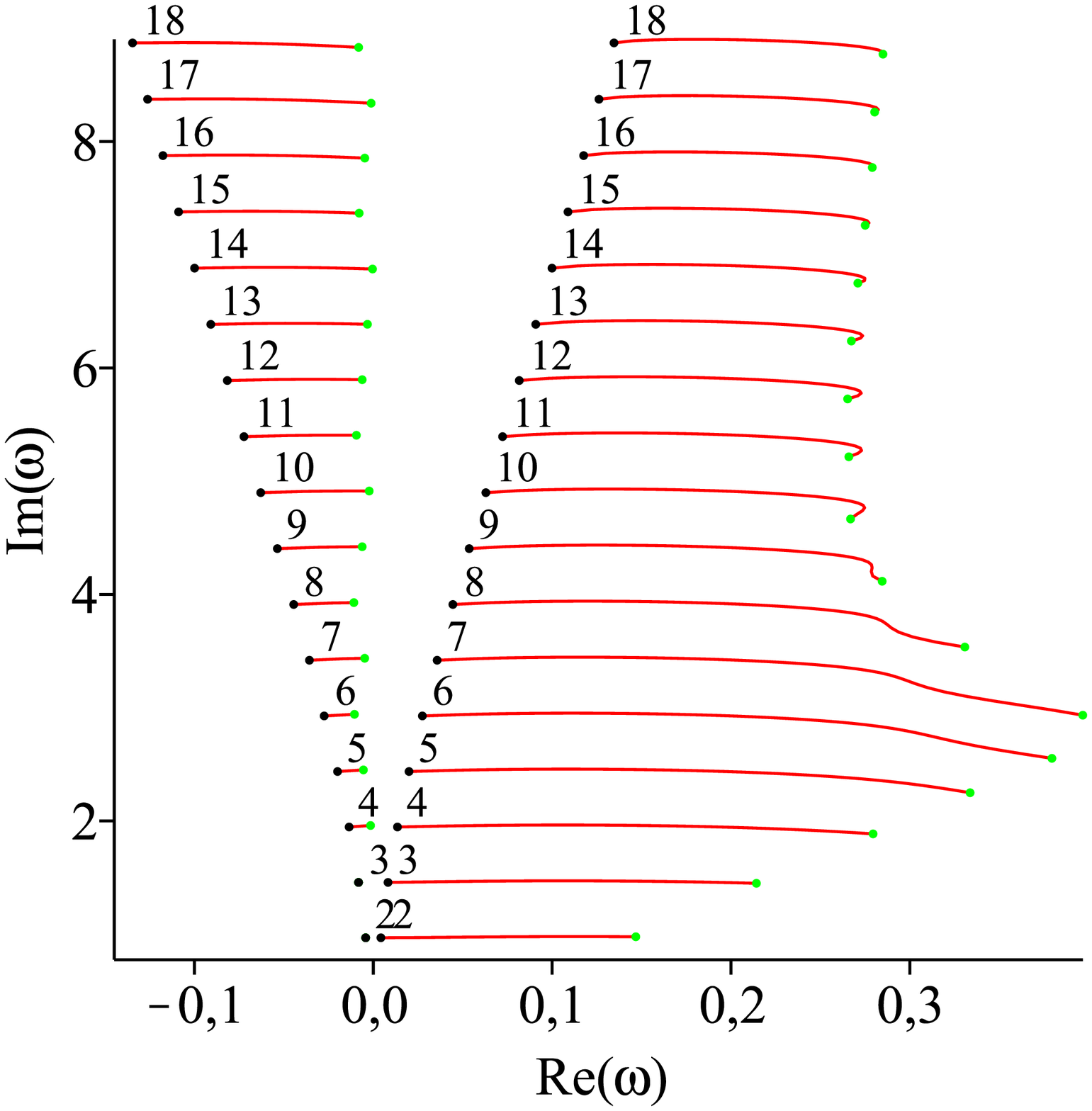}\\
\end{array}$
\end{center}
\caption{Complex plots in the case $m\!=\!-\!1$: a) $\omega^{+}_{n,-1}$, $n\!=\!3\!-\!14$ , b) $\omega^{+}_{n,-1}$, $n\!=\!2\!-\!16$, c) $\omega^{-}_{n,-1}$, $n\!=\!2\!-\!18$ . There are no loops for $\mid\! m \!\mid \!=\!1$}
\label{m1_loops}
\end{figure*}

\begin{figure*}[htb]
\begin{center}
$\begin{array}{lccr}
\hspace{-0.3in} \epsfxsize=1.89in
\epsffile{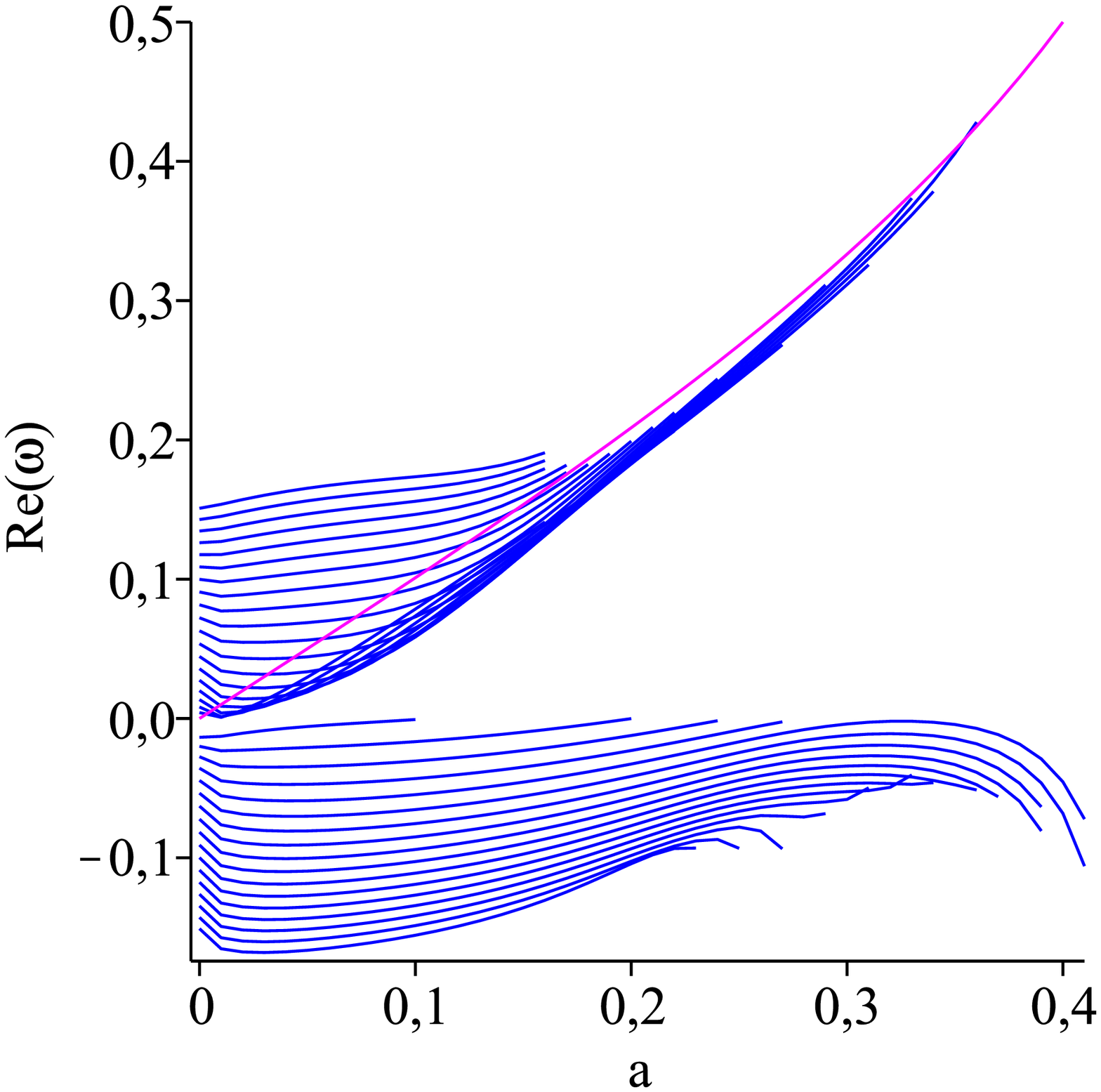} & \hspace{-0.4in}
    \epsfxsize=1.89in
    \epsffile{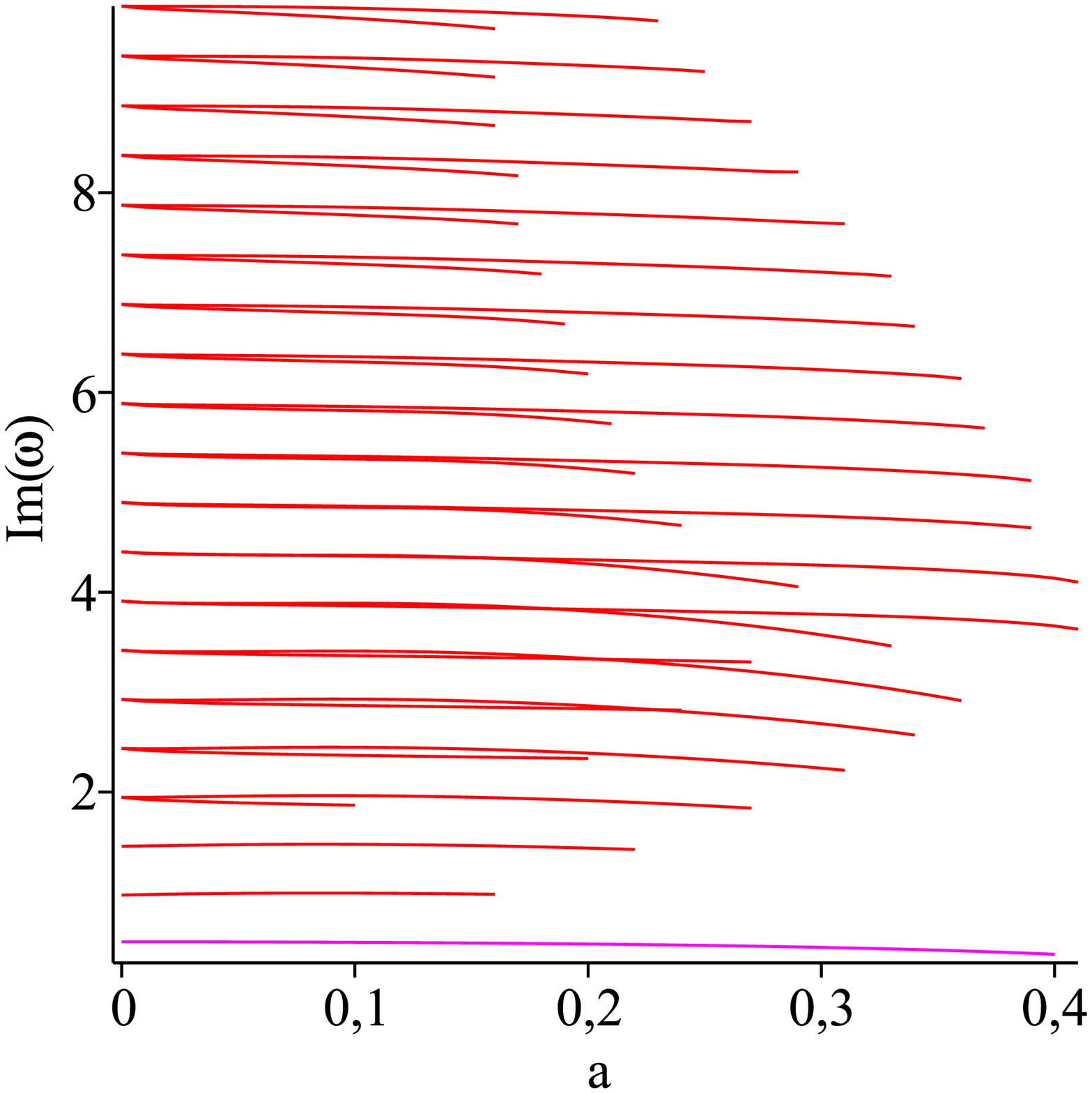} & \hspace{-0.4in}
\epsfxsize=1.89in
\epsffile{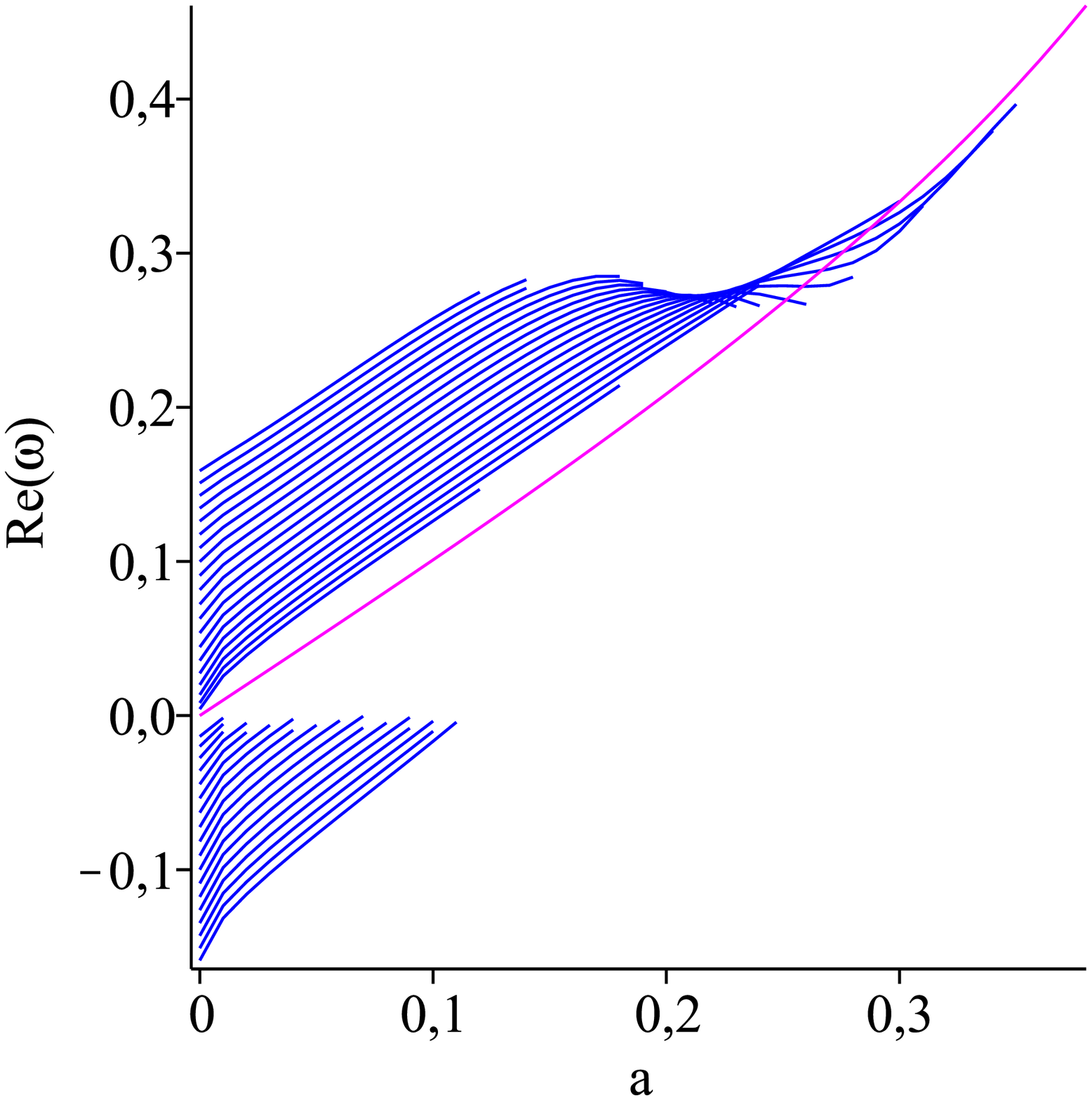} & \hspace{-0.4in}
    \epsfxsize=1.89in
    \epsffile{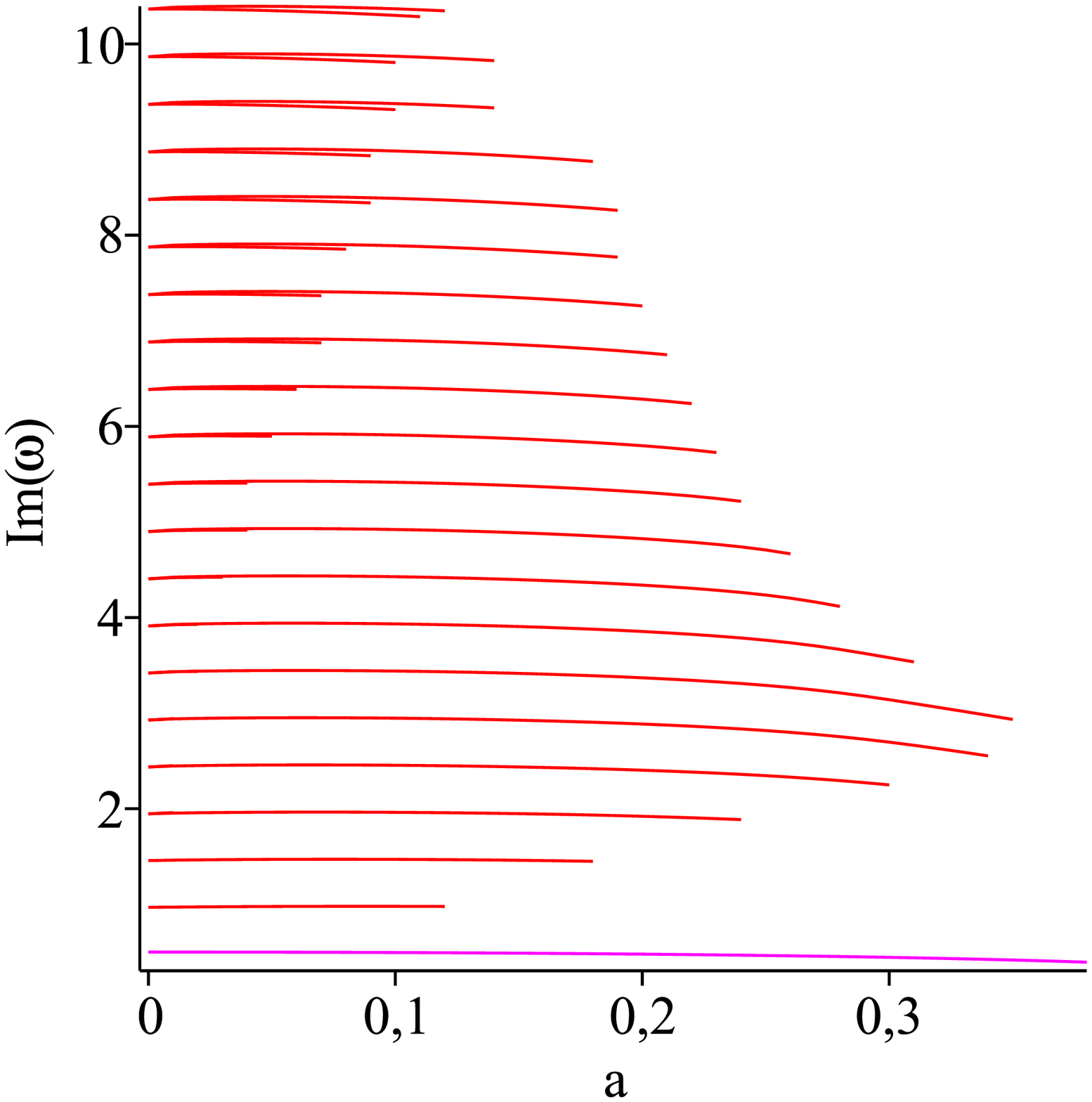}\hspace{-0.4in}\\
a)&b)&c)&d)\\
\end{array}$
\end{center}
\caption{The case $m=-1$, $n>0$: a) $\Re(\omega^{+}_{n,-1})$, b) $\Im(\omega^{+}_{n,-1})$, c) $\Re(\omega^{-}_{n,-1})$, d) $\Im(\omega^{-}_{n,-1})$. The magenta line marks the \textit{special} mode $n=1$ }
\label{m1sn}
\end{figure*}

%%%%%%%%%%%%%%%%%%%%%%%%

\subsection{Analysis of the results}

The spectra we presented demonstrates a clear transition at the bifurcation point $a=M$,
matching the transition from a rotating black hole to an extremal black hole demanded by the theory of black holes.
Since we can approach the point $a=M$ from both directions, it seems natural to speak also of extremal naked singularity,
for $a\rightarrow M$ but $a>M$.

From our figures it is clear that the behavior of the spectra when $a\rightarrow M$ from both sides is similar
-- the imaginary part of the critical frequency tends to zero at those points
and quickly rises to a (different) constant for $a\neq M$.
We couldn't find any frequencies with negative imaginary parts in this case;
thus, it seems that the primary jets from the KBH and the KNS are stable even in the regime of naked singularities.
This differs from the QNM case, in which QNM from naked singularities are unstable (\citeauthor{Press_Teukolsky}).

Another example of unstable spectra can be found in \citeauthor{Dotti} where completely
different boundary conditions for the overspinning Kerr space-time are considered.
There, the authors analyzed the central two-point connection problem on the singular interval $r \in (-\infty, \infty)$.
As we see, in our case the primary jets are stable.
There is no contradiction since we study a completely different physical problem.
Besides, the discrete levels of the spectra of the primary
jets from the KBH and the KNS appear to be smooth
in the whole interval of the bifurcation parameter $b$, except for the bifurcation point $b=1$.

Another surprise is that numerically $\Re(\omega_{n,m}) \equiv - m \Omega_{+}$ for $a<M$ for modes $n=0,1$.
Here $\Omega_{+}=a/2Mr_{_+}$ is the angular velocity of the event horizon.

It is curious that we can represent the same relation in the form $\Re(\omega_{n,m})=-m\omega_{cr}$, for $n=0,1$, where $\omega_{cr}$ is the critical frequency of superradiance for QNM  \citeauthor{superradiance},
\citeauthor{superradiance2}, \citeauthor{superradiance3}, \\ \citeauthor{superradiance5},
\citeauthor{superradiance6}, \\ \citeauthor{Teukolsky2}, \citeauthor{superradiance7}, \citeauthor{superradiance9},\\ \citeauthor{superradiance8}).

The complexity of the frequency $\omega_{1,m}$ whose real part equals $-m\omega_{cr}$
may show another essential physical difference between our primary jets-from-KBH
and jets-from-KNS solutions and standard QNM,
obtained using regular solutions of the angular equation Eq. \eqref{TAE}.

In our case, we observe two (lowest) modes ($\omega^{+}_{0,m}$, $\omega^{\pm}_{1,m}$
for $m>0$) whose real parts coincide with the critical frequency of QNM-superradiance.
However, only $\omega^{+}_{0,m}$ is almost real, having a very small imaginary part.
The magnitude of the imaginary parts of the other two, $\omega^{\pm}_{1,m}$, is of the same order as that of their real parts.
To the best of our knowledge this is the first time that an imaginary part of such quantity,
somehow related with the critical QNM-superradiance frequency, is discovered
solving BH boundary conditions in pure vacuum (i.e. without any mirrors, additional fields, etc).
Note that while for $a<M$ the small imaginary part of $\omega^{+}_{0,m}$
speaks of slow damping, for $a>M$ the imaginary part seems to grow quickly to become
comparable with the real part value, thus ensuring stability.
The two lowest modes in our primary jet spectra are the only ones, which we can trace from $a < M$ to $a > M$.
Although the lack of the higher modes ($n>1$) for $a>M$ could be due to problems with the numerical routines we use,
its persistence for every $m$ implies that there may be a possible deeper physical meaning behind it.

It is interesting to investigate in detail the relation of our results with
the superradiance phenomenon.

Although the considered primary jet-perturbations
to the Kerr metric in general damp quickly for $b\neq 1$ ($a\neq M)$,
this situation changes in the limit $b\rightarrow 1$ --
the imaginary part of $\omega$ then tends to zero near the bifurcation point,
both for $b\rightarrow 1-0$ and $b\rightarrow 1+0$.
In this case, the perturbation will damp very slowly with time while oscillating violently,
opening the possibility for interesting new primary jet-phenomena which
deserve additional detailed consideration.

The best fit for our numerical data for the lowest modes ($n=0,1$) turns out to be the exact formula
$(3.4a)$ in \citeauthor{Fiziev0908.4234}:
\begin{equation}
\hspace{15px} \omega_{n=0,1,m}=(-m + iN\sqrt{b^2-1}\,)\,\Omega_{+}, \,\,\,\,N=0,1.
\label{omega}
\end{equation}

\begin{figure*}[htbp]
$\begin{array}{lr}
\includegraphics[scale=0.35,keepaspectratio]{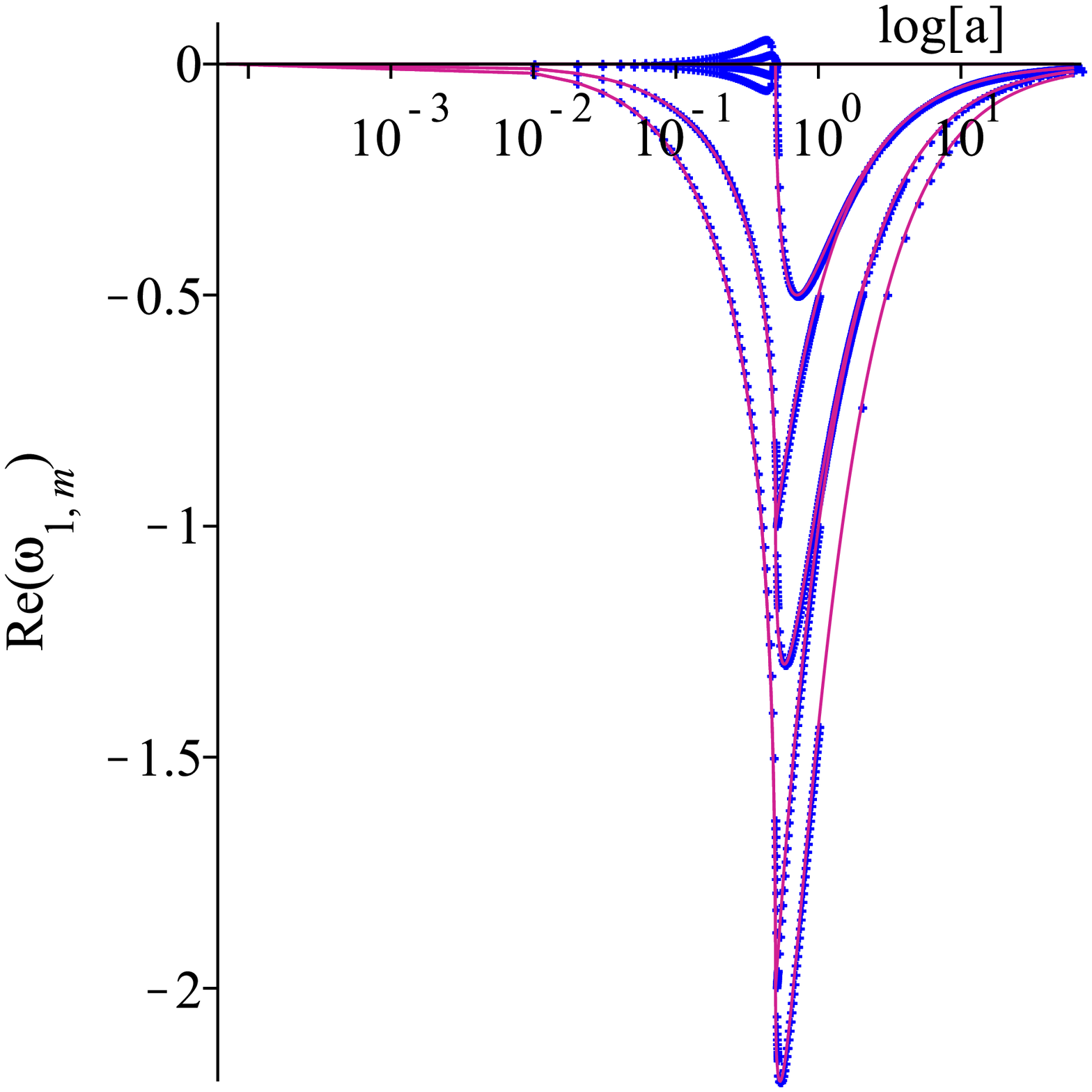} &
\includegraphics[scale=0.35,keepaspectratio]{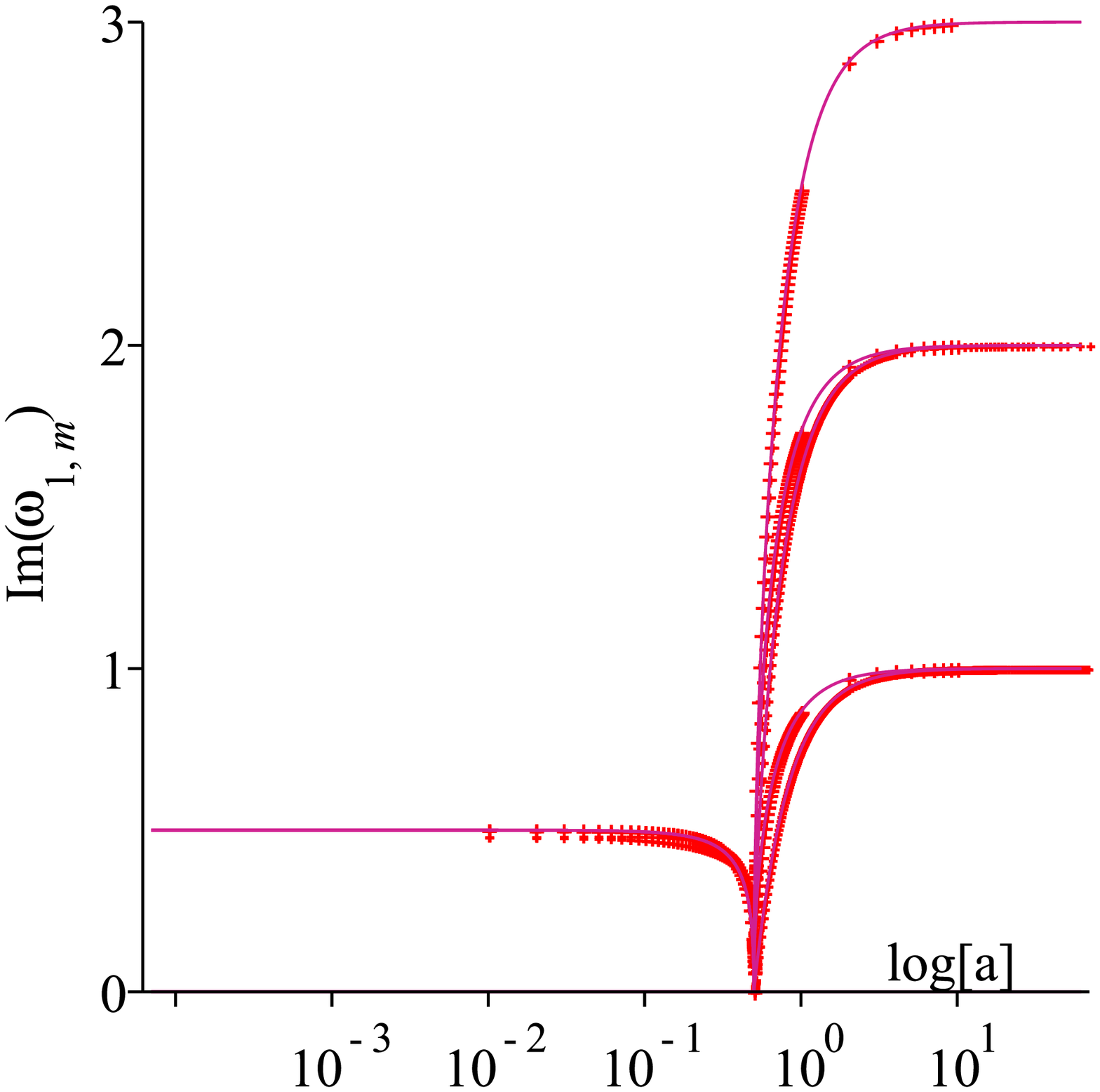}
\end{array}$
\caption{Comparison between our numerical results plotted with blue
(red) crosses and the analytical formula (violet lines) in the cases
$N=0,1$. The excellent fit is spoiled only for $m=0$, $a<0.5$}
\label{a:all}
\end{figure*}

Here, the formula is written in terms of angular velocity $\Omega_{+}$ and bifurcation parameter $b=M/a$
since these are more relevant to the problem at hand. The fit can be seen on Fig. \ref{a:all}.

Clearly this formula offers a very good fit to the data, except for
the case $m = 0$, $b > 1$ where it gives zeros, differing from our
numerical results. If we compare all other relevant points produced
in our numerical evaluations (i.e., the points for the two lowest $n
= 0, 1$ modes for $b > 1$ corresponding to $N = 0, 1$ in Eq.
\eqref{omega} and points for the only two modes available for $b <
1$, which corresponds to $N = 0, 1$ in Eq. \eqref{omega}, for $m =
0, \pm 1 , \pm 2$) with the values calculated according to formula
Eq. \eqref{omega} for the corresponding $a$ and $M$, the two sets
coincide with precision of at least 5 digits in the worst few cases,
and with at least 10 digits for most of the points. Usually, there
are some deviations for small $a$ ($b>>1$). The case $n=0$ for $a<M$
is not plotted, because numerically the real parts of $n=0$ and
$n=1$ coincide for $0.1<a<M$ , while the imaginary part of $n=0$ is
very small -- thus, the $n=0$ mode clearly obeys Eq. \eqref{omega}
for $N=0$. Taking into account that: 1) This exact formula is
analytically obtained from the properties of the confluent Heun
functions in the case of polynomial condition imposed on the {\em
the TRE} and not on the TAE;  and 2) The routines with which
\textsc{maple} evaluates  the Heun function, as well as their
precision are unknown, this match is extremely encouraging. Besides,
it poses once more the question about the physics behind Eq.
\eqref{omega}.

Eq. \eqref{omega} appears without derivation also in
\\ \citeauthor{Berti03} and \citeauthor{Berti_Kokkotas}, where the
authors imposed QNM boundary conditions meaning they worked with the
{\em regular} solutions of the TAE. In their paper, Eq.
\eqref{omega} was considered as a linear approximation to some
unknown nonlinear with respect to $\Omega_{+}$ relation. It is
written in terms of the temperature of the horizon, and according to
the authors, it fits very well the frequencies they obtained for the
QNM case $l=m=2 \quad (=\mid s \mid)$ and also it matches the
imaginary parts of the QNM ($n=1,2,3...$) when $m>0$ (for the other
cases it does not work).

In contrast, in our numerical results, Eq. \eqref{omega} describes
the lowest modes ($n = 0,1$) for all values of $m$ equally well
(except $m = 0, a < M$), while it fails to describe the modes with
$n > 1$ -- Eq.\eqref{omega}, for $N =n$ or $N = n + 1$ matches only
the relative magnitude of the modes and has no resemblance to their
highly non-trivial behavior. The fit does not improve even for the
highest mode ($n = 20$) we were able to obtain. Surprising
similarity between the two cases (primary jets and QNM) is that the real
parts of the higher modes ($n > 1$) in our results also seem to tend
to the lowest mode $\Re(\omega_{0,m})=\Re(\omega_{1,m}) =
-m\Omega_{+}$ . However, since in our case the frequencies come in
pairs, the frequencies with opposite to $\Omega_{+}$ sign do not
tend to $\Omega_{+}$  but they still seem to have a definite limit.

Despite the differences, the appearance of Eq. \eqref{omega} in the
two cases corresponding to different physical problems is
interesting and raises questions about the underlying physics it
represents. It also contributes to the validity of our approach,
since our results are comparable in some points with
\citeauthor{Berti03} who use a completely different but well
established in QNM calculations numerical approach (Leaver's method
of continued fractions).

Finally, it is worth recalling that only part of the modes with
$n>1$, $m\neq0$ presented here satisfy BHBC and represent
perturbations ingoing into the outer horizon $r_{+}$, thus
describing a perturbation of the KBH. The modes that violate BHBC
(on Fig. \ref{m1sn} between $\Re(\omega)=0$ and $\Re(\omega_{1,-1})$
-- the magenta line) describe perturbations outgoing from the
horizon, thus describing a perturbation of \textit{white hole}. It
is clearly interesting that the solutions of the equation $R_2(r)$
evolve with the rotation $a$ in a way that they first describe
ingoing waves and later describe outgoing waves, with some hints
that they will tend to $\Omega_{+}$, but it is impossible to make a
conclusion based on current numerical algorithms and without further
study of the whole white hole case. Although this is beyond the
scope of current article, it clearly demonstrates the complexity of
our approach and the richness of the possible results.

\section{Conclusion}

In this article, we presented new results of the numerical studies of our
model of primary jets of radiation from the central engine inspired by the properties of GRB.
We have already showed that the polynomial requirement imposed on the solutions to the TAE leads to collimated
jet-like shapes observed in the angular part of the solution, thus clearly offering a natural mechanism for
collimation due to purely gravitational effects (\citeauthor{Fiziev2010a}).

Continuing with the TRE, we imposed standard black hole
boundary conditions and we obtained highly nontrivial spectra
$\omega_{n,m}(a)$ for a collimated electromagnetic outflow from both the KBH and the KNS.
These novel spectra give a new theoretical basis for examination of the presence of the Kerr black hole
or the Kerr naked singularity in the central engine of different observed astrophysical objects.

We gave a detailed description of the qualitative change of the behavior of primary  jet
spectra under the transition from the KBH to the KNS and showed that this transition
can be considered as bifurcation of the family of ergo-surfaces of the Kerr metric.

Numerical investigation of the dependence of these primary
jet spectra on the rotation of the Kerr metric
is presented and discussed for the first time.
It is based on the study of the corresponding novel primary jet-boundary-problem formulated in the present article.
Here we use the exact solutions of the Teukolsky Master Equation for electromagnetic perturbations
of the Kerr metric in terms of confluent Heun's function.

The numerical spectra for the two lowest modes $n = 0,1$ (with some
exceptions) is best described by the exact analytical formula in
\citeauthor{Fiziev0908.4234}, though derived by imposing the polynomial
conditions on the solutions to the TRE instead to the TAE. Although
unexpected, this fit serves as confirmation of both our numerical
approach and the analytical result, and it is a hint for deeper
physics at work in those cases. Interestingly, the formula in
question, Eq. \eqref{omega}, works only for the lowest modes ($n =
0,1$) for {\em each} $m$ and correctly describes two frequencies
with coinciding for $a < M$ real parts and different imaginary parts
(zero and comparable to the real part, respectively). The modes with
higher $n$ are not described by it and have highly non-trivial
behavior -- they seem to have a definite limit which numerically
coincides with the angular velocity of the event horizon (and
critical frequency of QNM superradiance) in some of the cases.

The obtained primary jet spectra seem to describe
linearly stable electromagnetic primary jets from both the KBH and the KNS.
Indeed, the imaginary part of the complex primary jet
frequencies in our numerical calculations remains positive,
ensuring stability of the solutions in direction of time-future infinity and indicating
an explosion in the direction of time-past infinity.
This is surprising, since according to the theory of QNM of the KNS,
the naked-singularity regime should be unstable in time future.
There is no contradiction. The different conditions on the solutions to the TAE we use,
define a different physical situation -- we work with primary
jets from the KBH or the KNS, not with QNM,
and our numerical results suggest that these primary jets are stable even for the KNS.

We showed that the value $b=M/a=1$ describes bifurcation of ergo-surfaces of the Kerr metric.
Around this bifurcation point the imaginary part of the
primary jet frequency decreases to zero suggesting slower damping with time.
This could have interesting implications (including the loss of primary jet stability)
for rotating black holes
or naked singularities close to the extremal regime.
Besides, for $a>>M$ (i.e. $b\to 0$), or $a<<M$ (i.e. $b\to\infty$) the imaginary part of the
primary jet frequencies
remains approximately constant with respect to the bifurcation parameter $b$, being positive in both directions.

Although very simple, our model was able to produce interesting results whose
physical interpretation and applications require additional consideration.
It also demonstrates the advantage of the use of confluent Heun functions
and their properties in the physical problem at hand.

Whether this mechanism of collimation is working in GRBs, AGNs, quazars and other objects can be understood only
trough comparing the spectra of their real jets and our "primary jet spectra"
taking into account also the interference from the jet environment and from the different emission mechanisms.
For the real observation of the predicted here spectra is important the intensity of their lines. 
It depends on the power of the primary jet, created by the central engine, which is still an open problem. 
If there exist a natural gravitational mechanism of creation of very powerful primary jets, 
that jets may play the main role in observed astrophysical jets. 
Otherwise one can expect to see some traces of these primary jet spectra in the spectra of astrophysical jets, 
similar to the traces of standard QNM spectra in the exact numerical solutions of Einstein equations.          

It is remarkable that similar effects of specific collimation of flows of matter particles, described by
geodesics in rotating metrics like Kerr one are discovered independently of our study of specific jet-like
solutions of Teukolsky Master Equation. See the recent articles: \citeauthor{geodesics1} \citeyear{geodesics1}, \citeauthor{geodesics2} \citeyear{geodesics2} and the references therein.
The proper combination of these novel effects of collimation for both radiation and matter outflows from compact
objects may be important ingredient of the future theory of real astrophysical jets.

The approach based on the Teukolsky Master Equation seems to give more general picture of the collimation phenomena,
at least in the framework of perturbation theory of wave fields of different spin in Kerr metric.
In addition, our approach gives direct suggestions for observations, since the frequencies,
we have discussed in the present paper, in principle may be found in spectra of the real objects like jets
from GRBs, AGNs, quazars, e.t.c.
Their observation will present indisputable direct evidences for the real nature of the central engine,
since the corresponding spectra can be considered as fingerprints of that still mysterious object.

\section{Acknowledgements}
This article was supported by the Foundation "Theoretical and
Computational Physics and Astrophysics" and by the Bulgarian National Scientific
Fund
under contracts DO-1-872, DO-1-895 and DO-02-136.

The authors are thankful to the Bogolubov Laboratory of Theoretical Physics,
JINR, Dubna, Russia for the hospitality and good working conditions.
Essential part of the results was reported and discussed there during
the stay of the authors in the summer of 2009.

D.S. is deeply indebted to the leadership of JINR, for invaluable help to overcome consequences
of the incident during the School on Modern Mathematical Physics, July 20-29, 2009, Dubna, Russia.

\section{Author Contributions}
P.F. posed the problem, supervised the project and is responsible for the new theoretical developments and interpretations of the results.
D.S. is responsible for the highly nontrivial and time consuming numerical calculations and production of the numerous
computer graphs, as well as for the analysis of the results and their interpretation. Both authors discussed the results and their implications and commented at all stages. The manuscript was prepared by D.S. and edited and expanded by P.F..
Both authors developed numerical methods and their improvements, including testing of the new versions of the
\textsc{maple} package in collaboration with Maplesoft Company.

\makeatletter
\let\clear@thebibliography@page=\relax
\makeatother

\clearpage


\begin{thebibliography}{}
\bibitem[\protect\citeauthoryear{{Andersson}}{(1992)}]{QNM2} Andersson,
~N., {\em A numerically accurate investigation of black-hole normal modes},
Proc. Roy. Soc. London A\textbf{439} no.1905: 47-58 (1992)
%
\bibitem[\protect\citeauthoryear{{Antonelli et al. (2009)}}{2009}]{Antonelli}
Antonelli ~L.~A., Avanzo ~P.~D., Perna ~R. et al.,
{\em GRB090426: the farthest short gamma-ray burst?},
\\arXiv:0911.0046v1 [astro-ph.HE] (2009)
%
\bibitem[\protect\citeauthoryear{{Bardeen et al. (1972)}}{1972}]{Bardeen} Bardeen,~J.~M., Press ~W.~ H., Teukolsky ~S.~A.{\em Rotating Black Holes: Locally Nonrotating Frames, Energy Extraction, and Scalar Synchrotron Radiation}, ApJ, {\bf 178} , pp. 347-370 (1972);

%
\bibitem[\protect\citeauthoryear{{Barkov\& Komissarov (2008a)}}{(2008a)}]{Barkov}
Barkov ~M.~V., Komissarov ~S.~S.,{\em Central engines of Gamma Ray Bursts.
Magnetic mechanism in the collapsar model}, arXiv:0809.1402v1 [astro-ph] (2008);
%
\bibitem[\protect\citeauthoryear{{Barkov \& Komissarov (2008b)}}{(2008b)}]{Barkov2}
Barkov ~M.~V., Komissarov ~S.~S., {\em Close Binary Progenitors of Long Gamma
Ray Bursts}, arXiv:0908.0695v1 [astro-ph.HE] (2008)
%
\bibitem[\protect\citeauthoryear{{Berti \& Kokkotas (2003)}}{2003}]{Berti_Kokkotas}
 Berti ~E., Kokkotas ~K. ~D., {\em Asymptotic quasinormal modes of Reissner-Nordstr\"om and Kerr black holes}, Phys.Rev.D \textbf{68}
044027 (2003)
%
\bibitem[\protect\citeauthoryear{{Berti et al. (2003)}}{2003}]{Berti03} Berti
~E., Cardoso ~V., Kokkotas ~K. ~D. and Onozawa ~H., {\em Highly damped quasinormal modes of Kerr black holes}, Phys.Rev. D
\textbf{68} (2003) 124018 , arXiv:0307013v2 [hep-th] (2003)
%
\bibitem[\protect\citeauthoryear{{Berti (2004)}}{2004}]{QNM7} Berti E.,
%
{\em Black hole quasinormal modes: hints of quantum gravity?}, arXiv:0411025
[gr-qc] (2004)
%
\bibitem[\protect\citeauthoryear{{Berti et al. (2004)}}{2004}]{QNM8} Berti
~E., Cardoso ~V., Yoshida ~S.,{\em Highly damped quasinormal modes of Kerr black holes: A complete numerical investigation}, Phys.Rev. D {\bf 69}, 124018 (2004) 
%
\bibitem[\protect\citeauthoryear{{Berti et al.}}{(2009)}]{Berti09} Berti E., Cardoso V., Starinets A. O.,
{\em Quasinormal modes of black holes and black branes}, Class. Quant. Grav. \textbf{26} (2009) 163001, gr-qc/0905.2975 (2009)
%
\bibitem[\protect\citeauthoryear{{Blandford \& Znajek}}{(1977)}]{Blandford-Znajek}
 Blandford ~R.~D., Znajek ~R.~L., {\em Electromagnetic extraction of energy from Kerr black holes},  MNRAS {\bf
179}:
433 (1977)
%
\bibitem[\protect\citeauthoryear{{Blandford (2001)}}{2001}]{Blandford} Blandford
~R.~D.,
{\em Black Holes and Relativistic Jets}, arXiv:0110394 [astro-ph]
(2001)
%
\bibitem[\protect\citeauthoryear{{Borissov \& Fiziev}}{2009}]{RBPF}
Borissov ~R.~S., Fiziev ~P.~P., {\em Exact Solutions of Teukolsky Master
Equation with Continuous Spectrum}, arXiv: 0903.3617 [gr-qc] (2009)
%
\bibitem[\protect\citeauthoryear{{Burrows (2005)}}{2005}]{Burrows} Burrows
~D.~N.,Romano. ~P., Falcone ~A. et al., {\em Bright X-ray Flares in Gamma-Ray
Burst Afterglows} ,
Science, {\bf 309}1833-1835 (2005)
%
\bibitem[\protect\citeauthoryear{{Cardoso et al. (2004)}}{2004}]{superradiance8}
Cardoso ~V., Dias ~O.~J., Lemos ~J.~P., Yoshida ~S., {\em
    Black-hole bomb and superradiant instabilities}, Phys. Rev. D {\bf
70},  044039 (2004) 
%
\bibitem[\protect\citeauthoryear{{Chandrasekhar \!\!\& \!\!Detweiler}}{(1975)}]{QNM}
    Chandrasekhar ~S., Detweiler ~S., {\em The quasi-normal modes of the Schwarzschild black hole},  Proc. Roy. Soc.  London A{\bf 344}:
441-452
(1975)
%
\bibitem[\protect\citeauthoryear{{Chandrasekhar (1983)}}{1983}]{Chandrasekhar}
    Chandrasekhar ~S., {\em The mathematical theory of black holes}, , Clarendon Press/Oxford University Press (International Series of Monographs on Physics. Volume 69), (1983)

\bibitem[\protect\citeauthoryear{{Chicone \& Mashhoon}}{(2010)}]{geodesics1}
    Chicone ~C., Mashhoon ~B.  {\em Gravitomagnetic Jets}, \\ arXiv:1005.1420 (2010)
%
\bibitem[\protect\citeauthoryear{{Chicone et al.}}{(2005)}]{geodesics2}
    Chicone ~C., Mashhoon ~B., Punsly ~B., {\em Relativistic Motion of Spinning Particles in a Gravitational Field}, Phys.Lett. A {\bf 343} (2005) 1-7  arXiv:0504146 [gr-qc] (2005)
%
\bibitem[\protect\citeauthoryear{{Chirenti \& Rezzolla}}{(2007)}]{Chirenti_a}
    Chirenti ~C.~B.~M.~H., Rezzolla ~L., {\em How to tell a gravastar from a black hole}, CQG:{\bf 24},I.16: pp. 4191-4206;    arXiv:0706.1513v2 [gr-qc] (2007)
%
\bibitem[\protect\citeauthoryear{{Chirenti \& Rezzolla}}{(2008)}]{Chirenti_b}
Chirenti ~C.~B.~M.~H., Rezzolla ~L.,{\em Ergoregion instability in rotating gravastars}, PRD,:{\bf 78}, 080411;    arXiv:0808.4080v1 [gr-qc] (2008)
%
\bibitem[\protect\citeauthoryear{{Dado \& Dar (2009)}}{2009}]{cannonball}  Dado ~S., Dar ~A.,
{\em The cannonball model of long GRBs - overview}, AIP Conference Proceeding
1111: 333-343 (2009), arXiv:0901.4260
[astro-ph] (2009)
%
\bibitem[\protect\citeauthoryear{{ da Silva de Souza \& Opher (2009)}}{2009}]{BZ_field} da Silva de Souza ~R., Opher ~R., {\em Origin of $10^{15}-10^{16}$G Magnetic Fields in the Central Engine of Gamma Ray Bursts},   arXiv:0910.5258v1 [astro-ph.HE], (2009)
%
\bibitem[\protect\citeauthoryear{{Detweiler (1980)}}{1980}]{Detweiler} Detweiler ~S., {\em
    Black holes and gravitational waves. III - The resonant frequencies of rotating holes}, ApJ:{\bf 239}, 292-295, (1980)
%
\bibitem[\protect\citeauthoryear{{Dotti et al. (2008)}}{2008}]{Dotti}   Dotti ~G., Gleiser ~R.~J.,Ranea-Sandoval ~I.~F., Vucetich ~H., {\em Gravitational instabilities in Kerr spacetimes}, CQG:{\bf 25}, pp. 245012 (2008),    arXiv:0805.4306v3 [gr-qc] (2008)
%
\bibitem[\protect\citeauthoryear{{Dreyer et al. (2004)}}{2004}]{Dreyer} Dreyer ~O., Kelly ~B., Krishnan ~B., Finn ~L.~S., Garrison~D., Lopez-Aleman~R., {\em Black-hole spectroscopy: testing general relativity through gravitational-wave observations}, Class. Quantum Grav. {\bf 21} (2004) 787–803 , arXiv:0309007v1 [gr-qc] (2004)
%
\bibitem[\protect\citeauthoryear{{Evans et al.}}{(2007)}]{Evans}
    Evans ~P.~A., Beardmore ~A.~P., Page ~K.~L. et al., {\em An online repository of Swift/XRT light curves of ³-ray bursts}, A\&A {\bf 469}: 379-385 (2007),   arXiv:0704.0128v2 [astro-ph] (2007) 
%
\bibitem[\protect\citeauthoryear{{Fan (2009)}}{2009}]{Fan} Fan~Y-Z., {\em The spectrum of Gamma-ray Burst: a clue},     arXiv:0912.1884v1 [astro-ph.CO], (2009)
%
\bibitem[\protect\citeauthoryear{{Fermi/LAT Collaboration (2009)}}{2009}]{Fermi}
Fermi/LAT Collaboration, Ghisellini ~G.,  Maraschi ~L., Tavecchio ~F., 2009 {\em
Radio-Loud Narrow-Line Seyfert 1 as a New Class of Gamma-Ray AGN},
arXiv:0911.3485v1 [astro-ph.HE] (2009)
%
\bibitem[\protect\citeauthoryear{{Ferrari (1996)}}{1996}]{QNM3} Ferrari ~V.,  in
{\em Proc. of 7-th Marcel Grossmann Meeting},
                    ed Ruffini~R and Kaiser~M ( Singapoore: World Scientific),
World Scientific Publishing Co Pte Ltd: 537-559 (1996)
%
\bibitem[\protect\citeauthoryear{{Ferrari \& Gualtieri (2008)}}{2008}]{QNM11}
Ferrari~V., Gualtieri L., {\em Quasi-normal modes and gravitational wave astronomy}, Gen.Rel.Grav.{\bf 40}: 945-970 (2008), arXiv:0709.0657v2 [gr-qc] (2007)
%
\bibitem[\protect\citeauthoryear{{Ferrari (1997)}}{1997}]{QNM4} Ferrari~V., in
  {\em Black Holes and Relativistic Stars} ed R. Wald  The University of Chicago
Press: 1-23 (1997)
%
\bibitem[\protect\citeauthoryear{{Fiziev (2006)}}{2006}]{QNM9} Fiziev ~P.~P., {\em On the Exact Solutions of the Regge-Wheeler Equation in the Schwarzschild Black Hole Interior},
Class. Quant. Grav. {\bf 23}: 2447-2468 (2006), arXiv:0603003v4 [gr-qc]
%
\bibitem[\protect\citeauthoryear{{Fiziev (2007)}}{2007}]{QNM10} Fiziev
~P.~P., {\em Exact solutions of Regge-Wheeler equation}, Jour.  Phys. Conf. Ser. {\bf 66} 012016 (2007a), arXiv:0702014 [gr-qc]
%
\bibitem[\protect\citeauthoryear{{Fiziev (2009a)}}{2009a}]{Fiziev0902.1277}
Fiziev~P.~P., 2009 {\em Classes of Exact Solutions to Regge-Wheeler and
Teukolsky Equations}, arXiv:0902.1277 [gr-qc] (2009a)
%
\bibitem[\protect\citeauthoryear{{Fiziev (2009b)}}{(2009b)}]{Fiziev0904.0245}
Fiziev~P.~P., {\em Novel relations and new properties of confluent Heun's
functions and their derivatives of arbitrary order}, J. Phys. A: Math. Theor. {\bf 43} 035203 (2010),
arXiv:0904.0245 [gr-qc] (2009b)
%
\bibitem[\protect\citeauthoryear{{Fiziev (2009c)}}{2009c}]{Fiziev0906.5108}
Fiziev~P.~P., {\em Teukolsky-Starobinsky Identities - a Novel Derivation and
Generalizations}, Phys. Rev. D {\bf 80} 124001 (2009), arXiv:0906.5108 [gr-qc]
(2009c)
%
\bibitem[\protect\citeauthoryear{{Fiziev (2009d)}}{2009d}]{Fiziev0908.4234}
Fiziev~P.~P., {Classes of Exact Solutions to the Teukolsky Master Equation}, Class. Quant. Grav.{\bf 27}:135001, 2010, arXiv:0908.4234 [gr-qc] (2009d)
%
\bibitem[\protect\citeauthoryear{{Fiziev (2010a)}}{2010a}]{Fiziev2010a}
Fiziev~P.~P., {To the theory of astrophysical relativistic jets},
to be published (2010a)
%
\bibitem[\protect\citeauthoryear{{Fiziev \& Staicova (2009a)}}{2009a}]{FS1} Fiziev~P.~P.,
Staicova~D.~R., {\em A new model of the Central Engine of GRB and the Cosmic
Jets}, Bulg. Astrophys. Jour. \textbf{11}: 3,  arXiv:0902.2408  [atro-ph.HE] (2009)
%
\bibitem[\protect\citeauthoryear{{Fiziev \& Staicova (2009b)}}{2009b}]{FS2} Fiziev~P.~P.,
Staicova~D.~R., {\em Toward a New Model of the Central Engine of GRB}, Bulg. Astrophys. Jour.
\textbf{11}: 13,  arXiv:0902.2411 [astro-ph.HE] (2009)
%
\bibitem[\protect\citeauthoryear{{Foucart et al. (2010)}}{(2010)}]{1007.4203} Foucart ~F., Duez ~M. ~D., Kidder ~L.~E., Teukolsky ~S.~A., {\em Black hole-neutron star mergers: effects of the orientation of the black hole spin}, 	arXiv:1007.4203v1 [astro-ph.HE] (2010)
%
\bibitem[\protect\citeauthoryear{{Ghirlanda et al. (2009)}}{2009}]{short_long}
Ghirlanda ~G., Nava ~L., Ghisellini1  ~G., Celotti ~A.,  Firmani ~C.,
%
{\em Short versus Long Gamma-Ray Bursts: spectra, energetics, and luminosities
(2009)}, A \& A {\bf 496} Issue 3: 585-595 2009, arXiv:0902.0983 [astro-ph]
(2009)

\bibitem[\protect\citeauthoryear{{Granot (2006)}}{2006}]{jets} Granot~J.,
{\em Structure \& Dynamics of GRB Jets}, Talk given at the Conference
                                       "Challenges in Relativistic Jets"
Cracow, Poland, June 27, 2006,



http://www.oa.uj.edu.pl/2006jets/talks.html (2006)
%
\bibitem[\protect\citeauthoryear{{Kerr (1963)}}{1963}]{Kerr} Kerr ~R., {\em Gravitational Field of a Spinning Mass as an Example of Algebraically Special Metrics}, Phys.
Rev. L.,
{\bf 11}: 237-238 (1963)
%
\bibitem[\protect\citeauthoryear{{Kodama (2008)} }{2008}]{superradiance9} Kodama
~H., {\em Superradiance and Instability of Black Holes}, Prog. Theor. Phys. Suppl. {\bf 172} 11-20 (2008), \\   arXiv:0711.4184v2 [hep-th]
%
\bibitem[\protect\citeauthoryear{{Kokkotas \& Schmidt}}{(1999)}]{QNM5}
Kokkotas ~K.
~K., and Schmidt ~B. ~G., {\em Quasi-Normal Modes of Stars and Black Holes}, Living Rev. Relativity {\bf 2}: 2 (1999), arXiv:9909058v1 [gr-qc]
%
\bibitem[\protect\citeauthoryear{{Komissarov \!\&\! Barkov}}{(2009)}]{Komissarov}
Komissarov ~S.~S., Barkov ~M.~V.,{\em Activation of the Blandford-Znajek
mechanism in collapsing stars}, submitted to MNRAS, arXiv:0902.2881v4
[astro-ph.HE] (2009)
%
\bibitem[\protect\citeauthoryear{{K\"{o}enigl (2006)}}{2006}]{jets_d}
K\"{o}enigl~A., 2006 {\em Jet Launching - General Review} Talk given at the
Conference "Challenges in Relativistic Jets"
                            Cracow, Poland, June 27, 2006,

 http://www.oa.uj.edu.pl/2006jets/talks.html (2006)
%
\bibitem[\protect\citeauthoryear{{Komissarov (2008)}}{2008}]{Blandford-Znajek_d}
Komissarov~S.~S., 2008 {\em Blandford-Znajek mechanism versus Penrose process},
arXiv:0804.1912 [astro-ph] (2008)
%
\bibitem[\protect\citeauthoryear{{Krolik \!\&\! Hawley}}{(2009)}]{Krolik} Krolik
~J.~H., Hawley ~J. {\em General Relativistic MHD Jets}, arXiv:0909.2580v1
[astro-ph.HE] (2009)
%
\bibitem[\protect\citeauthoryear{{Leaver (1985)}}{1985}]{QNM1} Leaver~E.~W., {\em An analytic representation for the quasi-normal modes of Kerr black holes},
Proc. Roy. Soc. London A{\bf 402}: 285-298 (1985)
%
\bibitem[\protect\citeauthoryear{{Lee et al. (1999)}}{1999}]{Lee} Lee
~H.~K.,Wijers ~R.~A.~M.~J., Brown ~G.~E.{\em Blandford-Znajek process as a gamma
ray burst central engine}, ASP Conference Series,{\bf 190}, ed. Poutanen ~J. and
Svensson ~R:173,(1999), arXiv:9905373 [astro-ph] (1999)
%
\bibitem[\protect\citeauthoryear{{Lei et al.}}{(2008)}]{Lei} Lei
~W-H., Wang ~D-X., Zou ~Y-C., Zhang ~L., {\em Hyperaccretion after the
Blandford-Znajek Process: a New Model for GRBs with X-Ray Flares Observed in
Early Afterglows}, Chin. J. Astron. Astrophys. {\bf 8}: 404-410, arXiv:0802.0419
[astro-ph] (2008)
%
\bibitem[\protect\citeauthoryear{{LIGO Collaboration \& Hurley (2007)}}{2007}]{Andromeda}
    LIGO Scientific Collaboration, Hurley ~K., {\em Implications for the Origin of GRB 070201 from LIGO Observation}, ApJ {\bf 681}:1419-1428, 2008; arXiv:0711.1163v2 [astro-ph.HE] (2007)
%
\bibitem[\protect\citeauthoryear{{LIGO Collaboration \& Virgo Collaboration (2010)}}{2010}]{LIGO}
    LIGO Scientific Collaboration, Virgo Collaboration, {\em Search for gravitational-wave inspiral signals associated with short Gamma-Ray Bursts during LIGO's fifth and Virgo's first science run},  arXiv:1001.0165v1 [astro-ph.HE] (2010)
%
\bibitem[\protect\citeauthoryear{{Livio (2004)}}{2004}]{jets1}
    Livio ~M., {\em Astrophysical Jets Baltic Astronomy}, {\bf 13}: 273-279
(2004)
%
\bibitem[\protect\citeauthoryear{{L\"u et al. (2010)}}{2010}]{Lv} Lv ~H., Liang ~E., Zhang ~B. , Zhang ~B.,  {\em A New Classification Method for Gamma-Ray Bursts}, arXiv:1001.0598v1 [astro-ph.HE] (2010)
%
\bibitem[\protect\citeauthoryear{{Lyutikov (2009)}}{2009}]{Lyutikov} Lyutikov
~M., {\em Gamma Ray Bursts: back to the blackboard}, arXiv:0911.0349v2
[astro-ph.HE] (2009)
%
%
\bibitem[\protect\citeauthoryear{{Mao et al. (2010)}}{(2010)}]{1008.4899} Mao ~Z., Yu ~Y.-W., Dai ~Z.~G., Pi ~C. ~M. , Zheng ~X. ~P., {\em The termination shock of a magnetar wind: a possible origin of gamma-ray burst X-ray afterglow emission}, 	Astronomy and Astrophysics {\bf 518}, A27 (2010)		 arXiv:1008.4899v1 [astro-ph.HE] (2010)
%
\bibitem[\protect\citeauthoryear{{Marshall et al. (2006)}}{2006}]{jets_c}
Marshall~H.~L., Lopez~L.~A., Canizares~C.~R., Schulz N. S., Kane J. F., 2006
{\em Relativistic Jets from the
                            Black Hole in SS 433}, Talk given at the Conference
"Challenges in Relativistic Jets"
                            Cracow, Poland, June 27, 2006,

http://www.oa.uj.edu.pl/2006jets/talks.html (2006) 
%
\bibitem[\protect\citeauthoryear{{Maxham \& Zhang (2009)}}{2009}]{Maxham} Maxham
~A., Zhang ~B., {\em Modeling Gamma-Ray Burst X-Ray Flares within the Internal
Shock Model},
arXiv:0911.0707 [astro-ph.HE] (2009)
%
\bibitem[\protect\citeauthoryear{{Meliani \& Keppens (2010)}}{(2010)}]{1009.1224} Meliani ~Z., Keppens ~R., {\em Dynamics and stability of relativistic GRB blast waves}, arXiv:1009.1224v1 [astro-ph.HE] (2010)
%
\bibitem[\protect\citeauthoryear{{M\'{e}sz\'{a}ros (2001)}}{2001}]{Meszaros}
M\'{e}sz\'{a}ros ~P., {\em Gamma-Ray Bursts: Accumulating Afterglow Implications, Progenitor Clues, and Prospects}, Science {\bf 291}: 79-84 (2001), arXiv:0102255 [astro-ph]
%
\bibitem[\protect\citeauthoryear{{Mirabal et al. (2006)}}{2006}]{Mirabal}
Mirabal~N., Halpern~J.~P., An~D., Thorstensen ~J.~R., Terndrup ~D.~ M.,  {\em GRB 060218/SN 2006aj: A Gamma-Ray Burst and Prompt Supernova at z = 0.0335}, ApJ {\bf 643}: L99-L102 (2006)
%
\bibitem[\protect\citeauthoryear{{Mundt et al. (2009)}}{2009}]{binary} Mundt~R., Hamilton ~C.~M., Herbst ~W., Johns-Krull ~C.~M., Winn ~J.N., {\em Bipolar jets produced by a spectroscopic binary},    arXiv:0912.1740v1 [astro-ph.SR] (2009)
%
\bibitem[\protect\citeauthoryear{{Nagataki (2009)}}{2009}]{BZ process} Nagataki
~S.,{\em Development of General Relativistic Magnetohydrodynamic Code and its
Application to Central Engine of Long Gamma-Ray Bursts}, ApJ {\bf
704}:937-950, 2009; arXiv:0902.1908 [astro-ph.HE] (2009)
%
\bibitem[\protect\citeauthoryear{{Nollert (1999)}}{1999}]{QNM6} Nollert ~H-P., {\em TOPICAL REVIEW: Quasinormal modes: the characteristic `sound' of black holes and neutron stars},
Class. Quant. Grav.  {\bf 16}: R159-R216 (1999)
%
\bibitem[\protect\citeauthoryear{{Nysewander et al. (2009)}}{2009}]{GRB_two}
Nysewander ~M., Fruchter ~A.~S., Pe'er. ~A., {\em A Comparison of the Afterglows
of Short- and Long-Duration Gamma-Ray Bursts}, ApJ.{\bf
701}:824-836 (2009), arXiv:0806.3607 [astro-ph] (2009)
%
\bibitem[\protect\citeauthoryear{{Piran (1999)}}{1999}]{Piran}
Piran~T., {\em Gamma-Ray Bursts and the Fireball Model}, Physics Reports {\bf
314}: 575-667 (1999) , arXiv:9810256 [astro-ph] (1999)
%
\bibitem[\protect\citeauthoryear{{Press \& Teukolsky
(1973)}}{1973}]{Press_Teukolsky}  Press~W.~H., Teukolsly~S.~A., {\em Perturbations of a Rotating Black Hole. II. Dynamical Stability of the Kerr Metric},ApJ {\bf 185}:
649-673 (1973) 
%
\bibitem[\protect\citeauthoryear{{Press \& Teukolsky}}{(1972)}]{superradiance4} Press~W.~H., Teukolsky~S.~A., {\em Floating Orbits, Superradiant Scattering and the Black-hole Bomb}, Nature {\bf
238},Issue 5361: 211-212 (1972)
%
\bibitem[\protect\citeauthoryear{{Punsly \& Coroniti
(1990)}}{1990}]{Blandford-Znajek_b} Punsly ~B., Coroniti ~F.~V., {\em Ergosphere-driven winds}, ApJ {\bf 354}
: 583-615 (1990) 
%
\bibitem[\protect\citeauthoryear{{Rezzolla et al. (2010)}}{(2010)}]{1001.3074} Rezzolla ~L., Baiotti ~L.,  Giacomazzo ~B., Link ~D., Font ~J.~A., {\em Accurate evolutions of unequal-mass neutron-star binaries: properties of the torus and short GRB engines}, Class. Quantum Grav. {\bf 27} 114105 , arXiv:1001.3074v2 [gr-qc] (2010)
%
\bibitem[\protect\citeauthoryear{{\v{R}ipa et al. (2009)}}{2009}]{short_long_b}
\v{R}ipa~J.,M\'{e}sz\'{a}ros~A., Wigger~C., Huja~D, Hudec~R., Hajdas~W.,
                {\em Search for Gamma-Ray Burst Classes with the
RHESSI Satellite}, A \&A {\bf 498}: 399-406, arXiv:0902.1644 [astro-ph.HE]
(2009)
%
\bibitem[\protect\citeauthoryear{{ Shen et al. }}{(2009)}]{jets_t} Shen~R., Kumar ~P., Piran~T., {\em The late jet in gamma-ray bursts and its interactions with a supernova ejecta and a cocoon}  arXiv:0910.5727v3 [astro-ph.HE], (2009)
%
\bibitem[\protect\citeauthoryear{{Shahmoradi \& Nemiroff (2009)}}{2009}]{short_long_c} Shahmoradi ~A., Nemiroff~R.J.,{\em Hardness as a Spectral Peak Estimator for Gamma-Ray Bursts},  arXiv:0912.2148v1 [astro-ph.HE], (2009)
%
\bibitem[\protect\citeauthoryear{{Slavyanov \& Lay (2000)}}{2000}]{slav} Slavyanov and W.Lay, {\em Special Functions, A Uniﬁed Theory Based on Singularities}, Oxford Mathematical Monographs, New York (2000)
%
\bibitem[\protect\citeauthoryear{{Starobinskiy (1973)}}{1973}]{superradiance5}
Starobinskiy~A.~A., {\em Amplification of waves during reflection from a rotating "black hole"
}, Sov. Phys. JETP {\bf 64}: 49 (1973)
%
\bibitem[\protect\citeauthoryear{{Starobinskiy \& Churilov
(1973)}}{1973}]{superradiance6} Starobinskiy~A.~A., Churilov S. M., {\em Amplification of electromagnetic and gravitational waves scattered by a rotating black hole},1973 Sov.
Phys. JETP {\bf 65}: 3-11 (1973)
%
\bibitem[\protect\citeauthoryear{{Takeuchi et al. (2010)}}{(2010)}]{1009.0161} Takeuchi~S., Ohsuga ~K.,  Mineshige ~S., {\em A Novel Jet Model: Magnetically Collimated, Radiation-Pressure Driven Jet},	arXiv:1009.0161v1 [astro-ph.HE] (2010)
%
\bibitem[\protect\citeauthoryear{{Teukolsky (1972)}}{1972}]{Teukolsky}
Teukolsly~S.~A., {\em Rotating Black Holes: Separable Wave Equations for Gravitational and Electromagnetic Perturbations}, Phys.Rev.Lett. {\bf 29}: 1114-1118 (1972)
%
\bibitem[\protect\citeauthoryear{{Teukosky (1973)}}{1973}]{Teukolsky2}
Teukolsly~S.~A., {\em Perturbations of a Rotating Black Hole. I. Fundamental Equations for Gravitational, Electromagnetic, and Neutrino-Field Perturbations}, ApJ {\bf 185}: 635-648 (1973)
%
\bibitem[\protect\citeauthoryear{{Teukolsky \& Press
(1974)}}{1974}]{Teukolsky_Press}  Teukolsly~S.~A., Press W. H., {\em
    Perturbations of a rotating black hole. III - Interaction of the hole with gravitational and electromagnetic radiation}, ApJ {\bf 193}:
443 (1974)
%
\bibitem[\protect\citeauthoryear{{Vlahakis (2006)}}{2006}]{jets_b}
Vlahakis~N., {\em Jet Driving in GRB Sources},
                 Talk given at the Conference "Challenges in Relativistic Jets"
Cracow, Poland, June 27, 2006,

 http://www.oa.uj.edu.pl/2006jets/talks.html (2006) 
%
\bibitem[\protect\citeauthoryear{{Wald (1974)}}{1974}]{superradiance7} Wald~R.
M., {\em Energy Limits on the Penrose Process}, ApJ {\bf 191}: 231-233 (1974)
%
\bibitem[\protect\citeauthoryear{{Willingale et al. }}{(2009)}]{jets_a} Willingale ~R., Genet ~F., Granot ~J., O'Brien ~P.~T. {\em The spectral-temporal properties of the prompt pulses and rapid decay phase of GRBs},    arXiv:0912.1759v1 [astro-ph.HE], (2009) 
%
\bibitem[\protect\citeauthoryear{{Wheeler (1971)}}{1971}]{superradiance}
Wheeler~J., {\em Study Week on Nuclei of Galaxies}, ed D. J. K. O'Connell, North
Holland, Pontificae Academie Scripta Varia, {\bf 35}: 539 (1971)
%
\bibitem[\protect\citeauthoryear{{Zhang \& M\'{e}sz\'{a}ros
(2002)}}{2002}]{Zhang} Zhang~B.,
M\'{e}sz\'{a}ros~P., {\em Gamma-Ray Bursts with Continuous Energy Injection and Their Afterglow Signature}, ApJ {\bf566}: 712-722 (2002), arXiv:0108402 [astro-ph]
%
\bibitem[\protect\citeauthoryear{{Zel'dovich (1971)}}{1971}]{superradiance2}
Zel'dovich~Ya.~B., {\em Generation of Waves by a Rotating Body}, Sov. Phys. JETP Lett {\bf 14}: 180 (1971)
%
\bibitem[\protect\citeauthoryear{{Zel'dovich (1972)}}{1972}]{superradiance3}
Zel'dovich~Ya.~B., {\em Amplification of Cylindrical Electromagnetic Waves Reflected from a Rotating Body},  Sov. Phys. JETP {\bf 35}: 1085 (1972)
%
\bibitem[\protect\citeauthoryear{{GAS (2007))}}{2007}]{gas} talks given on the
subject:

http://tcpa.uni-sofia.bg/research
%
\end{thebibliography}
\end{document}